\documentclass[12pt]{article}
\usepackage{epsfig}
\usepackage{color}
\usepackage{epsf, cite,amssymb}
\usepackage{ifpdf}

\def\ie{{\it i.e.}}
\def\IZ{\relax\ifmmode\mathchoice
{\hbox{\cmss Z\kern-.4em Z}}{\hbox{\cmss Z\kern-.4em Z}}
{\lower.9pt\hbox{\cmsss Z\kern-.4em Z}} {\lower1.2pt\hbox{\cmsss
Z\kern-.4em Z}}\else{\cmss Z\kern-.4em Z}\fi}
\def\IR{\relax{\rm I\kern-.18em R}}

\def\one{{\hbox{ 1\kern-.8mm l}}}

\newcommand{\N}{{\cal N}}
\newlength{\bredde}
\def\slash#1{\settowidth{\bredde}{$#1$}\ifmmode\,\raisebox{.15ex}{/}
\hspace*{-\bredde} #1\else$\,\raisebox{.15ex}{/}\hspace*{-\bredde}#1$\fi}

\newcommand  {\Rbar} {{\mbox{\rm$\mbox{I}\!\mbox{R}$}}}

\newsavebox{\zzzbar}
\sbox{\zzzbar}
  {\setlength{\unitlength}{0.9em}
  \begin{picture}(0.6,0.7)
  \thinlines
  \put(0,0){\line(1,0){0.6}}
  \put(0,0.75){\line(1,0){0.575}}
  \multiput(0,0)(0.0125,0.025){30}{\rule{0.3pt}{0.3pt}}
  \multiput(0.2,0)(0.0125,0.025){30}{\rule{0.3pt}{0.3pt}}
  \put(0,0.75){\line(0,-1){0.15}}
  \put(0.015,0.75){\line(0,-1){0.1}}
  \put(0.03,0.75){\line(0,-1){0.075}}
  \put(0.045,0.75){\line(0,-1){0.05}}
  \put(0.05,0.75){\line(0,-1){0.025}}
  \put(0.6,0){\line(0,1){0.15}}
  \put(0.585,0){\line(0,1){0.1}}
  \put(0.57,0){\line(0,1){0.075}}
  \put(0.555,0){\line(0,1){0.05}}
  \put(0.55,0){\line(0,1){0.025}}
  \end{picture}}

\newcommand{\ena}{\end{eqnarray}}
\newcommand{\beqa}{\begin{eqnarray}}
\newcommand{\eeqa}{\end{eqnarray}}
\newcommand{\bea}{\begin{eqnarray}}
\newcommand{\eea}{\end{eqnarray}}

\newcommand{\eq}[1]{(\ref{#1})}
\newcommand{\fig}[1]{Figure~\ref{#1}}

\newcommand{\be}{\begin{equation}}
\newcommand{\ee}{\end{equation}}

\def\N{{\cal N}}

\newcommand{\half}{{1\over2}}
\newcommand{\Tr}{{\rm Tr}}
\usepackage{graphicx}

\def\be{\begin{equation}}
\def\ee{\end{equation}}
\def\beq{\begin{eqnarray}}
\def\eeq{\end{eqnarray}}
\def\a{\alpha}
\def\b{\beta}
\def\d{\delta}

\def\N{{\cal N}}

\def\({\left (}
\def\){\right )}
\def\[{\left [}
\def\[{\right ]}

\def\ba{\begin{eqnarray}}
\def\ea{\end{eqnarray}}

\def\a{\alpha}
\def\b{\beta}

\input amssym.def
\input amssym.tex
\begin{document}
\begin{titlepage}
\begin{flushright}
arXiv:0712.4180
\end{flushright}
\vfill
\begin{center}
{\LARGE\bf On the Quantum Resolution 
of Cosmological \\
Singularities 
using AdS/CFT } \\
\vskip 10mm

{\large Ben Craps,$^{1,2}$ Thomas Hertog$^{2,3}$ and Neil Turok$^{4}$}
\vskip 7mm

{\em $^1$ Theoretische Natuurkunde, Vrije Universiteit Brussel, \\
Pleinlaan 2, B-1050 Brussels, Belgium}
\vskip 3mm 
{\em 
$^2$International Solvay Institutes, Boulevard du Triomphe, \\
ULB--C.P.231, B-1050 Brussels, Belgium}
\vskip 3mm 
{\em 
$^3$ Institute for Theoretical Physics, KU Leuven, 3001 Leuven, Belgium\\}
\vskip 3mm 
{\em 
$^4$ DAMTP, CMS, Wilberforce Road, Cambridge, CB3 0WA, UK and\\
African Institute for Mathematical Sciences, 6-8 Melrose Rd, Muizenberg 7945, RSA} 
 
\vskip 3mm
{\small\noindent  {\tt Ben.Craps@vub.ac.be, Thomas.Hertog@fys.kuleuven.be, N.G.Turok@damtp.cam.ac.uk}}
\end{center}
\vfill

\begin{center}
{\bf ABSTRACT}\vspace{3mm}
\end{center}
The AdS/CFT correspondence allows us to map a dynamical cosmology to a dual quantum field theory living on the boundary of spacetime. Specifically, we study a five-dimensional model cosmology in type IIB supergravity, where the dual theory is an unstable deformation of $\N=4$ supersymmetric $SU(N)$ gauge theory on $\Rbar\times S^3$. A one-loop computation shows that the coupling governing the instability is asymptotically free, so quantum corrections cannot turn the potential around. The big crunch singularity in the bulk occurs when a boundary scalar field runs to infinity, in finite time. Consistent quantum evolution requires that we impose boundary conditions at infinite scalar field, {\it i.e.} a self-adjoint extension of the system. We find that quantum spreading of the homogeneous mode of the boundary scalar leads to a natural UV cutoff in particle production as the wavefunction for the homogeneous mode bounces back from infinity. However a perturbative calculation indicates that despite this, the logarithmic running of the boundary coupling governing the instability generally leads to significant particle production across the bounce. This prevents the wave packet of the homogeneous boundary scalar to return close to its initial form. Translating back to the bulk theory, we conclude that a quantum transition from a big crunch to a big bang is an improbable outcome of cosmological evolution in this class of five-dimensional models. 

\vfill
\end{titlepage}
\tableofcontents
\section{Introduction and Summary}

No theoretical framework for cosmology can claim to be complete until it resolves the cosmological singularity, most likely through quantum gravitational effects. We do not know, for example, whether the singularity represents the beginning of the universe and, if not, what happens to the thermodynamic arrow of time there. The no-boundary proposal \cite{Hartle:1983ai}, for example,  predicts the arrow of time reverses. Another, perhaps simpler possibility is that the physics responsible for resolving the singularity also describes dynamical evolution across it\cite{pbb,ekp,Khoury:2001bz}. 

Whichever of these options is correct has profound implications for cosmology. If the singularity was essentially the beginning, the horizon and flatness problems, and the problem of the origin of the observed density variations, seem to demand an early epoch of inflation\cite{inflation}. Although the no-boundary proposal provides some support for this, its formulation in a consistent quantum gravity framework remains a challenge.  If, on the other hand, the singularity was not the beginning and there was a preceding epoch of slow cosmological contraction (in Einstein frame) before it, non-inflationary solutions of the classic cosmological problems are possible \cite{ekp,cyclic,newekp}. Furthermore, in a universe where repeated cycles of evolution occur, technically natural mechanisms whereby the cosmological constant can relax to tiny values become viable \cite{lambda}. 

The AdS/CFT correspondence~\cite{Maldacena:1997re} has emerged as an extremely powerful tool for understanding quantum gravity, providing a non-perturbative definition of string theory in asymptotically anti-de Sitter (AdS) spacetimes, in terms of conformal field theories (CFTs) on their conformal boundaries. In this paper, we shall use the AdS/CFT correspondence to study the quantum dynamics near cosmological singularities. We study Maldacena's original $AdS_5 \times S^5$ example in type IIB supergravity, but with generalized boundary conditions on some of the negative mass squared scalars (saturating the Breitenlohner-Freedman bound). These generalized boundary conditions allow smooth, asymptotically AdS initial data to evolve into a big crunch singularity, namely a spacelike singularity that reaches the spacetime boundary in finite time \cite{Hertog:2004rz,Hertog:2005hu}. 

The dual description of these ``AdS cosmologies'' involves field theories with scalar potentials which are unbounded below and which drive certain boundary scalars to infinity in finite time. The AdS/CFT duality, therefore, relates the problem of resolving cosmological singularities to that of understanding the dynamics of quantum field theories of this type. The present paper describes an attempt to put forward a method that specifies a consistent rule for unitary quantum evolution in the boundary theory.

In our example, the dual field theory is a deformation of ${\cal N}=4$ Super-Yang-Mills (SYM) theory on $\Rbar \times S^3$ by an unbounded double trace potential $-f {\cal O}^2/2$, with ${\cal O}$ a trace operator quadratic in the adjoint Higgs scalars~\cite{Aharony:2001pa}. 
The dual field theory has several key properties that allow us to analyze its dynamics quantitatively: 
\begin{enumerate}
\item
The principal advantage of this five-dimensional setup, compared with similar four-dimensional cosmologies studied previously, is that the undeformed dual boundary theory is well-understood.%
\footnote{
After the present work appeared on the preprint archive, an improved understanding of the M2-brane theory \cite{Aharony:2008ug} has enabled us to study four-dimensional cosmologies as well \cite{Craps:2009qc}.
} 
Furthermore, the deformation is renormalizable and a one-loop computation shows that, because of the ``wrong'' sign of the deformation, the coupling $f$ that governs the instability is asymptotically free \cite{Witten:2001ua}, allowing perturbative field theory computations in the regime near the singularity, {\it i.e.}, for large ${\cal O}$, at least at small 't Hooft coupling. In particular, we can show that quantum corrections do not turn the potential around, so that it is really unbounded below.
\item
As the dual theory evolves towards the singularity, the field evolution becomes {\it ultralocal} on any fixed length scale, meaning that spatial gradients become dynamically unimportant. This means that one effectively has an infinite set of decoupled quantum mechanical systems, one at each spatial point, when one approaches the singularity.
\item
Since the conformal boundary has finite spatial volume, the unstable, homogeneous background mode evolves quantum mechanically. The quantum mechanical spread of its wave function will turn out to be crucial for the suppression of particle creation as the wavefunction for the homogeneous mode rolls down the potential and bounces back from infinity.
\item 
As the singularity is approached, the semiclassical approximation becomes increasingly accurate, both for the homogeneous background and for the fluctuations. This allows us to study the full quantum dynamics with some analytical precision.

\end{enumerate}

In the unstable dual theory, even if we start with the homogeneous background mode described by a localized wave packet, the wavefunction spreads to infinite scalar field in an arbitrarily short time. Unless suitable boundary conditions are imposed, probability will be lost at infinity. Therefore, in order to study quantum evolution in this theory at all, one has no choice but to impose unitary boundary conditions at large field values, {\it i.e.}, boundary conditions which restrict the Hilbert space to a subspace on which the Hamiltonian is self-adjoint. In quantum mechanics, such a restriction is known as a ``self-adjoint extension'' of the original ill-defined theory~\cite{Reed:1975uy,Carreau90}. In this paper, we will describe an attempt to extend this construction to the full boundary field theory.

The fact that the semiclassical approximation becomes more and more accurate near the singularity allows us to implement the self-adjoint extension using classical solutions and the method of images. The relevant solutions turn out to be generically complex near the singularity as a consequence of the quantum spread in the homogeneous mode. To leading order, therefore, the inhomogeneous modes evolve in a complex classical homogeneous background. The complexity of the background turns out to be essential to the resolution of the singularity, providing an ultraviolet cutoff on quantum particle creation. If the backreaction of particles on the homogeneous background were sufficiently small this procedure would specify a rule for unitary quantum evolution in the boundary theory, in the presence of bulk cosmological singularities. 

If the 't Hooft coupling in the boundary super-Yang-Mills theory is small, we have good control over the field theory. However, for small 't~Hooft coupling the bulk is in a stringy regime. In fact, we shall see that the form of the quantum effective potential remains valid at large 't~Hooft coupling, so at least the key dynamical feature of an unbounded negative potential is shared by both regimes. Further analysis is required, though, to fully treat the boundary theory in the regime where the bulk is well-described by supergravity.

With self-adjoint boundary conditions the homogeneous wave packet rolls down the potential and bounces back. In the bulk this behavior corresponds to a quantum transition from a big crunch to a big bang, as envisioned e.g. in ekpyrotic cosmology. However in order for the expectation value of  ${\cal O}$ to return close to its original form the backreaction of the particle production across the bounce must be sufficiently small. We shall find however that in the models we consider here, the logarithmic running of the boundary coupling governing the instability leads, despite the ultraviolet cutoff, to significant particle production. Our estimates indicate that for most of the wave packet the backreaction effect of this appears to prevent the homogeneous mode from rolling back up the potential. Hence we find that a transition from a big crunch to a big bang is rather improbable in this class of models. At the same time, however, our calculations shed light on what might be several key properties of boundary models that do predict a big crunch big bang transition in the bulk.

Let us now briefly review the status of other approaches to cosmological singularity resolution. The simplest string theory models with cosmological singularities are time-dependent orbifolds. Here, however, perturbative string theory tends to break down, with large gravitational backreaction leading to divergences in perturbative string scattering amplitudes \cite{perturbative}. It is natural to hope that a condensation of winding modes can improve the situation, but in general this is still unclear~\cite{BPR}. However, in a specific model it has been argued that the singularity is replaced by a winding tachyon condensate phase, and that the system can be analyzed within perturbative string theory~\cite{MS}. 

The ekpyrotic model \cite{ekp} involves similar time-dependent orbifold solutions to M-theory. Near the singularity, the only light modes in the theory are winding M2-branes \cite{perry,niz}, and the effective string coupling (in IIA or heterotic frame) vanishes. A study of the classical M2-brane dynamics suggests that the $\alpha'$ expansion must be replaced by an expansion in $1/\alpha'$~\cite{perry}, but fully quantum mechanical calculations have not yet been possible.

Other attempts to describe spacelike singularities using a non-perturbative {\it dual description} have also been made. Models with two spacetime dimensions have been described using $c=1$ matrix models \cite{oldmatrix}. Higher-dimensional models with light-like singularities have been studied in the framework of matrix theory \cite{BFSS}, which has led to the suggestion that spacetime may be replaced by non-commuting matrices near a singularity \cite{matrixbigbang}. The quantitative study of the dynamics of this regime is still work in progress, however. 

In earlier work, the AdS/CFT correspondence has also been used to study the singularity inside black holes, which is closely analogous to a cosmological singularity \cite{shenker}. Although some progress in this direction has been made, the fact that the singularity is hidden behind an event horizon clearly complicates the problem. The CFT evolution is dual to bulk evolution in Schwarschild time, so the CFT never directly ``sees'' the singularity. This should be contrasted with the model discussed in the present paper, in which the bulk singularity reaches the boundary in finite time and is thus directly visible in the boundary theory. Other AdS/CFT models of cosmological singularities include \cite{AdSCFTmodels}.

\begin{figure}
\begin{center}
\begin{picture}(0,0)
\put(82,74){$T=\infty$} 
\put(232,74){$T=T_{big \  bang}$}
\end{picture}
\epsfig{file=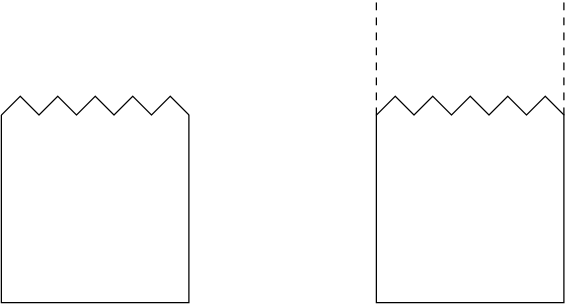, width=8cm}
\end{center}
\caption{The Penrose diagram of a large black hole in anti-de Sitter space is identical to that of an anti-de Sitter cosmology. In a black hole spacetime, however, time at infinity continues forever whereas in an AdS cosmology the singularity hits the boundary in finite time.}
\label{pdiagram}
\end{figure}
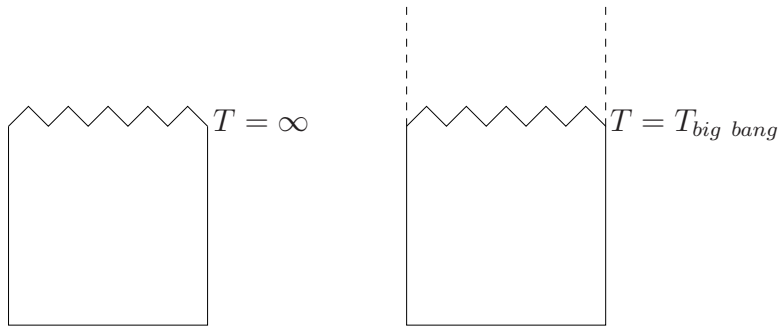

\begin{figure}
\begin{picture}(0,0)
\put(50,123){${\cal V}$} 
\put(210,78){$\langle {\cal O} \rangle$}
\put(301,123){${\cal V}$} 
\put(462,89){$\langle {\cal O} \rangle$}
\end{picture}
\epsfig{file=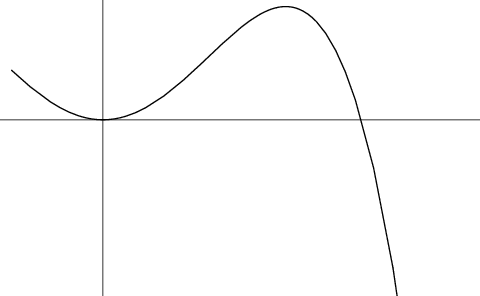, width=2.8in}\hfill
\epsfig{file=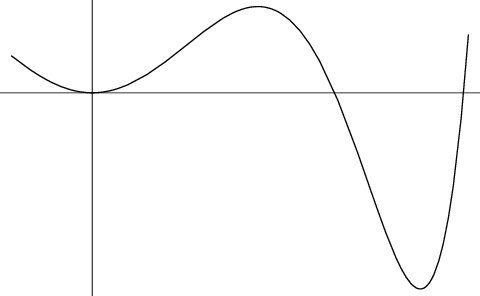, width=2.8in}
\caption{The dual description of AdS cosmologies involves a wave packet rolling down an unstable direction of the field theory potential (left) whereas the formation of large black holes in AdS is described as a thermalization process in a dual theory that has a ground state (right).} 
\label{dual}
\end{figure}
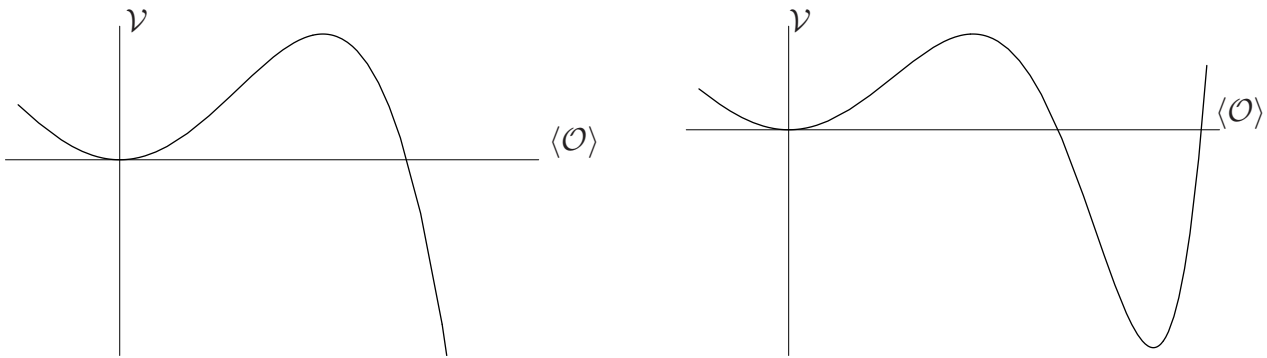

In the setup we consider, the difference between the quantum dynamics of cosmological singularities and the thermalization process that describes the formation of large AdS black holes is apparent in the dual theory. To compare the two situations, we show the Penrose diagram of gravitational collapse to a large black hole in anti-de Sitter space in Figure \ref{pdiagram}. This is identical to the Penrose diagram of an anti-de Sitter cosmology. Therefore from this point of view, there appears to be little distinction in the classical bulk theory between singularities inside black holes and cosmological singularities extending to the boundary in finite time. However the dual (quantum) description of both types of singularities appears to be qualitatively different: as we discussed above, AdS cosmologies are described in terms of unstable conformal field theories with steep potentials ${\cal V}$ that are unbounded below (Figure~\ref{dual}, left), whereas black holes are interpreted as thermal states requiring a vacuum state in the dual theory (Figure~\ref{dual}, right). 

If one ``regularizes'' the unbounded potential in Figure~\ref{dual} (left), for example by adding higher order terms to obtain a potential like the one shown in Figure~\ref{dual} (right), one finds this changes the evolution in the bulk near the big crunch, in the upper corners of the Penrose diagram. This is because the regularization affects the bulk boundary conditions. In particular one finds this causes the cosmological singularity to turn into a large (stable) black hole with scalar hair \cite{Hertog:2005hu}, where the bulk scalar field turned on is dual to the operator ${\cal O}$ in the boundary theory. This is a new type of black hole which does not exist (in a stable form) for the original bulk boundary conditions. It has a natural interpretation in the dual theory as an oscillatory excitation about the global negative minimum of ${\cal V}$ \cite{Hertog:2005hu} (whereas the usual Schwarschild-AdS black holes correspond to thermal states around the standard vacuum at $\langle {\cal O} \rangle =0$). This new black hole with scalar hair is the natural end state of evolution in the bulk corresponding to a wave packet rolling down a regularized potential.

Considering a series of dual theories where the global minimum is taken to be more and more negative, one finds that the horizon size of the black holes with scalar hair -- keeping the mass constant -- increases. In the limit where the global minimum goes to minus infinity the hairy black holes become infinitely large and we recover the original cosmological solutions. Since for the black hole case (with bounded potentials) one expects that the boundary system will eventually thermalize, it seems plausible  that the late time behaviour in the cosmological case (with unbounded potentials) is similar. From a cosmological perspective, the dynamics at intermediate times may be more interesting, however, if one can find models where a wave packet rolling down the potential bounces back a number of times before the system thermalizes and settles down. 

We emphasize, however, that although the dual description in terms of a bounded ${\cal V}$ describing the formation of hairy black holes appears to share many qualitative features with the cosmologies considered here, it does not provide us with a precise and consistent model of singularities. This is because the higher order terms needed to regularize the potential generally lead to a non-renormalizable boundary theory. In contrast, in this paper we focus on a class of five-dimensional cosmologies for which the dual field theory is a renormalizable deformation of ${\cal N}=4$ SYM on $\Rbar \times S^3$ by an unbounded double trace potential $-f {\cal O}^2/2$.

The outline of this paper is as follows. In Section~2, we introduce the bulk cosmologies of interest. We review how modifications of the standard AdS boundary conditions allow smooth initial data to evolve into a big crunch singularity, and focus on a specific example for which the dual field theory analysis will be tractable. We also show that the bulk cosmological solution to IIB supergravity is, after a duality taking us to type IIA frame, qualitatively similar to that describing compactified Milne (or colliding orbifold planes) in M-theory.  In Section~3, we discuss the dual field theory, an unstable double trace deformation of ${\cal N}=4$ SYM theory, deriving the effective potential both at weak and strong 't Hooft coupling.  Section~4 discusses self-adjoint extensions in quantum mechanics, as well as an attempt to extend this idea to quantum field theory. In Section~5, we discuss the quantum evolution of the homogeneous background, which exhibits a quantum spread because the field theory lives on a finite volume space. We develop a method for implementing self-adjoint extensions in the semiclassical expansion, using complex classical solutions and the method of images. We also explain how this method extends to potentials with branch points, such as the quantum effective potential of interest in this paper. Section~6 focuses on the inhomogeneous modes, in particular on the question whether abundant particle creation prevents the scalar field from running back up the potential after the bounce. We find that for most of the range of the Schr\"odinger wavefunction's argument $\phi_f$ carrying significant probability, the quantum spread of the homogeneous background provides an ultraviolet cutoff on the wavelength of produced particles. In Section~7, we find nevertheless that the backreaction of the produced particles on the homogeneous mode appears to be strong enough to prevent the wavefunction from rolling back up the potential. Appendix~A contains more details on the bulk theory. In Appendix~B, we discuss a number of technical details related to the field theory effective potential. Appendix~C  discusses stress tensor correlators in the boundary theory, which would be an important ingredient in relating a more realistic version of our model to cosmological observables.

\setcounter{equation}{0}
\section{Anti-de Sitter Cosmology}
\subsection{Setup}
Our starting point is ${\cal N}=8$ gauged supergravity in five dimensions \cite{Gunaydin:1984qu,Gunaydin:1985cu,Pernici:1985ju}, which is thought to be a consistent truncation of ten-dimensional type IIB supergravity on $S^5$. The spectrum of 
this compactification involves 42 scalars parameterizing the coset $E_{6(6)}/USp(8)$. We concentrate on the subset of scalars that parameterizes the coset $SL(6,R)/SO(6)$. From the higher-dimensional viewpoint, these arise from different quadrupole distortions of $S^5$. The relevant part of the action involves five scalars $\alpha_i$ and takes the form \cite{Freedman:1999gk}
\be\label{act}
S =  \int \sqrt{-g} \left [\frac{1}{2}  R -
\sum_{i=1}^{5}\frac{1}{2}(\nabla \alpha_{i})^2 -V(\alpha_{i})\right ],
\ee
where we have chosen units in which the coefficient of the Ricci scalar is ${1\over 2}$, {\it i.e.}, the 5d Planck mass is unity. The potential for the scalars $\alpha_{i}$ is given in terms of a 
superpotential ${\cal W}(\alpha_i)$ via
\be \label{superpot}
V = \frac{1}{R_{AdS}^2} \sum_{i=1}^5 
    \left( \frac{\partial {\cal W}}{\partial\alpha_i} \right)^2 - 
    \frac{4}{3R_{AdS}^2} {\cal W}^2.
\ee
${\cal W}$ is most simply expressed as
\be
{\cal W} =- \frac{1}{2\sqrt{2}} \sum_{i=1}^6 e^{2\beta_i},
\ee
where the $\beta_i$ sum to zero, and are related to the five $\alpha_{i}$'s
with standard kinetic terms as follows \cite{Freedman:1999gk},
\be\label{alphamatrix}
\pmatrix{ \beta_1 \cr \beta_2 \cr \beta_3 \cr \beta_4 \cr \beta_5 \cr 
\beta_6 } = 
\pmatrix{ 1/2 & 1/2 & 1/2 & 0 & 1/2\sqrt{3} \cr
              1/2 & -1/2 & -1/2 & 0 & 1/2\sqrt{3} \cr
              -1/2 & -1/2 & 1/2 & 0 & 1/2\sqrt{3} \cr
              -1/2 & 1/2 & -1/2 & 0 & 1/2\sqrt{3} \cr
              0 & 0 & 0 & 1/\sqrt{2} & -1/\sqrt{3} \cr
              0 & 0 & 0 & -1/\sqrt{2} & -1/\sqrt{3} }
\pmatrix{ \alpha_1 \cr \alpha_2 \cr \alpha_3 \cr \alpha_4 \cr \alpha_5 }.
\ee

The potential reaches a negative local maximum when all the scalar fields $\alpha_{i}$ vanish. This is the maximally supersymmetric AdS state, corresponding to the unperturbed $S^5$ in the type IIB theory.
At linear order around the AdS solution, the five scalars each obey the free wave equation with a mass that saturates the Breitenlohner-Freedman (BF) bound $m^2_{BF}=-(d-1)^2/4R_{AdS}^2$
\cite{Breitenlohner82} in five dimensions,
\be
m^2= -4/R_{AdS}^2.
\ee
Nonperturbatively, the fields couple to each other and it is generally not consistent to set only some of them to zero. However it is possible to truncate this theory further \cite{Gunaydin:1985cu} to gravity coupled to a single $SO(5)$-invariant scalar $\varphi$ by setting%
\footnote{
There are several inequivalent ways in which this theory can be further truncated to a single scalar as only matter field \cite{Freedman:1999gk}.
An alternative option that was studied in \cite{Hertog:2004rz,Hertog:2005hu} is to take $\beta_{i}=\varphi/\sqrt{3}$ for $i=1,..,4$ and $\beta_5=\beta_6=-2\varphi/\sqrt{3}$, which corresponds to taking only $\alpha_5 \neq 0$. This choice preserves an $SO(4) \times SO(2)$ symmetry.
} 
$\beta_{i}=\varphi/\sqrt{30}$ for $i=1,..,5$ and $\beta_6=-5\varphi/\sqrt{30}$. The action (\ref{act}) then reduces to
\be \label{lagr}
S = \int \sqrt{-g}\left[ \frac{1}{2} R -\frac{1}{2}(\nabla \varphi )^2 
+\frac{1}{4R^2_{AdS}}\(15e^{2\gamma \varphi}+10e^{-4\gamma \varphi}-e^{-10\gamma\varphi} \)\right ],
\ee
with $\gamma = \sqrt{2/15}$.

\subsection{Boundary Conditions}

We will work mainly in global coordinates in which the $AdS_5$ metric takes the form
\be \label{adsmetric}
ds^2_0 = 
R_{AdS}^{2}\left(-(1+r^2 )dt^2 +\frac {dr^2}{ 1+r^2} + r^2 d\Omega_{3}\right).
\ee
In all asymptotically AdS solutions, the scalar $\varphi$ decays at large radius as
\be \label{asscalar}
\varphi (r)= \frac{\alpha \ln r}{ r^2} + \frac{\beta }{r^2},
\ee
where  $\a$ and $\b$ generally depend on the other coordinates. 

For the dynamics of the theory to be well-defined it is necessary to specify boundary conditions at $r=\infty$ on the fields. This amounts to specifying a relation between $\a$ and $\b$ in (\ref{asscalar}).
For example, one can take $\a=0$, leaving $\b$ totally unspecified. This is the usual boundary condition, which preserves the full AdS symmetry group and which has empty AdS as its stable ground state
\cite{Gibbons83,Townsend84}. Alternatively, one can adopt boundary conditions of the form
\be \label{genbc}
\a = -\frac{\d W }{ \d \b},
\ee
where $W(\b)$ is an essentially arbitrary real smooth function.%
\footnote{
In general, boundary conditions specified by an arbitrary function $W$ can be imposed in anti-de Sitter gravity coupled to a tachyonic scalar with mass $m^2$ in the range $m_{BF}^2 \leq m^2 < m^2_{BF}+R_{AdS}^{-2}$ (see e.g.\cite{Henneaux:2006hk}).
} Theories of this type have been called designer gravity theories \cite{Hertog:2004ns}, since their dynamical properties depend significantly on the choice of $W$. We will see that under the AdS/CFT duality, this function $W$ appears as an additional potential term in the action of the dual field theory. 

Boundary conditions of the form (\ref{genbc}) generically break some of the asymptotic AdS symmetries, but they are invariant under global time translations. The conserved energy associated with this is well-defined and finite \cite{Hertog:2004dr,Henneaux:2004zi}, but its expression depends on the function $W$. 

This can be seen as follows. When $W$ is nonzero, the scalar field falls off more slowly than usual. This backreacts on the asymptotic behavior of the $g_{r\mu}$ metric components, which causes the usual gravitational surface term of the Hamiltonian to diverge. This divergence is exactly canceled, however, by an additional scalar contribution to the surface terms. The total charge can therefore be integrated (provided one has specified a functional relation between $\a$ and $\b$). Hence one arrives at a finite expression for the conserved mass, which generally contains an explicit finite contribution from the scalar field which depends on $W$. Whether or not the energy admits a positive mass theorem,%
\footnote{
See \cite{Amsel06} for recent work on the stability of theories of this type.
} 
however, depends on the choice of the function $W$.
 
Below we will be interested in solutions with the following scalar field boundary conditions 
\be\label{bdycond}
\a_f=f\b,
\ee
where $f$ is an arbitrary constant.
The corresponding asymptotic form of the $g_{rr}$ component of the metric is given by
\be \label{5-grr}
g_{rr}=\frac{1}{r^2}-\frac{1}{r^4}-\frac{2f^2 \b^2}{3r^6}(\ln r)^2-
\frac{4f \b^2}{3r^6}\ln r +\frac{f^2\b^2}{3r^6}\ln r +O(1/r^6).
\ee
The conserved mass of spherically symmetric configurations with these boundary conditions 
reads
\be \label{mass}
M=2\pi^2 R_{AdS}^2\left[\frac{3 }{ 2} M_0+\b^2\left(1-\frac{1}{2}f\right)\right],
\ee
where $M_0$ is the coefficient of the $O(1/r^6)$ correction to the $g_{rr}$ component of 
the metric.

\subsection{AdS Cosmologies}

We now construct a class of asymptotically AdS big bang/big crunch cosmologies that are solutions of (\ref{lagr}) with boundary conditions (\ref{bdycond}) on the scalar field, with $f >0$. This is a straightforward generalization of the four-dimensional cosmologies discussed in
\cite{Hertog:2004rz,Hertog:2005hu}.

\begin{figure}
\begin{center}
\epsfig{file=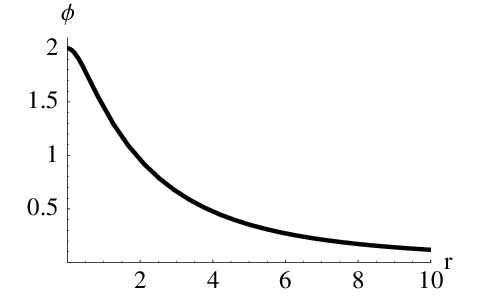, width=8cm}
\end{center}
\caption{Regular initial data $\varphi (r)$ that evolve to a big crunch singularity for boundary conditions with $f =0.1$.}
\label{fig-scalar}
\end{figure}

A particularly simple example of an open FLRW cosmology can be found from the evolution of an initial scalar field profile $\varphi(r)$ obtained from an $O(5)$-invariant Euclidean instanton\footnote{See e.g. \cite{Hertog:2004rz,Hertog:2005hu,Haro07} for a discussion of similar four-dimensional cosmologies.}. Indeed, in Appendix A we show that all boundary conditions (\ref{bdycond}), for $f>0$, admit precisely one such instanton solution.\footnote{Instantons also exist for negative $f$ provided $f < -{\cal O}(1)$. However, as we explain below, we do not expect initial data obtained from slicing these instantons across the four sphere to evolve to a big crunch.} The slice through the instanton obtained by restricting to the equator of the four sphere defines time symmetric initial data for a Lorentzian solution with mass $M= -\pi^2 R_{AdS}^2f^2\b^2/4$. The instanton, therefore, specifies {\it negative mass} initial data in this theory.%
\footnote{
As mentioned earlier, the mass of initial data obtained from instantons depends on the asymptotic behavior of the fields. For the AdS-invariant boundary conditions discussed in Appendix~A (see eq. 
\ref{asads}) one finds the instanton initial data have exactly zero mass, in line with their interpretation as the solution $AdS_5$ decays into \cite{Hertog:2004rz}.
} 
In Figure~\ref{fig-scalar} we show the initial scalar field profile obtained in this way for $f=0.1$.

Analytic continuation of the Euclidean geometry yields a Lorentzian solution that describes the evolution of these initial data under AdS-invariant boundary conditions\footnote{For our model these are given by (\ref{asads}).} \cite{Coleman80}. The origin of the Euclidean instanton then becomes the lightcone emanating from the origin of the Lorentzian solution. Outside the lightcone, the scalar field is constant along four-dimensional de Sitter slices of AdS and the scalar remains bounded in this region. On the light cone we have $\varphi=\varphi(r=0)$ and $\partial_{t}\varphi=0$ (since $\varphi_{,r}=0$ at the origin in the instanton). Inside the lightcone, the $SO(4,1)$ symmetry ensures that the solution evolves like an open FLRW universe,
\be \label{ametric}
ds^2 = -dt^2+ a^2(t) d\sigma_4,
\ee
where $d\sigma_4$ is the metric on the four-dimensional unit hyperboloid. Under time evolution, $\varphi$ rolls down the negative potential. This causes the scale factor $a(t)$ to vanish in finite time, producing a singularity that extends to the boundary of AdS in finite global time. A coordinate transformation in the asymptotic region outside the light cone between the usual static coordinates (\ref{adsmetric}) for $AdS_5$ and the $SO(4,1)$ invariant coordinates (see Appendix~A) shows that $\b \sim\b(t=0)/(\cos t)^2$. Hence $\b$ is now time dependent and  blows up as $t \rightarrow \pi/2$, when the singularity hits the boundary.

\begin{figure}
\begin{center}
\epsfig{file=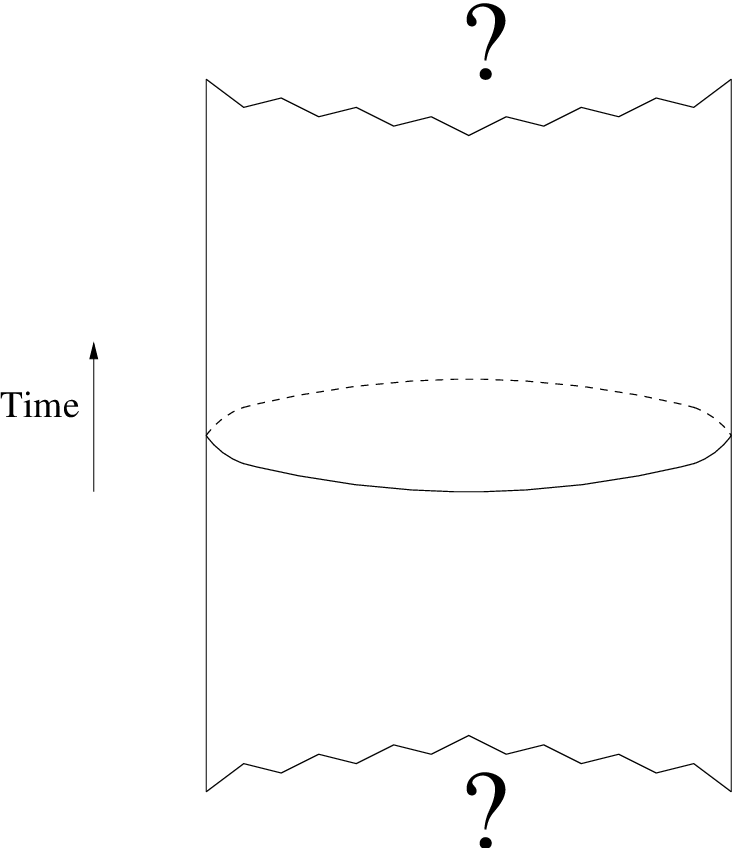, width=6cm}
\end{center}
\caption{Anti-de Sitter cosmology. To predict what happens at the singularities one must turn to the dual field theory description.}
\label{fig-adscosmo}
\end{figure}

The boundary conditions (\ref{bdycond}) differ from the AdS-invariant boundary conditions (\ref{asads}).
This means we cannot obtain the Lorentzian solution simply by analytic continuation from an $O(5)$-invariant instanton. Instead one must evolve the initial data numerically. However, for sufficiently small values of $f>0$ , our boundary conditions are nearly AdS-invariant except when $\a$ becomes large, and causality then restricts its effect. In particular, since the evolution of the instanton data for AdS-invariant boundary conditions has trapped surfaces, a singularity will still form in the central region and the effect of the modification of the boundary conditions on the evolution will only be appreciable in the corners of the conformal diagram where the singularity hits the boundary at infinity. Hence it is reasonable to expect that a singularity must form under evolution with $\a_f$ boundary conditions when $f$ is small. This restricts us to positive $f$ since for negative $f$ we find instantons only for $f <-{\cal O}(1)$. It would now appear possible, however, for the singularity to be enclosed inside a large black hole instead of extending out to infinity. Since the instanton initial data have slightly negative mass, $M R_{AdS}^{-2} \sim -{\cal O}(1)$, these black holes must necessarily have scalar hair. In Appendix~A we numerically integrate the field equations to verify whether $\a_f $ boundary conditions admit static, spherically symmetric black hole solutions with scalar hair outside the horizon. We find a one-parameter set of spherical hairy black holes, which can be characterized by their conserved mass. However, it turns out that for $f>0$ the hairy black holes are always more massive than a vacuum Schwarzschild-AdS black hole of the same size. Hence the singularity that develops from the spherical negative mass initial data defined by the instanton cannot be hidden behind an event horizon. It is therefore plausible that it extends all the way to the boundary, cutting off all space.


\subsection{Ten-dimensional Viewpoint}

$D=5,\ \N=8$ gauged supergravity is believed to be a consistent truncation of ten-dimensional IIB supergravity on $S^5$. This means that it should be possible to lift our five-dimensional solution to ten dimensions. At the linearized level, the $D=5$ scalar fields which saturate the BF bound correspond to quadrupole modes on $S^5$. Since in our example, the $D=5$ scalar field we retain diverges at the classical singularity, one expects that the sphere will become highly squashed.

Even though it is not known how to lift a general solution of $D=5,\ \N=8$ supergravity to ten dimensions, solutions that only involve the metric and scalars saturating the BF bound can be lifted to ten dimensions~\cite{Cvetic00}. 
The ten-dimensional solution involves only the metric and the self-dual five-form.
To describe them, we first introduce coordinates on $S^5$ so that the metric on the unit sphere 
takes the form ($0 \le \xi \le \pi$)
\be
d\Omega_5 = d\xi^2 + \sin^2\xi d\Omega_4.
\ee
Letting $f = e^{\gamma \varphi}$ and $\Delta^2 = f\sin^2\xi + f^{-5} \cos^2\xi$,
the full ten-dimensional metric is 
\be
ds^2_{10} = \Delta ds_5^2 + f^4\Delta d\xi^2  + (f\Delta)^{-1} \sin^2\xi d\Omega_4,
\ee
which preserves an $SO(5)$ symmetry of the five-sphere, as expected.
The five-form is given by
\be
G_5 = U \epsilon_5 +6\sin\xi \cos\xi f^{-1} *df\wedge d\xi,
\ee
where $\epsilon_5$ and $*$ are the volume-form and dual in the five-dimensional solution and
\be
U= -3f^2 \sin^2\xi +f^{-10}\cos^2\xi - f^{-4}-4f^{-4}\cos^2 \xi.
\ee

In the homogeneous region of the asymptotic $AdS_5$ space,
the metric can be written in Robertson-Walker form (\ref{ametric}).
Near the singularity both the potential and the curvature are unimportant in the
Friedmann-Lema\^itre equations, and we have $a(t) \propto (t_s -t)^{1/4}$ and 
$\varphi (t) = -(\sqrt{3}/2)\ln (t_s -t)$. Therefore, over most of the $S^5$ 
near the singularity, the metric approaches
\be
ds^2_{10} =  \sin\xi  [-(t_s-t)^{-1/2\sqrt{10}} dt^2 
+ (t_s-t)^{1/2-1/2\sqrt{10}} d\sigma^2_4 + (t_s-t)^{-9/2\sqrt{10}} d\xi^2
+(t_s-t)^{3/2\sqrt{10}}  d\Omega_4 ].
\ee
Introducing a new time coordinate $T = (t_s-t)^{\mu/4\sqrt{10}}$, where $\mu=4\sqrt{10}-1$, this takes a simple Kasner-like form
\be
ds^2_{10}  =  \sin\xi  [-\(\frac{\mu+1}{\mu}\)^2 dT^2 + T^{(2\sqrt{10}-2)/\mu} d\sigma^2_4
 +T^{-18/\mu} d\xi^2 +T^{6/\mu} d\Omega_4 ].
\ee
The anti de Sitter space and four of the dimensions of the $S^5$ shrink to zero, with Kasner exponents of approximately 0.19 and 0.26 respectively, while the fifth dimension of the $S^5$, labelled by $\xi$, blows up with a Kasner exponent of $\approx -0.77$ so the $S^5$ becomes spindle-shaped. Following the analysis of \cite{Tseytlin:1991xk}, we can T-dualize the $\xi$ dimension to obtain a homogeneous spacetime solution of type IIA string theory. The duality-invariant dilaton is $2 \phi - \sum_i\lambda_i$, where the scale factor of the $i$'th spatial dimension is $a_i = e^{\lambda_i}$. Under T-duality, $\lambda_\xi \rightarrow -\lambda_\xi$ and the string-frame Kasner exponent for the $\xi$ dimension becomes $+0.77$. The dilaton was static in the original IIB frame, but in the T-dual theory we have $e^\phi \sim T^{0.77}$ so the string coupling tends to zero at the singularity. The new metric is now contracting in all directions, and hence qualitatively similar to the isotropic cosmological solution to the low energy effective action for string theory in string frame, with $a_i \propto T^{1 \over 3}$ for $i=1,\dots 9$, and $e^\phi \propto T$. This background solution in type IIA theory corresponds to the solution to 11-dimensional M-theory in which the M-theory dimension and time form compactified Milne spacetime, with the other dimensions are static~\cite{Khoury:2001bz,niz}. It will be very interesting to see whether we can find an unstable mode within the generalized $AdS^5 \times S^5$ setup which, near the singularity, corresponds precisely to the collapse of the M-theory dimension and which could, once the appropriate boundary deformation is identified, be used to model an end-of-the-world brane collision in 11 dimensions, in the heterotic model. 

More generally, it is clear that the cosmological solution we focus on here is only one of many possible cosmologies allowed by generalized boundary conditions on $AdS_5 \times S^5$. The AdS/CFT correspondence can thus be used as a ``laboratory'' for the study of the nonperturbative counterparts (in both $\alpha'$ and $g_s$) of a large class of cosmological solutions to the low-energy effective actions for string theories. Every nontrivial solution possesses a spacelike singularity but, by mapping the theories in each case into an unstable dual quantum field theory, it might be possible to resolve (some of) these singularities and to describe the passage of model universes through them.

\setcounter{equation}{0}
\section{The Boundary Theory: a Double Trace Deformation of ${\cal N}=4$ Super-Yang-Mills Theory}\label{sec:boundary}
In the previous section, we have seen that our bulk theory with boundary conditions \eq{bdycond} allows smooth, asymptotically AdS initial data to evolve in finite time into a big crunch singularity that extends all the way to the boundary. Now we discuss the dual CFT counterpart of this phenomenon. First, we review that the boundary conditions \eq{bdycond} with $f>0$ correspond to adding an unstable potential to the boundary field theory. Then we argue that, in this particular model, the quantum effective potential shares the property of being unbounded below.

For the usual $\a=0$ boundary conditions on the bulk scalars, the dual field theory is $\N=4$ super Yang-Mills theory. The bulk scalars that saturate the BF bound in AdS correspond in the gauge theory to the operators $c\,\Tr[\Phi^i \Phi^j - (1/6) \delta^{ij} \Phi^2]$, where $\Phi^i$ are the six scalars in $\N=4$ super Yang-Mills and $c$ is a normalization factor that will be fixed momentarily. The $SO(5)$-invariant bulk scalar $\varphi$ that we have kept in \eq{lagr} couples to the operator 
\be \label{operator}
{\cal O}=c\,\Tr\left[\Phi_1^2 - {1\over5} \sum_{i=2}^6 \Phi_{i}^2\right].
\ee
According to the AdS/CFT correspondence, this means the following. In the asymptotic behavior \eq{asscalar}, the function $\alpha$ of the coordinates along the boundary plays the role of a source for ${\cal O}$ in the field theory: the field theory action has a term $\int d^4x\,\alpha(x){\cal O}(x)$. The usual boundary conditions set this source to zero, meaning that the boundary field theory is the undeformed $\N=4$ super Yang-Mills theory. On the other hand, the function $\beta$ plays the role of the expectation value of ${\cal O}$ in the field theory; different $\beta$ correspond to different quantum states of the boundary theory.\footnote{For a detailed discussion of the relation between the bulk field and the Yang-Mills operator, see for instance \cite{Bianchi:2001kw}. For issues specific to AdS/CFT in Lorentzian signature, see for instance \cite{Balasubramanian:1998sn}.}

In general, imposing nontrivial boundary conditions $\a (\b)$ in the bulk corresponds to adding a multi-trace interaction $W({\cal O})$ to the CFT action, such that after formally replacing ${\cal O}$ by its expectation value $\b$ one has \cite{Witten:2001ua,Berkooz:2002ug} 
\be \label{multitrace}
\a = -\frac{\d W }{ \d \b}.
\ee
Adding a source term to the action can be considered as a special case, where $W$ is a single-trace interaction linear in ${\cal O}$.
The boundary conditions (\ref{bdycond}) that we have adopted correspond to adding a double trace term to the field theory action
\be\label{Ocubed}
S = S_0 - W({\cal O}) = S_0+\frac{ f }{ 2} \int {\cal O}^2.
\ee
The operator ${\cal O}$ has dimension two, so the extra term is marginal and preserves conformal invariance, at least classically (we shall see that conformal invariance is broken quantum mechanically). 

In the previous section we have taken the constant $f$ to be small and positive in the bulk. 
The term we have added to the CFT action, therefore, corresponds to a negative potential.  Since the
energy associated with the asymptotic time translation in the bulk can be negative,  the dual field theory should also admit negative energy states and have a spectrum unbounded below. This shows that the usual vacuum must be unstable, and that there are 
(nongravitational) instantons which describe its  decay. After the tunneling, the field rolls down the potential and becomes infinite in finite time. This provides a qualitative dual explanation for the fact that the function $\beta$ of the asymptotic bulk solution (\ref{asscalar}) diverges as $t\rightarrow \pi/2$, when the big crunch singularity hits the boundary. Since $\beta$ is interpreted as the expectation value of ${\cal O}$ in the dual CFT, this shows that to leading order in $1/N$, $\langle {\cal O}\rangle$ diverges in finite time.%
\footnote{
In fact, we shall see later that when we consider our theory on $\Rbar\times S^3$, the expectation value of ${\cal O}$ in a state described by a wavepacket diverges well before the center of the wavepacket reaches infinity, so in our deformed theory on a finite volume space it is inappropriate to phrase the dynamics in terms of expectation values. What happens in that case is that the bulk of the wavepacket reaches infinity in finite time.
}

So the big crunch spacetime in the bulk theory corresponds in the boundary theory to an operator rolling down an unbounded potential in finite time. It is important to know whether quantum corrections preserve the unbounded nature of the potential in the boundary theory. While this was unclear for the AdS$_4$ model that was the main focus of earlier work \cite{Hertog:2004rz, Hertog:2005hu, Elitzur:2005kz, Banks:2006hy}, we shall now argue that for our model we indeed have an unbounded quantum effective potential.
  
For this purpose, we first briefly summarize the renormalization properties of a double trace deformation \eq{Ocubed} of ${\cal N}=4$ super-Yang-Mills theory; a more detailed discussion can be found in Appendix~B. As explained in \cite{Witten:2001ua}, the computation of amplitudes at order $f^2$ involves matrix elements of
\be
{f^2\over8}\int d^4x d^4y {\cal O}^2(x){\cal O}^2(y).
\ee
From conformal invariance, $\langle{\cal O}(x){\cal O}(y)\rangle=v/|x-y|^4$ on flat $\Rbar^4$ (where the constant $v$ depends on the normalization factor $c$ in \eq{operator} as well as on $N$). This leads to a short distance divergence that renormalizes $f$ and survives in the large $N$ limit:
\be\label{shortdistance}
{f^2\over2}\int d^4x d^4y {\cal O}(x){\cal O}(y)\langle{\cal O}(x){\cal O}(y)\rangle\sim \pi^2 f^2 v \ln\Lambda \int d^4x {\cal O}^2(x),
\ee
with $\Lambda$ an ultraviolet cutoff. This leads to a one-loop beta function for $f$, which does not receive higher loop corrections in the large $N$ limit \cite{Witten:2001ua}. As in \cite{Witten:2001ua}, we now fix the normalization constant $c$ (and thus $v$) by demanding that the beta function coefficient should be one. At least for small 't~Hooft coupling, it is easy to see that $v\sim c^2N^2$ for large $N$; therefore $c=a/N$ with $a$ a numerical constant. The coupling $f$ can then be kept fixed in the large $N$ limit.  The existence of a non-vanishing beta function means that the conformal invariance of ${\cal N}=4$ super-Yang-Mills theory is broken quantum mechanically by the double trace deformation.

We will be interested in an approximation to the quantum effective action that is valid for a large range of field values, in particular for large field values. An appropriate framework is that of \cite{Coleman:1973jx}, where the standard Feynman diagram expansion is resummed and the theory is organized in a derivative expansion and an expansion in the number of loops (see Appendix~B). The one-loop effective potential is given by \cite{Banados:2006de}
\be\label{effpot}
V({\cal O})=-{f_\mu\over 2}{\cal O}^2+{f_\mu^2{\cal O}^2\ln({\cal O}/\mu^2)\over4},
\ee
where $\mu$ is a renormalization scale, $f_\mu$ a renormalized coupling, and the counterterms have been chosen such that there is no constant and no linear term in ${\cal O}$, and such that
\be
V(\mu^2)=-{f_\mu\over2}\mu^4.
\ee
The renormalization group equation can be obtained by demanding that $V({\cal O})$ be independent of $\mu$:
\be\label{eqrenO}
\mu{df_\mu\over d\mu}=-f_\mu^2,
\ee
which shows that the normalization of ${\cal O}$ implicit in \eq{effpot} is indeed such that the beta function coefficient is one. 
In \eq{eqrenO}, we have ignored a contribution from $d/d\mu$ hitting the $f_\mu^2$ in the second term of \eq{effpot}, which is justified as long as $|f_\mu\ln({\cal O}/\mu^2)|\ll 1$. 
Equation \eq{eqrenO} is solved by
\be
f_\mu={1\over\ln(\mu/\tilde M)},
\ee
with $\tilde M$ an arbitrary scale (this implements dimensional transmutation). Choosing $\mu^2={\cal O}$, {\it i.e.}, the renormalization scale is set by the value of the field ${\cal O}$, the Coleman-Weinberg potential can then be written as
\be\label{CWphiO}
V({\cal O})=-{{\cal O}^2\over \ln({\cal O}/\tilde M^2)}.
\ee
Now suppose that for some value ${\cal O}_0$, the coupling is small,
\be
0< f_{{\cal O}_0}\ll 1,
\ee
then $\tilde M^2<{\cal O}_0$ and \eq{CWphiO} is trustworthy ({\it i.e.}, higher order corrections can be ignored) for any ${\cal O}$ such that ${\cal O}>{\cal O}_0$. As a result, we can conclude that in this case 
\be
V({\cal O})\rightarrow -\infty\ \ \ {\rm for}\ \ {\cal O}\rightarrow\infty.
\ee
Other quantum corrections are small in the large ${\cal O}$ regime we will be interested in, as described at the end of Appendix~B.

As was shown in \cite{Witten:2001ua}, these renormalization properties have a precise counterpart in the bulk theory. 
First, define a new coordinate $0\le u<1$ by $r=2u/(1-u^2)$. In terms of this coordinate, the conformal boundary of AdS is at $u=1$. Near a point on the boundary, the metric takes the form
\be
ds^2={R_{AdS}^2\over z^2}(-dt^2+dz^2+dx^i\,dx^i),
\ee 
where $z=1-u\approx 1/r$ and $x^i$ replace the coordinates of the three-sphere (which is approximately flat when zooming in on one point) \cite{Polchinski:1999yd}.
With the boundary condition $\alpha=f\beta$, the behavior of the field $\phi$ near the boundary $z=0$ is
\be\label{bndyphi}
\phi=\beta z^2(-f\ln z+1).
\ee 
It was shown in \cite{Susskind:1998dq} that the near-boundary (small $z$) region of the bulk theory corresponds to the UV of the dual field theory. Therefore, to study this UV regime, we introduce a new coordinate 
\be
\tilde z = {z\over\epsilon}
\ee
with $\epsilon\ll 1$; this new coordinate is well-suited to studying the near-boundary (small $z$) region of AdS. In terms of $\tilde z$, the boundary behavior \eq{bndyphi} reads
\be
\phi=\tilde\beta \tilde z^2(-\tilde f\ln \tilde z+1)
\ee
with
\be \label{renorm}
\tilde f={f\over 1- f\ln\epsilon}
\ee
and $\tilde\beta=(1- f\ln\epsilon)\beta$.
Interpreting $\epsilon$ as a ratio of renormalization scales, $\epsilon=\mu/\tilde\mu$, and interpreting $f\equiv f_\mu$ and $\tilde f\equiv f_{\tilde\mu}$ as the coupling defined at the scales $\mu$ and $\tilde\mu$, respectively, \eq{renorm} implies the following relation between the couplings at different scales:
\be\label{relation}
f_{\tilde\mu}={f_\mu\over 1- f_\mu\ln(\mu/\tilde\mu)},
\ee
which is consistent with the renormalization group equation \eq{eqrenO}. The reason that this bulk computation (valid for large 't Hooft coupling) agrees with the perturbative field theory computation (valid for small 't Hooft coupling) is the fact that the beta function is one-loop exact for large $N$ \cite{Witten:2001ua}.%
\footnote{
To relate our version of the above argument to Witten's, given in Ref.~\cite{Witten:2001ua}, note that our conventions are related to his by $z=r_W, \alpha=-\alpha_W, f=-f_W$. In the original argument, the boundary condition is written as
\be
\phi=\beta z^2(-f\ln(\mu z)+1),
\ee 
where $\mu$ is an arbitrary scale introduced to define the logarithm. One can choose a different mass scale $\tilde\mu$ if one ``renormalizes'' the field $\beta$ and the coupling $f$ in such a way that the bulk field $\phi$ is left invariant:
\be
\tilde\beta[-\tilde f\ln(\tilde\mu z)+1]=\beta[-f\ln(\mu z)+1],
\ee  
which implies the relation \eq{relation} between the couplings at different scales.}

In what follows we will mostly concentrate on the steepest negative direction of the effective potential. Fluctuations in orthogonal directions in field space acquire a positive mass and we will see these are suppressed. For the $SO(5)$-invariant operator we consider, the most unstable direction comes from the $-\Phi_1^4$ term in \eq{CWphiO} (see \eq{operator}). We focus on the dynamics of $\Phi_1$ as it rolls along a fixed direction in $su(N)$: 
\be
\Phi_1(x)=\phi(x)U
\ee
with $U$ a constant Hermitian matrix satisfying ${\rm Tr} U^2=1$, so that $\phi$ is a canonically normalized scalar field. The Coleman-Weinberg potential \eq{CWphiO} for this scalar is then given by
\be\label{CWphi}
V(\phi)=-{\lambda_0\over4}{\phi^4 \over \ln\left({\phi\over NM}\right)} \equiv -{\lambda_0\over4}{\phi^4 \over l}\equiv -{\lambda_\phi\over4}\phi^4,
\ee
where 
\be\label{lambda0}
\lambda_0={2a^2\over N^2},\ \ \ \ \ \ M^2={\tilde M^2\over Na},\ \ \ \ \ \ \lambda_\phi={\lambda_0\over l}
\ee
with $a$ the numerical constant implicitly defined after \eq{shortdistance}.

\setcounter{equation}{0}
\section{Unbounded Potentials, Self-Adjoint Extensions and Ultra-Locality}
\label{UP}

We have seen that our field theory description involves the potential \eq{CWphi}, which is unbounded below. This implies that the quantum field theory has no ground state. While such unstable theories are usually considered unphysical, we want to explore whether quantum mechanical evolution can be defined for them in a consistent way. A clue is provided by quantum mechanics (as opposed to quantum field theory) with unbounded potentials. As we shall review momentarily, if a potential allows a wavepacket to move off to infinity in finite time, one can nevertheless define unitary quantum evolution by imposing appropriate boundary conditions at infinity. Technically, one restricts the domain of allowed wavefunctions to those on which the Hamiltonian is self-adjoint -- this is called a ``self-adjoint extension''. In this section, we begin to investigate the possibility that the quantum field theory we are interested in might also possess a self-adjoint extension. In Subsection~\ref{subsect:SA-QM}, we review unbounded potentials in quantum mechanics. In Subsection~\ref{subsect:SA-QFT} we show explicitly how the field theory dynamics become ``ultralocal" as the singularity is approached, thus making it plausible that the quantum field is described by an independent set of identical quantum mechanical systems, one for each spatial point. When trying to use this to define self-adjoint extensions point by point, it turns out, however, that the ultralocality does not hold beyond the singularity, invalidating the attempt. Therefore, in subsequent sections, we will limit ourselves to considering a self-adjoint extension for the homogeneous field mode only, and treat the inhomogeneous modes perturbatively around the homogeneous background.

\subsection{Quantum Mechanics in a $-\lambda x^p/4$ Potential for $p>2$}\label{subsect:SA-QM}

We are interested in a theory of a scalar field which is classically conformal invariant but unstable. Let us first emphasize the generality of this setup, in a holographic context.  We consider a classically conformal-invariant scalar field theory on a $d-$dimensional conformal boundary of the form $\Rbar\times S^{d-1}$ where $\Rbar$ is time. The action is 
\be
\int d^d x \sqrt{-g_d} \left(-{1\over 2} (\partial \phi)^2 - \frac{1}{2}\xi_d R_d \phi^2 +{1\over 4} \lambda_p \phi^{\,p}  \right),
\ee
where $R_d$ is the Ricci scalar, and the coupling $\lambda_p$ is an arbitrary constant.  Conformal invariance requires $p = 2+ 4/(d-2)$ and $\xi_d=(d-2)/\left(4(d-1)\right)$. For any $d>2$, $p$ is greater than 2 and provided $\lambda_p$ is positive, $\phi$ will run to infinity in a finite time.  When focusing on the behavior near the singularity, where the $\phi^{\,p}$ term dominates in the potential, we shall neglect the $R_d \phi^2$ term. 

The simple fact that the spatial volume of the conformal boundary is finite is very important for the quantum behavior of the boundary theory. Consider the quantum description of the homogeneous mode $\overline{\phi}$ of the scalar field. Its kinetic term in the action is $V_{d-1} \int dt {1\over 2} \dot{\overline{\phi}}^2$ where the volume of space, $V_{d-1}$, acts as the ``mass'' of $\overline{\phi}$. In the infinite volume limit, this ``mass'' becomes infinite,  and $\overline{\phi}$ undergoes no quantum spreading: it becomes a classical variable.
(This is the essential reason why spontaneous symmetry breaking in possible in quantum field theory but impossible in quantum mechanics). When $V_{d-1}$ is finite, as here, the homogeneous mode $\overline{\phi}$ undergoes quantum spreading. It is convenient to canonically normalize the homogeneous mode,  setting $x=(V_{d-1})^{1\over 2} \phi_0$ and $\lambda=(V_{d-1})^{1-p/2} \lambda_p$, which is constant (for now, we are ignoring the running of the coupling constant). We then have a unit mass quantum mechanical particle with coordinate $x$ and potential 
\be\label{QMpotential}
V(x)=-{1\over 4} \lambda x^{\,p}.
\ee
In this subsection, we summarize the operator approach to the quantum mechanics of such potentials \cite{Reed:1975uy} as reviewed in \cite{Carreau90}.%
\footnote{
For a related recent discussion, see \cite{Fredenhagen:2003ut}.
} 
In Subsection~\ref{sax}, we shall implement the self-adjoint extension using complex solutions to the classical equations of motion, which will be useful for describing the evolution of Gaussian wavepackets.

A classical particle rolling down the potential \eq{QMpotential}, with $p>2$,  reaches infinity in finite time. The same is true for a quantum mechanical wavepacket, and this at first sight appears to lead to a loss of probability, {\it i.e.}, to non-unitary evolution. However, if a self-adjoint Hamiltonian could be defined for this system, unitary quantum mechanical evolution would be guaranteed. As we shall now review, this can be done by carefully specifying an appropriate domain for the Hamiltonian
\be\label{QMHamiltonian}
\hat H=-\half{d^2\over dx^2}+V(x).
\ee 
Since the WKB approximation becomes increasingly accurate at large $x$, we can use it to study the generic behavior of energy eigenfunctions there. 
The two WKB wavefunctions for fixed energy $E$  are proportional to
\be\label{WKB}
\chi_E^\pm(x)=[2(E+\lambda x^p/4)]^{-1/4}\exp\left(\pm i\int^x\sqrt{2(E+\lambda y^p/4)}dy\right),
\ee
where the lower limit of the integral may be chosen arbitrarily.

The Hamiltonian is self-adjoint if for any wavefunctions $\phi_1,\phi_2$ in its domain, $(\hat H\phi_1,\phi_2)=(\phi_1,\hat H\phi_2)$. Using integration by parts, one sees that this is equivalent to
\be\label{BC}
\left[{d\phi_1^*\over dx}\phi_2-\phi_1^*{d\phi_2\over dx}\right]_{x=\infty}=0.
\ee
This can be arranged if for each energy $E$ we select the linear combination of the two WKB wavefunctions which behaves like
\be
\psi_E^\alpha(x)\sim  x^{-p/4} \cos\left({\sqrt{2\lambda}\,x^{p/2+1}\over p+2}+\alpha\right)
\label{asymform}
\ee
at large $x$, where $\alpha$ is an arbitrary constant phase. (The angle $0\leq \alpha \leq \pi$ labels a one-parameter family of inequivalent self-adjoint Hamiltonians.) If $E$ is positive, for example, we set
\be\label{WKBcomb}
\psi_E^\alpha(x)=[2(E+\lambda x^p/4)]^{-1/4}\cos\left(\int_0^x\sqrt{2(E+\lambda y^p/4)}dy+\varphi_E^\alpha\right)
\ee
with
\be
\varphi_E^\alpha=\alpha-\int_0^\infty\left[\sqrt{2(E+\lambda y^p/4)}-\sqrt{\lambda y^p/2}\right]dy,
\label{phasefact}
\ee
which tends to the required form at large $x$. A similar construction can be given for negative $E$ \cite{Carreau90}.

As a consequence of the fact that every energy eigenfunction tends to the same, energy-independent form (\ref{asymform}), equation \eq{BC} is satisfied if $\phi_1=\psi_E^\alpha$ and $\phi_2=\psi_{E'}^\alpha$ with the same value of $\alpha$. 
The domain of the ``self-adjoint extension'' of $\hat H$ labelled by $\alpha$ can now be defined as all wavefunctions
\be
\phi(x)=\int dE\tilde\phi(E)\psi_E^\alpha(x)
\ee
that satisfy $\int dE |\tilde\phi(E)|^2<\infty$, so $\phi$ is square integrable, and $\int dE E^2 |\tilde\phi(E)|^2<\infty$, so that $\hat H\phi$ is also square integrable, {\it i.e.}, it is also a normalizable wavefunction. Under these conditions, the inner product $(\phi_1,\hat H\phi_2)$ makes sense and, as we have just checked, it equals $(\hat H\phi_1,\phi_2)$ so that $\hat H$ is self-adjoint. 

One can interpret the parameter $\alpha$ as follows \cite{Carreau90}. If one placed a ``brick wall" at large $x$, it would force the energy eigenfunctions to vanish there. However, since for finite $E$ the de Broglie wavelength becomes independent of energy at large $x$, displacing the wall by a half-integral number of de Broglie wavelengths would, in this regime, have no effect. Hence only the location of the brick wall modulo an integer number of half-wavelengths matters physically, and this is the information contained in the phase $\alpha$. 

Physically, one can interpret these self-adjoint extensions as follows. A right-moving wavepacket that moves to infinity is always accompanied by a left-moving ``reflected'' wavepacket that runs back up the hill. The time it takes for a right-moving wavepacket to run to infinity and for the left-moving wavepacket to run back up the potential hill can be shown to be the same as a classical particle would take to fall to infinity and climb up the potential again after being reflected at infinity. If the potential is bounded for $x<0$, the Hamiltonian allows a continuum of scattering states. It also has an infinite number of bound states with quantized (negative) energies,%
\footnote{
We note in passing that there is another approach to quantum mechanics in potentials which are unbounded below, based on a $PT$ symmetry \cite{Bender04}. This approach is motivated by analytic continuation from the harmonic oscillator and it results in a positive energy spectrum. In contrast, for all self-adjoint extensions described above,  there is an infinite set of negative energies.  Since we know the bulk theory has negative energy solutions (the instanton discussed in Section~2 provides one example), the approach using self-adjoint extensions seems more appropriate.
} 
the values of which depend on $\alpha$. However, for the scattering problem -- the bounce off the singularity -- which we study, to a very good approximation the phase $\alpha$ enters only as an overall phase in the final wavefunction and hence has no physical consequence. 

In Section~6, we shall compute the Schr\"odinger wavefunctional $\Psi\left(\phi({\bf x})\right)$ for the full quantum field, decomposed into its homogeneous and inhomogeneous parts, $\phi({\bf x})= \overline{\phi} + \delta \phi({\bf x})$. The homogeneous part $\overline{\phi}$, discussed in Section~5, behaves like the coordinate $x$ considered in this section. 
The inhomogeneous part $\delta \phi({\bf x})$ is well-described at early times in terms of its Fourier modes, describing a set of harmonic oscillators which we shall take to be in their incoming ground state. 
We will then look for a range of final values of $\overline{\phi}$ where it is consistent to treat the inhomogeneous modes $\delta \phi({\bf x})$ to quadratic order while ignoring their backreaction on $\overline{\phi}$.  


\subsection{Ultralocality and Self-Adjoint Extensions in Quantum Field Theory}\label{subsect:SA-QFT}

To address the question whether these methods can be extended to quantum field theory, we shall first consider a simplified model that shares the same finite time singularity, namely the quantum field theory of a single scalar field $\phi$ with a negative quartic potential. In the context of the actual dual quantum field theory, this approximation amounts to concentrating on the scalar that parameterizes the steepest negative direction: the gauge-invariant magnitude of $\Phi_1$. We shall attempt to extend the method of self-adjoint extensions to this quantum field theory. First, we show that when approaching the singularity, the scalar field evolution becomes ultra-local: spatial gradients become unimportant and the quantum field theory can be thought of as a collection of independent, identical quantum mechanical systems, one for each point in space. As a consequence, one could try to implement the self-adjoint extension method point by point. Unfortunately, though, it turns out that the ultralocality underlying this construction breaks down as soon as the field reaches infinity at any spatial point. Therefore, what we shall do in subsequent sections is apply the method of self-adjoint extensions to the homogeneous mode of the field, and treat inhomogeneous fluctuations around this background perturbatively. For consistency, it will then be needed that the backreaction of the inhomogeneous modes on the homogeneous mode is small, which we investigate in Section~7.

To begin with, we shall study the classical, conformally-invariant field theory on $\Rbar\times S^3$,
\be
{\cal L}=-{1\over2}(\partial \phi)^2+{\lambda_4\over 4}\phi^4-{1\over 12} R_4 \phi^2,
\ee
where $R_4$ is the Ricci scalar. We are interested in studying 
generic solutions, {\it i.e.}, those specified by the appropriate number of arbitrary functions of space, in the vicinity of the singularity, \ie, the locus where $\phi$ reaches infinity, which we shall assume to be a space-like surface $\Sigma$.%
\footnote{The form of 
$\Sigma$ will obviously depend on the specific solution one considers, \ie, on the initial conditions for the field $\phi$. One may wonder whether $\Sigma$ will be a smooth, spacelike surface for reasonable initial conditions, and in particular for those assumed in the remainder of the paper, where the inhomogeneous field modes start out in the adiabatic vacuum. These quantum mechanical initial conditions generate a classical solution by the phenomenon of ``mode freezing,'' essentially in the same way that classical perturbations are generated from quantum fluctuations in inflationary cosmology. For those solutions, we shall verify in Subsection~6.6 that the surface $\Sigma$ is indeed smooth and space-like, justifying the assumption we are presently making. Note that this conclusion is not in contradiction with the conclusion we shall reach in the present section, namely that the field will evolve in an ultra-local way towards the singularity. While ultralocality implies that field fluctuations $\delta\phi$ grow as the singularity is approached, it does not cause wild fluctuations in the surface $\Sigma$, as we shall show in Subsection~6.6.   
} 
The field equation is
\be
\label{phieq}
\Box \phi = - \lambda_4 \phi^3 +{1\over 6} R_4 \phi.
\ee
The curvature of the three-sphere plays an important role in determining the initial conditions for the scalar field, but as the field runs down towards the singularity the spatial curvature rapidly becomes irrelevant. Since we are concerned with the behavior of the field near the singularity, in this section we shall ignore the last term in (\ref{phieq}).

It is also convenient to change variables to $\chi=(2/\lambda_4)^{1\over 2} \phi^{-1}$ so that the equation of motion assumes the form
\be
\label{chieq}
\chi \partial^2 \chi -2 (\partial \chi)^2 = 2,
\ee
and the zero-energy, homogeneous background solution is just $\chi= t_*-t$, for $t<t_*$, 
with $t_*$ an arbitrary constant. 

\begin{figure}
\begin{center}
\epsfig{file=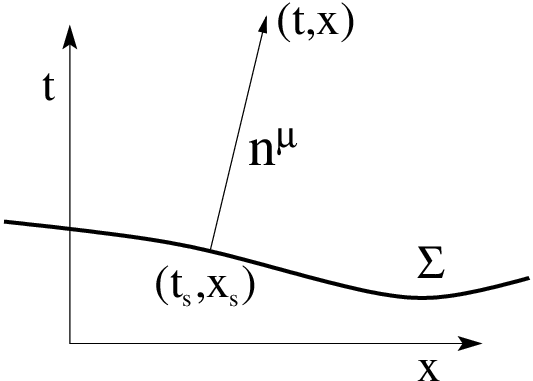, width=10cm}
\end{center}
\caption{The singular hypersurface $\Sigma$, assumed spacelike, upon which $\phi$ is infinite and $\chi$ is zero.}
\label{surface}
\end{figure}

We would like to construct the generic, spatially inhomogeneous solution to (\ref{chieq}) as 
a Taylor series in the proper time away from the singularity, which, as mentioned above, we assume to be a spacelike surface $\Sigma$. 
The coordinates of the embedding Minkowski  spacetime (recall, we are neglecting the space curvature of the $S^3$) are $t, {\bf x}$; the locus of the singular hypersurface $\Sigma$ is $t=t_s({\bf x}_s)$, ${\bf x}={\bf x}_s$. 
As long as $\Sigma$ is not too strongly curved, for every spacetime point $x^\mu$ there is a unique corresponding point $x^\mu_s$ on $\Sigma$, connected to $x^\mu$ by a geodesic normal to $\Sigma$.
The unit normal to $\Sigma$ is
\be
\label{normal}
n^\mu= \left(\gamma, \gamma {\bf \nabla} t_s\right), 
\qquad \gamma \equiv {1\over \sqrt{1- ({\bf \nabla}t_s)^2}},
\ee 
where ${\bf \nabla}$ indicates the gradient with respect to ${\bf x}_s$. It is then natural to change coordinates from $(t,{\bf x})$ to $\tau$, the proper time from $\Sigma$, and ${\bf x}_s$, the corresponding point on $\Sigma$, given by $x^\mu-x^\mu_s = n^\mu \tau$, or
\be 
\label{tauuse} t=t_s({\bf x}_s)+\gamma \tau, \qquad {\bf x} = {\bf x}_s + \gamma \tau {\bf \nabla }t_s,
\ee
 where $\tau = \sqrt{(t-t_s)^2-({\bf x} -{\bf x}_s)^2}$ is (the absolute value of) the proper time from $\Sigma$.

To construct the scalar field equation we need the metric in the new coordinates. From the Minkowski line element in $t,{\bf x}$, and (\ref{tauuse}), we find the line element in $(\tau, {\bf x}_s)$ coordinates is $ -d\tau^2 + h_{ij} dx_s^i dx_s^j$, where
\ba 
\label{spatmet}
h_{ij}&=&h_{ij}^{(0)} + 2 K_{ij} \tau + K_{ik} h^{kl}_{(0)} K_{lj} \tau^2 \cr
h_{ij}^{(0)}&\equiv&\delta_{ij}-\partial_i t_s \partial_j t_s, \qquad K_{ij} \equiv \gamma \partial_i \partial_j t_s.
\ea
Here, $t_s({\bf x}_s)$ is the locus of the singular hypersurface, whose metric is $h^{(0)}_{ij}$ and inverse metric is $h^{ij}_{(0)}$. All space derivatives $\partial_i$ are taken with respect to ${\bf x}_s$, at fixed $\tau$. The matrix $K_{ij}$ is the extrinsic curvature (or second fundamental form) of the singular hypersurface $\Sigma$. (Since the embedding space is flat, the intrinsic curvature of $\Sigma$ can also be expressed in terms of $K_{ij}$, by the Gauss relation $R^i_{jkl}(\Sigma) = K_{jk} K^i_l-K_{jl} K^i_k$, and is thus not an independent quantity.)

In these coordinates, the field equation (\ref{chieq}) reads
\be
\label{chieqnew}
2 \chi_{,\tau}^2 - \chi h^{-{1\over 2}}( h^{1\over 2} \chi_{,\tau})_{,\tau} 
-2 h^{ij} \partial_i \chi \partial_j \chi + \chi h^{-{1\over 2}}( h^{1\over 2} h^{ij} \chi_{,i})_{,j} =2,
\ee
where $h$ is the determinant and $h^{ij}$ the inverse of $h_{ij}$. By substituting (\ref{spatmet}) into (\ref{chieqnew}), we obtain an asymptotic expansion for $\chi$, 
\ba 
\label{chiexp}
\chi&=&  \tau + {1\over 6} K_1 \tau^2 + {1\over 18}(K_1^2-3K_2) \tau^3
+{1\over 4} (K_3 -{13\over 18} K_2 K_1 +{7\over 54} K_1^3 - {1\over 6} \nabla_s^2 K_1) \tau^4\cr  
&&+{\rm ln} \tau \left[ {1\over 5} \left(-7 K_4+{29\over 3} K_1K_3 +{11\over 3} K_2^2 -{67\over 9} K_1^2 K_2 +{34\over 27} K_1^4 
-{1\over 18} (\partial K_1)^2\right.\right.\cr
&&\left.\left.+{2\over 9} K_1 \nabla_s^2 K_1 -{1\over 6} \nabla_s^2 K_2 \right)\tau^5+ C_6\tau^6 +\dots \right] + {\lambda_4\rho(x_s)\over10}\, \tau^5 + D_6 \tau^6 +\dots ,
\ea
where $K_1= h_{(0)}^{ij} K_{ij}= h^{-{1\over 2}} \partial_\tau h^{1\over 2}|_{\tau=0}$, $K_2= {\rm Tr}( K^2)$, and so on, and traces are taken using the metric $h_{(0)}^{ij}$. The higher coefficients $C_n({\bf x}_s)$, $n\geq 6$, are all determined in terms of $t_s({\bf x}_s)$, and the higher coeffients $D_n({\bf x}_s)$, $n\geq 6$, are determined in terms of the second arbitrary function $\rho({\bf x}_s)$. The reason for the entrance of the logarithm, and the second arbitrary function, at order $\tau^5$ is seen by examining the first two terms of (\ref{chieqnew}), at order $\tau^4$. At this order, it is necessary for consistency to introduce a term proportional to $\tau^5 \ln \tau$ into the series expansion, and the coefficient of $\tau^5$ is then left undetermined. 

Let us comment on the general form of the full nonlinear solution (\ref{chiexp}). First, if we re-express $\tau$ in terms of the global time coordinate, 
\be
\tau= |t-t_s({\bf x}_s)|+O((\nabla t_s)^2),
\label{tauexp}
\ee
(\ref{chiexp}) can be viewed as an expansion in powers of ${\bf \nabla} t_s$ as well as in $(t-t_s) {\bf \nabla}_s$. However, if we linearize in the two arbitrary functions $t_s({\bf x})$ and $\rho({\bf x})$, we obtain the solution
\be
\chi (t,{\bf x}) = -t+t_s({\bf x})+ {1\over 6} t^2 {\bf \nabla}^2 t_s({\bf x}) -{1\over 24}t^4 {\bf \nabla}^4 t_s ({\bf x})+\dots - {\lambda_4\rho({\bf x})\over10}\,t^5 +\dots,
\label{chilin}
\ee
which is actually completely regular at $t=0$. Within this linearized approximation, the two arbitrary functions are easily interpreted. We decompose the scalar field into a homogeneous component $\overline{\phi}$, treated nonlinearly, and an inhomogeneous component $\delta \phi({\bf x})$ which we treat only at the linearized level. Then $t_s({\bf x})$ is the {\it time delay} to or from the singularity, given in linear theory by the limit of $\delta \phi /\dot{\overline{\phi}}$ as $t$ tends to zero. And $\rho ({\bf x})$ is the linearized {\it perturbation in the Hamiltonian density}, evaluated at the singularity: setting
\be
\label{enpert}
\delta {\cal H}= \dot{\overline{\phi}} \dot{\delta \phi} + V_{,\phi}(\overline{\phi})\delta \phi = \partial_t\left({\delta \phi \over \dot{\overline{\phi}}} \right)\dot{\overline{\phi}}^2,
\ee
we find that as $t\rightarrow 0$, $\delta {\cal H} $ tends to $\rho({\bf x})$. Note that spatial gradients do not appear in $\delta {\cal H}$ at linearized order because the background solution $\overline{\phi}= -(2/\lambda_4)^{1\over 2} t^{-1}$ is spatially homogeneous.

Second, (\ref{chiexp}) is clearly an expansion in $\tau {\bf \nabla}_s$, so that gradients become less and less significant in the dynamical evolution as the singularity $\tau=0$ is approached: the evolution of $\chi$ becomes {\it ultralocal} in this limit. To see this, consider the ``velocity-dominated'' version of the equation of motion \eq{chieq}, {\it i.e.}, the equation of motion with the gradient terms omitted. The solution of the velocity-dominated equations of motion is
\be
\chi^{(0)}= t_s({\bf x})-t+{\lambda_4\rho({\bf x})\over10}\,(t_s({\bf x})-t)^5.
\ee
Therefore, we see that the full solutions (\ref{chiexp}) differ from the velocity-dominated ones only by terms that are unimportant near the singularity, so that the gradient terms are indeed dynamically unimportant there.

Finally, note that the non-analytic terms in (\ref{chiexp}), involving $\ln \tau$, enter only at quadratic and higher order in the perturbations. If these nonlinear terms are constructed perturbatively by expanding in the linearized perturbation, then no new functions enter: the nonlinear terms are completely dictated by the equations of motion. Hence, a matching condition within linear theory completely determines the matching of the full nonlinear solution, as long as the 
latter is well-described by linear perturbation theory. 

To summarise, we have exhibited a generic class of classical solutions which are  singular on an arbitrary spacelike hypersurface $\Sigma$. The solutions are constructed as an expansion in proper time $\tau$ times spatial gradients: if we follow a solution with spatial structure on some wavenumber $k$, gradients become unimportant in the evolution when $|k \tau|$ falls below unity. At this point, the evolution becomes ultralocal so that the field values at different spatial points decouple. This behavior is relevant to the full quantum theory, because the semiclassical expansion becomes exact as the singularity is approached, so that the quantum wavefunction can be constructed from classical solutions. 

In contructing the quantum theory, it would then seem natural to use the same self-adjoint extension (or boundary condition at large $\phi$) at every spatial point, to evolve the quantum field across the singularity. However, a problem with this attempt becomes clear by studying a simple toy model with discretized space, where the field is represented by particles in a potential at each point of space and the gradient interaction by springs connecting neighboring particles. As soon as one of the particles hits infinity, the gradient terms dominate the dynamics, showing that ultralocality only holds before the singularity is reached. Therefore, we will follow a different approach, where we use a self-adjoint extension for the homogeneous field mode and treat the inhomogeneous modes perturbatively around the corresponding background.

Specifically, we shall implement the self-adjoint extension using the method of images, so the full wavefunction will be constructed from two classical solutions with one, roughly speaking, starting out ``behind" the singularity. A key question is whether the UV regime of the quantum field theory can be consistently treated, or whether the presence of structure on smaller and smaller scales will lead to divergences that invalidate perturbation theory. The remainder of this paper will largely be devoted to analysing this question. 

\setcounter{equation}{0}
\section{Quantum Evolution of the Homogeneous Component}

We want to compute the Schr\"odinger wavefunctional $\Psi\left(t,\phi({\bf x})\right)$ for the full quantum field $\phi({\bf x})$ in the semiclassical expansion. We write the field as the sum of its homogeneous and inhomogeneous parts, $\phi= \overline{\phi}+\delta \phi({\bf x})$, and assume that each Fourier mode of the inhomogeneous component starts out in its quantum ground state. As we shall see later, the inhomogeneous quantum fluctuations are suppressed by powers of $\lambda_\phi$, so it makes sense to first calculate the wavefunction for homogeneous component $\overline{\phi}$, with the inhomogeneous modes ignored.

As emphasized in Subsection~\ref{subsect:SA-QM}, $\overline{\phi}$ is quantum mechanical because the boundary theory lives on a finite space. This simple fact plays an essential role: the quantum spread in $\overline{\phi}$ means that the bulk of the wavefunction stays away from large $\overline{\phi}$. In the semiclassical expansion, this will translate into the fact that the relevant classical trajectories are generically complex and generically avoid the singularity at infinity in the complex $\overline{\phi}$-plane.

In this section, we show how the self-adjoint extension described in Subsection~\ref{subsect:SA-QM} can be implemented in the semiclassical expansion. Since the leading term in this expansion becomes exact near the singularity, the calculation should accurately capture the physics there. We impose the required boundary condition on the wavefunction at infinite $\overline{\phi}$ by the method of images, using a reflection symmetry of the Hamiltonian to satisfy the boundary condition with the sum of two semiclassical wavefunctions. The resulting wavefunction bounces from infinity and returns close to its starting point.  The quantum spread in the wavefunction is represented by a small imaginary part in each of the relevant complex classical solutions which is, for generic final values of $\overline{\phi}$, nonzero at all times and which dominates at times corresponding to the classical encounter with the singularity. 

Having determined the wavefunction $\Psi(\overline{\phi})$ in the approximation where the inhomogeneous modes were ignored, the next step is to construct the full wavefunction $\Psi\left(t_f, \overline{\phi}, \delta \phi({\bf x})\right)$ at the final times of interest, ignoring the backreaction of $\delta \phi$ on $\overline{\phi}$. We again employ the semiclassical expansion, finding the relevant (spatially inhomogeneous) complex classical solution which incorporates the initial condition that the inhomogeneous fluctuation modes of $\delta \phi$ are in their incoming adiabatic ground state.  Working to quadratic order in $\delta \phi$ in the action and ignoring backreaction, the solution for $\delta \phi$ is determined from the relevant linearized field equation.  The small imaginary part in the complex solution for the background $\overline{\phi}$, representing its quantum spread, now plays an essential role: it provides a UV cutoff on the production of high $k$ modes. 


The combined inhomogeneous classical solution yields the semiclassical wavefunction $\Psi\left(t_f, \overline{\phi}, \delta \phi({\bf x})\right)$. One can then verify for what values of $\overline{\phi}$ this provides a consistent approximation scheme. We shall find that the perturbative calculation is reliable only for values of $\overline{\phi}$ far away from the real classical solution corresponding to the initial center and momentum of the Gaussian wavepacket. By contrast,
for values of $\overline{\phi}$ carrying most of the final probability it appears the imaginary part of the relevant complex classical solution is too small for backreaction of particles to be negligible, and the approximation breaks down. As the final value for $\overline{\phi}$ is taken near to the classically predicted value the imaginary part of the complex solution disappears. If backreaction is neglected the quantum particle production of inhomogeneous modes diverges in this limit.

Hence the relevant band of final $\overline{\phi}$ without strong backreaction from particle production is parametrically small in the class of models we consider. Since we are working in a theory where unitarity has been built in, we conclude that it is improbable for the system to bounce with parametrically small particle production. 

We begin this section with a review of the use of complex classical solutions to construct the solution of the time-dependent Schr\"odinger equation in a semiclassical expansion.

\subsection{Complex Classical Solutions and Quantum Mechanics}

Our discussion of the evolution of the quantum field $\phi$ will rest upon the use of complex classical solutions. These are used to obtain the first semiclassical approximation to the Schr\"odinger wavefunction, which then forms the basis of the semiclassical expansion. The method is ideally suited to the present problem because the first semiclassical approximation becomes exact as the singularity is approached.

Consider a quantum mechanical particle with Hamiltonian $H(x,p)= {1\over 2} p^2 +V(x)$. At the initial time $t=t_i$ we prepare the system in a Gaussian wavepacket centred on $x=x_c$ and with momentum $p=p_c$. Our initial condition may be written as
\be
{x\over 2 L} + i {p L\over \hbar} = {x_c\over 2 L} + i {p_c L\over \hbar} \equiv \alpha, \qquad t=t_i,
\label{eq1}
\ee
because this equation, when regarded as an eigenvalue equation with $p= -i \hbar d/dx$, yields for the wavefunction
\be
\Psi(x,t_i) \sim e^{i {p_c x\over \hbar}} e^{-{(x-x_c)^2\over 4 L^2}}.
\label{eq2}
\ee
The corresponding probability density is a Gaussian with variance $\Delta x^2=L^2$. Equation (\ref{eq1}) can also be written as
\be
{x\over \Delta x} + i {p \over \Delta p} = {x_c\over \Delta x} + i {p_c \over \Delta p},
\label{eq1a}
\ee
because for a Gaussian, $\Delta x \Delta p= \hbar/ 2$, saturating the uncertainty relation. 

For a given choice of $\Delta x=L$, a set of Gaussian wavepackets with $x_c$ and $p_c$ running over all real values can be thought of as providing a smeared description of classical phase space $(x,p)$, with the minimal ``smearing area'' $\Delta x \Delta p$. 

The normalized states labeled by $x_c$ and $p_c$, 
\be
\Psi^L_{x_c,p_c}(x)= {1\over (2 \pi L^2)^{1\over 4}}  e^{i {p_c x\over \hbar}} e^{-{(x-x_c)^2\over 4 L^2}}\ ,
\label{normcl}
\ee
are sometimes called ``coherent squeezed states'' and the complex number $\alpha$ defined in (\ref{eq1}) is the coherent state parameter. For given $L$, these states form an {\it overcomplete} basis: they are not orthogonal, but any state can be expressed as a linear combination of them. To see this, note that 
\be
\int {d x_c dp_c \over 2 \pi \hbar} \Psi^{L*}_{x_c,p_c}(x)\Psi^L_{x_c,p_c}(x')= \delta(x-x'),
\label{idc}
\ee
gives the resolution of the identity operator. An arbitrary wavefunction $\Psi(x)$ can be expressed as a linear combination of the $ \Psi^L_{x_c,p_c}$ by multiplying each side of (\ref{idc}) by $\Psi(x')$ and integrating over $x'$. Clearly, if we determine the time evolution of Gaussian wavepackets, for some $L$, we will have determined the time evolution of any state.

For a harmonic oscillator with frequency $\omega$, {\it i.e.}, $V={1\over 2} \omega^2 x^2$, the semiclassical approximation is exact. Furthermore, in this case there is a natural choice for $L$ given by the width of the ground state $L^2=\hbar/(2\omega)$. The left hand side of (\ref{eq1}) is then just the annihilation operator $a=\sqrt{\omega/(2\hbar) }\,(x+ip/\omega)$ obeying $\left[a, a^\dagger \right]=1$, and in fact this was the reason for adopting the normalization of the coherent state parameter $\alpha$ in (\ref{eq1}).
The initial condition corresponding to the oscillator being in its ground state is simply $a=0$: we shall adopt this initial condition for the inhomogeneous modes of the scalar field on the $S^3$ which, at times well away from the bounce, are accurately described as harmonic oscillators with adiabatically varying frequency $\omega$. More generally, one can consider initial and final states which are eigenstates of $a$ with eigenvalue $\alpha$, just as in (\ref{eq1}). These eigenstates are given by $|\alpha\rangle \propto e^{\alpha a^\dagger} |0\rangle$.  The transition amplitude between such states $\langle \beta,{\rm out}|\alpha ,{\rm in} \rangle$ provides the generating function of the S-matrix between arbitrary n-particle Fock states.~\cite{fadeevslavnov}  Therefore the formalism of complex classical solutions extends naturally to a computation of the full S-matrix, in the semiclassical expansion. 

The semiclassical expansion may be derived as a formal expansion in powers of $\hbar$: substituting
\be
\Psi(x,t) = A(x,t) e^{i S(x,t)/\hbar}
\label{eq3}
\ee
into the time-dependent Schr\"odinger equation 
\be 
i\hbar {\partial \Psi \over \partial t}= \left(-{\hbar^2 \over 2} {\partial^2 \over \partial x^2} +V(x)\right) \Psi,
\label{eq3a}
\ee 
at leading order in $\hbar$ we obtain the Hamilton-Jacobi equation for $S$, namely
\be
{\partial S \over \partial t} = -\left({1\over 2} \left({\partial S \over \partial x}\right)^2 +V(x)\right) = -H(x,{\partial S \over \partial x}).
\label{hj}
\ee
This equation is solved by the classical action, {\it i.e.}, $S=S_{Cl}$ being the action evaluated on the appropriate classical path $x_{Cl}$. Since our initial condition (\ref{eq1}) is complex, we shall in general be interested in complex solutions of the classical equations of motion. The classical action is 
\begin{equation}
S_{Cl}(x,t)=\int_{{t}_i}^t dt' \left(p_{Cl} \dot{x}_{Cl}-H(x_{Cl},p_{Cl})\right)(t')+B(x_i,p_i)
\label{classact}
\end{equation}
for a classical path satisfying the initial condition (\ref{eq1}) at time $t_i$ and the final condition $x_{Cl} =x$ at the final time $t$. (We often abbreviate the final coordinate $x_f$ as $x$ and the final time $t_f$ as $t$: the context should make our meaning clear.)
It is necessary to add the boundary term $B$ to ensure that the stationarity of the action $\delta S=0$ under all variations $\delta x(t) $ respecting the given boundary conditions is equivalent to the classical equations of motion. Our initial condition (\ref{eq1}) implies $ \delta x_i+ 2 i \delta p_i L^2/\hbar =0$, and the correct boundary term is found to be
\be 
B(x_i,p_i)= {p_i^2 L^2 \over i \hbar}. 
\label{bt}
\ee 

Now consider the value of the classical action $S_{Cl}$ when the final time $t$ and the final position $x$ are varied, $t \rightarrow t+\delta t$, $x \rightarrow x+\delta x$. The variation in the corresponding classical path is $x_{Cl}(t)\rightarrow x_{Cl}(t)+\delta x_{Cl}(t)$, and the variation of the classical action is
\begin{equation}
\delta S_{Cl}= \delta t (p_{Cl} \dot{x}_{Cl}-H(x_{Cl},p_{Cl}))(t) + \left[ p_{Cl} \, \delta x_{Cl} \right]^t_{{t}_i}+\delta B
\label{vars}
\end{equation}
where we used integration by parts and Hamilton's equations for the classical solution, $\dot{x}_{Cl}= (\partial H /\partial p)_{Cl}$, $\dot{p}_{Cl}=-(\partial H /\partial x)_{Cl}$. The value of the final coordinate at the final time $t+\delta t$ has the first order variation
\be 
\delta x=\delta x_{Cl}(t+\delta t)=\delta x_{Cl}(t) +\dot{x}_{Cl} \delta t,
\label{varxf}
\ee 
hence
\begin{equation}
\delta S_{Cl}= -H \delta t + p \delta x -p_i \delta x_i +\delta B = -H \delta t + p \delta x, 
\label{vars2}
\end{equation}
because $B$ was chosen precisely to ensure that $\delta B-p\delta x_{Cl}$ vanishes at $t=t_i$ for our chosen initial condition. Thus we have 
\be
{\partial S_{Cl} \over \partial t} = -H(x,p), \qquad {\partial S_{Cl}\over \partial x} = p, 
\label{seqs} 
\ee
and the wavefunction (\ref{eq3}), with $S=S_{Cl}$,  satisfies  both the initial condition (\ref{eq1}) and the time-dependent Schr\"odinger equation (\ref{hj}) to leading order in $\hbar$.

For completeness, let us give the equation for the time evolution of the prefactor $A(x,t)$: 
\be
{\partial A \over \partial t}+p {\partial A \over \partial x}= -{1\over 2} {\partial p \over \partial x} A +{i \hbar \over 2 }{\partial^2 A \over \partial x^2}. 
\label{aeq}
\ee
With the neglect of the last, higher order term, this equation is straightforwardly solved by integrating $A$ along the classical trajectories of the system. For example, if $A$ is independent of $t$, then the solution is just $A \propto p^{-{1\over 2}}$.

As a simple illustration, consider a free particle, with $H={1\over 2} p^2$. In this case the Lagrangian is quadratic and the second variation of $S$ with respect to $x$, {\it i.e.}\ $(\partial p/\partial x)$, depends only on $t$. Here, as can be seen from (\ref{aeq}), it is consistent to set $(\partial A / \partial x)$ zero. The $x$-dependence of the wavefunction then arises solely from the classical action $S_{Cl}$. 
Since the equation of motion is trivial, the classical solution satisfying our chosen boundary conditions is easily found,
\be
x(t) = x_f+(t-t_f) {p_c+{i \hbar \over 2 L^2} (x_f-x_c) \over 1+ {i \hbar \over 2 L^2} (t_f-t_i)}, 
\label{eq5}
\ee
where the subscript $f$ denotes final quantities. The classical action (\ref{classact}), with (\ref{bt}), is then
\be
S_{Cl}(x_f,t_f) = {L^2\over i \hbar} {\left(p_c+{i \hbar \over 2 L^2} (x_f-x_c)\right)^2\over 1+{i \hbar \over 2 L^2} (t_f -t_i)}.
\label{eq6}
\ee
Substituting this expression in (\ref{aeq}), we find
\be 
A(t) = { N \over \sqrt{1+i {\hbar t \over 2 L^2} }} 
\label{asolution}
\ee 
with $N$ a constant determined by normalising the wavefunction, $N = 
e^{-p_c^2 L^2/\hbar^2}/(2 \pi L^2)^{1\over 4}$. Equation (\ref{eq3}) then gives the familiar free-particle solution to the time-dependent Schr\"odinger equation, describing a spreading wavepacket. 

There are several points worth noting. First, the relevant classical solution is generally complex, and only real in the special case where $x_f$ lies precisely on the classical trajectory, $x_f=x_c+p_c (t_f-t_i)$. Second, in general the solution does not start at $x_c$, nor is its momentum $p_c$. The initial value of $x$ at $t=t_i$ is
\be
x_i= x_c + {x_f-x_c -p_c(t_f-t_i) \over 1+ {i \hbar \over 2 L^2 } (t_f-t_i)},
\label{eq7}
\ee
and value of the conserved momentum is
\be
p_i=p= p_c + {i \hbar \over 2 L^2 } {x_f-x_c -p_c (t_f-t_i) \over 1+ {i \hbar \over 2 L^2 } (t_f-t_i)}.
\label{eq8}
\ee
If $x_f$ is chosen to lie ahead of (or behind) the classical solution then the imaginary part of $x_i$ is negative (or positive) and the complex classical  trajectory lies in the lower (or upper) half of the complex $x$-plane. The initial condition (\ref{eq1}) implies that $p_i-p_c$ is directed at right angles to $x_i-x_c$ in the complex $x$-plane. Finally, the magnitude of the imaginary part of $x_i -x_c$ is set by the difference between $x_f$ and the value predicted by the classical solution, $x_f-x_c -p_c (t_f-t_i)$, times the dimensionless number $(\hbar/ L^2 ) (t_f-t_i) $ which is the fractional spreading of the wavepacket, $\Delta L/L \sim \Delta p (t_f-t_i) /L$ where $\Delta p \sim \hbar/L$ is the initial spread in momentum implied by the uncertainty relation. If the fractional spreading of the wavepacket is small over the relevant interval of time, then for typical values of $x_f$ the imaginary part of the relevant classical solution will also be small.

We can easily generalize to a free particle confined to the positive half-line by a brick wall located at $x=0$, forcing the wavefunction to vanish there. We can satisfy this boundary condition with free particle wavefunctions by the method of images. Assume that the particle starts out described, to a good approximation, by a Gaussian wavefunction obeying (\ref{eq1}), with $x_c$ large and positive and $p_c$ negative. To ensure the wavefunction vanishes on the wall, we subtract the mirror image wavefunction obtained by reflecting the initial conditions, $x_c\rightarrow -x_c$, $p_c\rightarrow -p_c$. 
Squaring the complete wavefunction, one finds the probability density
\be
|\Psi|^2(x,t)= C \left(e^{-{(x-x_c -p_c (t-t_i))^2\over 2 \sigma^2}}
+ e^{-{(x+x_c +p_c (t-t_i))^2\over 2 \sigma^2}} 
-2 
e^{-{x^2+(x_c +p_c (t-t_i))^2\over 2 \sigma^2}} \cos \left( {2 x\hat{p_c}L^2\over \hbar \sigma^2}\right) \right),
\label{eq9}
\ee
with $C= 1/\sqrt{2 \pi \sigma^2}$, $\sigma^2= L^2+{\hbar^2 (t-t_i)^2 \over 4 L^2}$ the growing dispersion of the Gaussian and $\hat{p}_c(t) = p_c- x_c \hbar^2 (t-t_i) / (4 L^4)$. We minimize the spread of the wavefunction over the time $t-t_i$ of interest if we choose $L=\sqrt{\hbar (t-t_i) /2}$.

At early times, the probability density (\ref{eq9}) is well-approximated by the first term alone, namely a slowly spreading Gaussian centered on the first classical trajectory. When the wavepacket hits the wall, at $t-t_i \sim -x_c/p_c$, interference between the incident and reflected waves leads to strongly oscillatory behavior in both space and time. It is this regime which will be most closely analogous to the bounce from the singularity in our problem. As time runs forward, the second classical solution, obeying the mirror image initial conditions, now determines the dominant piece of the wavefunction. The probability once again assumes a smooth Gaussian form but now moving away from the wall. Similar behavior will be encountered below, when we deal with ``reflection from infinity" in our problem.

The semiclassical formalism we are using emphasizes the similarities between the quantum and classical problems, but their differences emerge in two interesting and important ways. First, more than one classical solution can be relevant to determining the wavefunction. In the brick wall problem, two are needed and it is the interference between them which forces the wavefunction to vanish at the wall. Second, the relevant classical solutions are generically complex, and, in the brick wall problem, generically avoid the location of the wall in the complex $x$-plane. 

\subsection{The Self-Adjoint Extension via Complex Classical Solutions}
\label{sax}

In our problem, the homogeneous mode of the scalar field is described by the action for a quantum mechanical particle,
\be
S = \int dt \left({1\over 2} \dot{x}^2 -V(x) \right),
\label{qpart}
\ee
in which $V(x)$ satisfies two general properties. First, it is unbounded below and falls faster than quadratically at large $x$ so generic real trajectories lead to finite-time singularities. Second, $V'(x)/\left(-V(x)\right)^{3\over 2}$ tends to zero at large $x$ so the semiclassical approximation becomes exact there. As discussed in 
Subsection~\ref{subsect:SA-QM}, under these two conditions the singularity can be quantum mechanically resolved by restricting the allowed stationary states to those proportional to a fixed real function of $x$ at large $x$. For example, for $V= - {1\over 4} \lambda x^4$ the allowed energy eigenfunctions behave as
\be
\psi_E^\alpha(x)\sim  x^{-1} \cos\left(\sqrt{\lambda \over 2}\,{x^3\over 3 \hbar}+\alpha \right), \qquad x \rightarrow \infty,
\label{asymform1}
\ee
with $\alpha$ an arbitrary constant phase labeling the particular self-adjoint extension. 

For the case we are interested in, the use of the energy eigenfunction basis turns out to be an inessential complication. We are interested in the time evolution of a non-stationary state in which the homogeneous component of the field $\overline{\phi}$ is localized around some small value in propagating the state across the quantum ``bounce." In principle, we could compute the quantum evolution by expressing the initial state in terms of energy eigenfunctions, but it is far more convenient to describe the process using Gaussian wavepackets. As we shall now show, the self-adjoint extension may be implemented directly using complex classical solutions, for a class of potentials including those of interest here. 

The first step is to represent the theory (\ref{qpart}) in the complex $x-$plane, viewed as a Riemann sphere including the point at infinity. The simplest potentials to deal with are those which are meromorphic in the region containing the singularity and symmetric under reflection through the singularity, $x \rightarrow -x$. The potential $V(x)=-{1\over 4} \lambda x^4$ satisfies these requirements and we shall start by considering this case.  

Consider an initial Gaussian wavepacket with coordinates $(x_c,p_c)$and spread $\Delta x=L$. The initial condition (\ref{eq1}) provides one complex equation for $x_i$ and $p_i=\dot{x}_i$. A second equation is obtained from energy conservation: since $e={1\over 2} \dot{x}^2 -{1\over 4} \lambda x^4$ is constant, we have 
\be 
t_f-t_i= \int_{x_{i}}^{x_f} {d x \over \sqrt{2(e+{1\over 4} \lambda x^4)}}.
\label{sole}
\ee
For each value of $x_f$, and at fixed $t_f$, equations (\ref{eq1}) and (\ref{sole}) provide two complex equations for the unknowns $x_i$ and $\dot{x}_i$ (or equivalently, $e$), which completely specify the first classical solution $x_1(t)$. 

By analogy with the brick wall problem, we construct a second, ``image," classical solution $x_2(t)$ by reflecting the initial condition (\ref{eq1}) through infinity,
\be
{x_{2,i}\over 2 L} + i {p_{2,i} L\over \hbar} = -\left({x_c\over 2 L} + i {p_c L\over \hbar}\right), \qquad t=t_i,
\label{2ics}
\ee
but imposing the same final condition $x_{2,f}=x_{1,f} =x_f$ at time $t_f$. As for the brick wall, we expect the corresponding semiclassical wavefunction to start out ``behind" the singularity, moving towards it, and to emerge from it.

Now we want to show that the full semiclassical wavefunction for the self-adjoint extension is just the sum of two terms 
\be
\Psi^\alpha(x,t)= A_1(x,t) e^{i S_1/\hbar} + e^{-2i \alpha} A_2(x,t) e^{i S_2/\hbar},
\label{psieqa}
\ee
each determined by the appropriate classical solution. As in the brick wall problem, the boundary conditions are satisfied by the method of images, due to the interference between the two terms. The main difference here is that whereas we had to take $\alpha=\pi/2$ to ensure the wavefunction vanished at the brick wall, in the present problem we can take any real $0\leq \alpha < \pi$ and still ensure the Hamiltonian is self-adjoint.

To show that the Hamiltonian is self-adjoint, we simply compute the wavefunction (\ref{psieqa}) in the limit of large $x_f$. In this limit, the second classical solution becomes the reflected image of the first, $x_2(t)=-x_1(t)$, and this leads to the desired asymptotic behavior. Explicitly, we express the action in canonical form, 
\be
S=\int p \,dx - e (t_f-t_i) + {p_i^2 L^2\over i \hbar}. 
\label{canact}
\ee
Because the two solutions are related by simple reflection in the limit of large $x_f$, they involve opposite momenta in this limit: in the first solution $p_1= +\sqrt{2(e_1+ {1\over 4} \lambda x^4)}$, whereas $p_2= -\sqrt{2(e_2+ {1\over 4} \lambda x^4)}$. Substituting into (\ref{canact}) and integrating by parts to extract the leading dependence for large $x_f$, we find
\ba 
S_1 &\rightarrow & \sqrt{\lambda\over 2} {x_f^3\over 3} +{e_0\over 3}(t_f-t_i)+{\cal O}(x_f^{-1}) \cr
S_2 &\rightarrow & -\sqrt{\lambda\over 2} {x_f^3\over 3} +{e_0\over 3}(t_f-t_i)+{\cal O}(x_f^{-1}), \qquad x_f \rightarrow \infty
\label{acts12} 
\ea
where $e_0$ is the limiting energy of either solution as $x_f$ tends to infinity, and $e_{1,2}=e_0+{\cal O} (x_f^{-1})$, as can be straightforwardly shown from equation (\ref{sole}). The semiclassical wavefunction (\ref{psieqa}) thus behaves as
\be
\Psi^\alpha(x,t) \sim e^{-i\alpha} e^{i {1\over 3} e_0 t/\hbar} x^{-1} \cos\left(\sqrt{\lambda \over 2}\,{x^3\over 3 \hbar}+\alpha \right), \qquad x \rightarrow \infty.
\label{asymform1a}
\ee
Since this is proportional to (\ref{asymform1}), it is obvious that $\Psi^\alpha {d\over dx} \psi_E^\alpha-\psi_E^\alpha {d\over dx} \Psi^\alpha$ vanishes as  $x\rightarrow \infty$, for any energy eigenfunction $\psi^\alpha_E$. Thus our time-dependent wavefunction $\Psi^\alpha(x,t)$ lies within the domain of the self-adjoint extension.

\subsection{Dealing with Branch Cuts in $V(x)$ at Complex $x$}
\label{bcsect}

The above discussion is straightforwardly generalised to potentials whose analytic continuations are meromorphic in the region of the complex plane containing the singularity and which are symmetric under reflection through it. Some additional considerations are needed to deal with potentials, like that of interest in this paper, which possess branch points in the relevant region of the complex plane. In earlier versions of this work on the prepring archive, an elaborate prescription was given for how to construct the image wavefunction given Gaussian initial conditions for the original wavefunction. This involved complex solutions crossing branch cuts, and corresponding adjustments of the initial conditions for the image wavefunction. It turns out, however, that most of these relatively complicated details will not be important for the computations we will do in the present paper, so we have chosen to omit them here. What is important, is that near the final coordinate $x_f$ the branch of the potential $V$ should be chosen such that $V$ is real at $x_f$, so that Schr\"odinger's equation is satisfied with the correct potential.

The reason why the details do not matter too much, is that at times well beyond the crunch the total wavefunction is dominated by the image wavefunction only. In Section~6, we will check whether our attempted self-adjoint extension is invalidated by strong backreaction of created particles in the background of this image wavefunction. Since the answer will turn out to be yes, it is not too important precisely which ``original'' wavefunction our image wavefunction corresponds to.

\subsection{Complex Solutions at Fixed Coupling}

In this subsection, we want to explicitly construct the complex classical solution describing a homogeneous bounce off the singularity at $x=\infty$. We shall in the first instance ignore the logarithmic running of the quartic coupling $\lambda_\phi$, treating it as a constant. After a field and coupling redefinition to remove factors of the spatial volume, the action is just
\be 
S= \int dt \left({1\over 2} \dot{x}^2+{\lambda \over 4}\, x^4\right).
\label{sx}
\ee
Furthermore, we can remove $\lambda$ by setting $\chi= \sqrt{2/\lambda}\,x^{-1}$, so
\be 
S= {1\over \lambda }\int dt {1\over \chi^4} \left(\dot{\chi}^2+1 \right).
\label{sxa}
\ee
We want to find classical solutions satisfying (\ref{eq1}), which reads 
\be 
- 2 i L^2 {\dot{\chi}_i\over \chi_i^2} +{1\over \chi_i} = -2 i L^2 {\dot{\chi}_c\over \chi_c^2} +{1\over \chi_c}, \qquad t=t_i,
\label{icschi}
\ee
and $\chi = \chi_f$, real, at $t=t_f$. Energy conservation reads
\be
\dot{\chi}^2=1+2 e \chi^4,
\label{energy} 
\ee
with $e$ a constant. Integrating, we obtain
\be
t_f -t_i= \int_{\chi_i}^{\chi_f} {d \chi \over \sqrt{1+2 e \chi^4}}.
\label{timeen} 
\ee
Note that $L$ is the width of the Gaussian in the variable $x$: the width in the homogeneous component of the field, $\overline{\phi}$, is $\Delta {\overline{\phi}}= L/\sqrt{V_3}$. Eliminating $\dot{\chi}_i$ in favour of $e$ using (\ref{energy}), equations (\ref{timeen}) and (\ref{icschi}) provide two complex equations for two complex unknowns, $\chi_i$ and $e$.  

We shall solve the equations in the regime most relevant to the description of the final state after a bounce. For this, we need the second, ``image" solution, with initial conditions that $\chi_c$ is large and negative and $\dot{\chi}_c \approx 1$ so that we are close to the zero-energy scaling solution. We want to evolve to a final $\chi_f$ which is large and positive so that we have crossed the singularity and returned to the configuration where the homogeneous component $\overline{\phi}$ is close to its initial position.

We shall look for complex classical solutions close to the real classical solution $\chi=t$, hence the imaginary part of $\chi$ will be small. Setting $\chi_i=r_i+is_i$, we first calculate the real and imaginary parts of (\ref{icschi}).  To linear order in $s_i$ these read
\ba
s_i&\approx& - {2 L^2\over\hbar} (\dot{r}_i- \dot{r}_c {r_i^2\over r_c^2}), \cr
{1\over r_i} - {1\over \chi_c} &\approx& - {2 L^2\over\hbar} \left( {\dot{s}_i\over r_i^2} - 2 {s_i \dot{r}_i \over r_i^3} \right),
\label{icsm} 
\ea
where $r_c \equiv \chi_c$ (which is real). Since the energy of the solutions we are interested in is small, it is reasonable to expand the square root in (\ref{timeen}) in $e \chi^4$, assumed small. To leading order in $e$ we obtain
\be
t_f-t_i \approx \chi_f-\chi_i -{e\over 5} (\chi_f^5-\chi_i^5).
\label{tle} 
\ee
Now we make several further approximations, again related to our assumption that the complex solutions of interest are close to the real classical solution. First, we assume $s$ is small throughout, so we need only work to first order in $s$. Second, we assume that $\chi_i$ is close to $\chi_c$, so that $r_i= \chi_c(1-\delta)$, with $\delta \ll 1$. And finally, we assume that since $\delta$, $e$ and $s$ are all small, we can ignore any term involving products of them. 

The initial conditions (\ref{icsm}) now give 
\ba
s_i &\approx& - {4 L^2\over\hbar} \delta - {L^2\over\hbar} 2 e_0 \chi_c^4,\cr
\delta &\approx& {2 L^2\over\hbar} \left({2 s_i\over r_c^2} - e_1 r_c^3\right),
\label{veqri} 
\ea
where $e= e_0+i e_1$, with $e_0$ and $e_1$ real. In the same approximation, (\ref{tle}) reads
\ba 
t_f-t_i &\approx& \chi_f-\chi_c -{e_0\over 5} (\chi_f^5-\chi_c^5) +\chi_c \delta, \cr
s_i&\approx& - {e_1\over 5} (\chi_f^5-\chi_i^5).
\label{tlap}
\ea 

Equations (\ref{veqri}) and (\ref{tlap}) provide four equations for four unknowns, $s,\delta, e_0$ and $e_1$.
We consider a final $\chi_f$ close to $-\chi_c$, so we are looking at the final wavefunction around the initial value of $\overline{\phi}$, to first order in small quantities. 
We find
\ba
s_i&=& - {5 L^2\over\hbar} {{\cal J} \over \chi_c} {1  \over 1 + L^4 (\hbar\chi_c)^{-2}}, \cr
\delta&=&- {L^2 \over \hbar\chi_c^2 } \,s_i,\cr
e_0&=&-{5 \over \chi_f^5 -\chi_c^5} \left({\cal J} - \chi_c \delta\right),\cr
e_1&=& - {5 \over \chi_f^5-\chi_c^5}
\,s_i, \label{sollin} 
\ea
where ${\cal J} = t_f-t_i - (\chi_f-\chi_c)$ measures the deviation of the argument of the wavefunction from the classical zero-energy solution. 

Of particular interest to us later will be the minimum value of $|\chi|$ attained, since that will act as a UV cutoff for the quantum production of inhomogeneous fluctuations. Evaluating (\ref{timeen}), with $t_i$ replaced by $t$, we find $s$ is quintic in $r$ and 
\be
s_{min}= -{e_1\over 5} \chi_f^5 = {1\over 2} s_i, 
\label{smina}
\ee
with $s_i$ given by (\ref{sollin}). 

Let us now translate this result into the situation of
interest. We want to consider an initial Gaussian wavepacket with nearly zero energy, but over the barrier, {\it i.e.}, centered on $\phi_c \sim \lambda_\phi^{-1/2} R_{AdS}^{-1}$. We parameterize the initial width $\Delta \phi$ of the Gaussian in terms of the width $\Delta \phi_{min}$ which minimizes the spread of the wavepacket over the entire duration of the putative bounce, a time of order $R_{AdS}$. It makes sense to perform the calculation in terms of the variable $\chi$, defined so that $\chi = |t|$ is the classical zero-energy scaling solution. In the presence of the running logarithm we thus define $\chi = \int_\phi^\infty d \phi (2/\lambda_\phi)^{1/ 2} \phi^{-2}$. The final variance in $\chi$ is then approximately given by $\Delta \chi_i^2 + R_{AdS}^2 (\Delta \dot\chi_i)^2$. Here $\Delta\dot\chi_i \sim (\Delta\dot\phi)\lambda_\phi^{-1/2} \phi^{-2} = (\Delta\pi_\phi) V_3^{-1} \lambda_\phi^{-1/2} \phi^{-2} \sim (\Delta \phi)^{-1} V_3^{-1} \lambda_\phi^{-1/2} \phi^{-2} \sim \lambda_\phi R/(\Delta \chi_i)$. Thus the final variance is $\Delta \chi_i^2 +\lambda_\phi^2 R_{AdS}^4/(\Delta \chi_i)^2$, minimized by $\Delta \chi_i \sim \lambda_\phi^{1/2} R_{AdS} \ll \chi_i \sim R_{AdS}$. The corresponding value of $\Delta \phi_{min}$ is $R^{-1}_{AdS} \ll \phi_c$. 

We shall write the width of our initial wavepacket $\Delta \phi = W \Delta \phi_{min}$, where $W$ is a dimensionless parameter. 
From (\ref{smina}), the value of $s_{min}$ is, for typical values of $\chi_f$ where the wavefunction has most of its support, of order $W^2 R_{AdS} \Delta \chi_i/\chi_i (1/1+W^4) \sim W^3 \lambda_\phi^{1/2} R_{AdS} (1/1+W^4)$. Hence the minimal value of $|\chi|$ is largest for wave packets of minimal width $W \sim {\cal O}(1)$. Since the minimum value of $|\chi|$ will act as a UV cutoff for quantum particle production across the bounce, we will consider minimal width wavepackets in the remainder of this paper.


Before we calculate particle production in this background we want to study the large-$\overline{\phi}$ behavior of the wavefunction at very short times, in order to justify our claim that the quantum spreading cannot be ignored, in principle, for arbitrarily short times and hence the dynamical evolution cannot even be discussed without imposing a unitary boundary condition at $\overline{\phi} =\infty$. 

\subsection{Behavior of the Wavefunction at Large $\overline{\phi}$ and Small $t$}
\label{shortt}

One of the unusual features of the system we are considering is that even though the quantum evolution of any state is perfectly well-defined, neither the expectation value of the homogeneous mode $\overline{\phi}$, nor any of its correlators, exist. Hence the usual description of quantum field theory in terms of correlation functions of fields, does not, strictly speaking, make sense. Instead, one has to deal directly with the wavefunction itself. That the correlation functions are in general ill-defined may be seen from (\ref{asymform}), for example. Even though the energy eigenfunctions are all normalizable since $|\Psi|^2 \sim x^{-2}\cos^2(A x^3 +\alpha)$ at large $x$, the expectation of $x=\overline{\phi} V_3^{1\over 2}$ is a meaningless logarithmically divergent integral $\sim \int (d x/x) \cos^2(A x^3 +\alpha)$ in any one of them.

At first sight, this behavior appears puzzling. There is no problem in setting up a Gaussian wavepacket at time $t=0$. In such a wavepacket, the expectation value of $x$ or any integral power of $x$ is well-defined. Likewise, the time derivative $(d/dt)\langle x \rangle = \langle \dot{x} \rangle = \langle p \rangle$, is also perfectly well-defined. Using the Heisenberg equations of motion, one can express the $n$'th time derivative of $\langle x \rangle$ as an expectation value of a polynomial in $x$ and $p$, whose value is perfectly well-defined at the initial time. Hence, formally, $\langle x \rangle$ is a function whose value and time derivatives to any order are finite at $t=0$. Nevertheless, $\langle x \rangle$ does not actually exist, for any $t>0$. 

We can understand what happens by calculating the semiclassical wavefunction using complex classical solutions. Consider the case where we are given a Gaussian wavepacket for $x$ whose initial coordinate $x_c$ is small, and initial speed $\dot{x}_c$ is given by the zero-energy scaling solution. We want to study the form of the wavefunction at very large $x_f$, at small positive final times $t_f$. (We set $t_i=0$ in this subsection.) 

\begin{figure}
\begin{center}
\epsfig{file=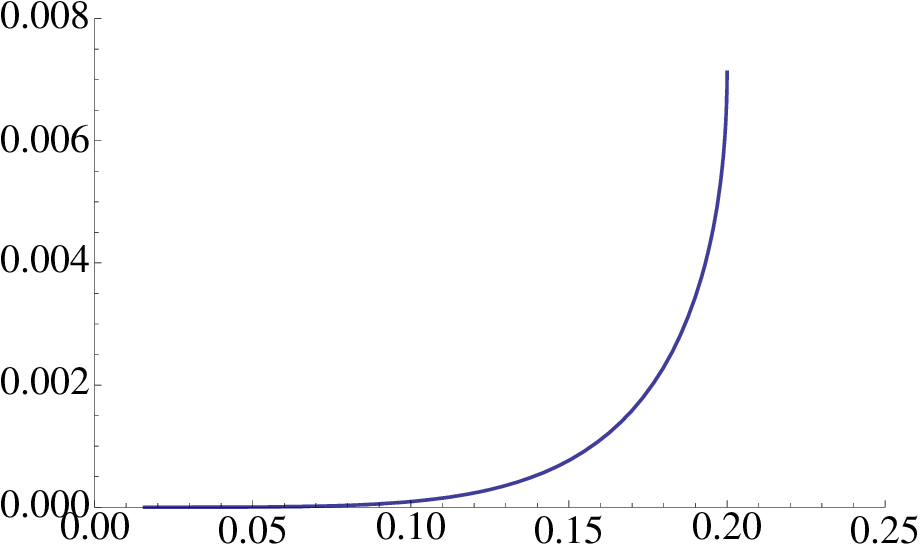, width=10cm}
\end{center}
\caption{Complex classical solution describing the large-$\phi$, small time behaviour of the wavefunction for the homogeneous background. The trajectory of the variable $\chi$ is plotted in the complex $\chi$-plane.}
\label{chifig}
\end{figure}

As in the previous subsection, we change variables to $\chi= (2/\lambda)^{1/2} x^{-1}$, for which the initial conditions are
given by (\ref{icschi}) with $\chi_c \ll 1$ and $\dot{\chi}_c =-1$.
Again, conservation of energy yields 
\be 
\dot{\chi}=-\sqrt{1+e \chi^4},
\label{energya}
\ee
with $e$ a constant. The sign is chosen because the solution we are interested in runs from some large initial $\chi$ to a very small real final $\chi_f$, after a short time $t_f$. The form of the solution is illustrated in a specific numerical example in Figure~\ref{chifig}.

Recall that the classical solution does not need to start at $\chi_c$. If we take $\chi_c$ large, the second term on the right hand side of (\ref{icschi}) dominates. We now solve for the initial velocity,
\be 
\dot{\chi}_i= {1\over 2 i L^2} \chi_i \left(1-{\chi_i\over \chi_c}\right),
\label{initd}
\ee
which is small and nearly negative imaginary if $\chi_i$ is nearly real. In order to achieve consistency with (\ref{energya}), we must take $\chi_i$ close to the branch point in the right hand side of (\ref{energya}), $\chi_i \approx \chi_B=(-e)^{-{1\over 4}}$. If $e=e_0+i e_1$ with $e_0$ large and negative, and $e_1$ small and positive, then $\chi_B \approx (-e_0)^{-{1\over 4}} (1- ie_1/(4 e_0))$ has a small positive imaginary part. Setting  $\chi=\chi_B+\delta \chi$, (\ref{energya}) reads 
\be 
\dot{\chi}= -2 (-\delta \chi)^{1\over 2} (-e)^{3\over 8}.
\label{branch}
\ee
Thus, if $\chi$ starts out just to the right of $\chi_B$, {\it i.e.}, with $\delta \chi$ small and positive, then from (\ref{branch}), $\dot{\chi}_i$ is negative imaginary as required. As $\chi$ runs downwards towards the real $\chi$-axis, from (\ref{branch}), $\dot{\chi}$ turns towards the origin so that $\chi$ heads to small, real values. Solving (\ref{initd}) and (\ref{branch}) for $\delta \chi$, we find
\be 
\chi_i = (-e)^{-{1\over 4}} \left(1-{1\over 16 L^4 e}\right),
\label{ichi}
\ee
just to the right of the branch point. 

If the real part of the energy $e_0$ is large and negative, as we have assumed, then $\chi_i$ is small. We are assuming that $\chi_f$ is even smaller. But then (\ref{timeen}) becomes
\be
t_f-t_i \approx \int_0^{\chi_i} { d \tilde{\chi} \over \sqrt{1-(\tilde{\chi}/\chi)^2}} = \chi_i \int_0^1{dz \over \sqrt{1-z^4}}
= \kappa \chi \label{timer} 
\ee
where $\kappa=\sqrt{\pi} (\Gamma ({5\over 4})/ \Gamma ({3\over 4})= 1.31103\dots$ is the complete elliptic integral of the first kind with parameter $i$. From this value of $\chi_i$, (\ref{initd}) gives $\dot{\chi}_i$. Neglecting terms down by $\chi_i/\chi_c$, and using (\ref{timer}) we obtain
\be 
\dot{x} = -\sqrt{2/\lambda} {\dot{\chi}_i \over \chi_i^2} = {i \kappa\over (2 \lambda)^{1/2} L^2 t_f}, 
\label{inferdotphi}
\ee
at the initial time. Armed with this classical solution, we can compute the classical action, with the boundary term (\ref{bt}).
The integral is dominated by  times $t$ near $t_*$, the time when the solution, if continued past $t_f$, would encounter the singularity:
\be 
S_{Cl}= \int_0^{{t}_f} dt {1\over \lambda (t_*-t)^4} +{i L^2 \dot{x}_i^2 \over i \hbar}.   
\label{actcla}
\ee
The integral is re-expressed in terms of $x_f$ using 
$t_*= t_f +(2/\lambda)^{1/2} x_f^{-1}$; the boundary term is computed from (\ref{inferdotphi}). We obtain 
\be 
{i S_{Cl} \over \hbar } \approx i { (\lambda/2)^{1/2} x_f^3 \over 3 \hbar} - {\kappa^2 \over 2 \hbar^2 \lambda L^2 t_f^2}.
\label{finalsres}
\ee

Now we can read off the behavior of the wavefunction at very large $x_f$ for small final times $t_f$. Including the contribution from the second classical solution, where the initial conditions were reflected through $\chi=0$, we find for the total wavefunction at large $x$ and small times $t_f$,
\be
\Psi(x,t_f) \sim x^{-1} {\rm exp} \left(-{\kappa^2\over 2 \hbar^2 \lambda L^2 t_f^2}\right)   {\rm cos} \left(\sqrt{\lambda\over 2} {x^3 \over 3 \hbar} +\alpha \right).  
\label{psibehavt} 
\ee
Note that the exponential factor $\sim e^{- c t_f^{-2}},$ with $c $ a constant, is indeed a function whose value and time derivatives of arbitrary order vanish at $t=0$. Nevertheless, the function is nonzero for any nonzero $t$. Thus, although correlators of $x$ or $p$ are well defined at $t=0$, for any positive $t$ the probability acquires a tail at large $x$ proportional to $x^{-2}$ times an oscillatory factor which averages to ${1\over 2}$. Thus for any positive time $t$, the expectation value of $x^n$ with $n\geq 1$ does not exist.

\subsection{Complex Solutions for $\phi$ with a Running Coupling}
\label{runcosec} 

As we shall see later, a critical factor in our calculations will be the  running of the quartic coupling $\lambda$: although this is a weak effect, without it, the particle production would vanish at linearized order.  We start from the effective potential obtained in \eq{CWphi}, 
\be
S = \int d^4x \left(-{1\over 2} (\partial \phi)^2 + {\lambda_\phi\over 4} \phi^4\right),
\label{logactphi}
\ee
and we again ignore the curvature of the $S^3$ because this is unimportant near the singularity.  We are interested in the regime where the logarithm $l$ in the denominator of $\lambda_\phi$ is large: in this regime, the logarithm varies slowly with $\phi$ and we can solve the equations of motion in an expansion in inverse powers of $l$. 

As before, we first change variables to $\chi= (2/\lambda_0)^{1\over 2} \phi^{-1}$, obtaining the action
\be 
{\cal S} = {1\over \lambda_0} \int {d^4x \over \chi^{4}} \left(-(\partial \chi)^2 + l^{-1} \right).
\label{logactchi}
\ee
In order to minimize the effect of the logarithm, we redefine the field variable, setting
\be
\label{chideflog}
d\chi_l =l^{{1\over 2}}(\chi) \,d \chi,
\ee
where $l={\rm ln}\left((2/\lambda_0)^{1\over 2}/(NM\chi)\right)$. This is solved by
\be
\label{chidef}
\chi_l= l^{1\over 2} h(l)\, \chi; \qquad h(l)= 1+{1\over 2 l} -{1\over 4 l^2} +{3\over 8 l^3} \dots.
\ee 
\begin{figure}
\epsfig{file=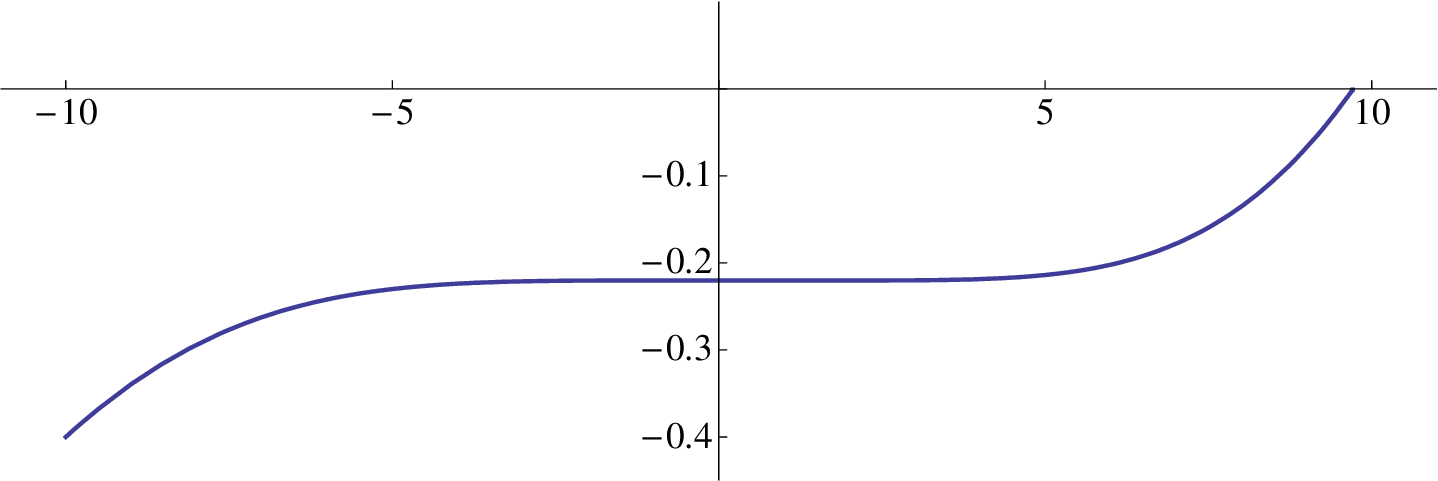, width=14cm}
\caption{Complex classical background solution in which particle production is computed, including the effect of the running coupling. The parameters used were $NM=10^{-5}$, $L=1$, $t_i=-10$, $r_i= 10^{-5}$, $s_i=-0.4$, $t_f=8.4$.}
\label{hombg}
\end{figure}

We obtain the following action for $\chi_l$,
\be
{\cal S} ={1\over \lambda_0}  \int {d^4x \,l \, h^4(l) \over  \chi_l^{4} }\left(-(\partial \chi_l)^2 + 1\right),
\label{newactdf}
\ee
in which form it is clear that at small $\chi$, one obtains a large classical action and therefore one can expect the semiclassical approximation to become accurate. The equation of motion for $\chi_l$ is
\be
- \chi_l \partial^2 \chi_l +2 (\partial \chi_l)^2 + 2 = -2 g(l)\left((\partial \chi_l)^2 + 1 \right), 
\label{eomf}
\ee
where $g(l)= h(l)-1- {1\over 4} l^{-1} h(l)={1\over 4} l^{-1}(1-{3\over 2} l^{-1} +{7 \over 4} l^{-2}+ \dots)$.

From (\ref{eomf}) one observes that $\chi_l= \pm(t_s-t)$ with constant $t_s$ still solves the equation, just as before. By redefining the field as we have, we have removed the logarithmic corrections from the background solution. We could proceed to write the general solution in the vicinity of a spacelike singularity by simply modifying (\ref{chiexp}) so that each coefficient of $\tau^n, \, n\geq 2$ and $\tau^n {\rm ln} \tau,\,  n\geq 5$, becomes a series in inverse powers of $l$. Near the singularity, $l$ diverges so these logarithmic corrections become less and less significant.

Figure \ref{hombg} shows an example of a complex solution to equation
(\ref{eomf}) describing the homogeneous mode $\chi_l(t)$. This solution corresponds to the ``image" solution which starts out behind the singularity at $\chi=0$ and runs to a positive real final value $\chi_f$. Notice that the behavior of the complex solution is very simple in the vicinity of the origin: $\chi_l$ is well-approximated in this region as $\chi_l= t-i\epsilon$ with $\epsilon$ a constant. In the next section, we shall discuss the evolution of the inhomogeneous modes in backgrounds of this form. 

The set of complex ``image" solutions for different final values of $\chi_f$ can be used to construct the wavefunction of the homogeneous mode, at late times $t_f$. The action (\ref{newactdf}) has a small imaginary part that determines the probability for the field to have value $\chi_f$ at time $t_f$.



\setcounter{equation}{0}

\section{Quantum Evolution of the Inhomogeneous Modes}

Having discussed the quantum description of the homogeneous mode of the scalar field, we turn to the quantum evolution of the inhomogeneous modes,
\be 
\phi(t,{\bf x}) = \sum_{\bf k} \phi_{\bf k}(t) Y_{\bf k} ({\bf x}),
\label{modes}
\ee
where $Y_{\bf k} ({\bf x})$ are the spherical harmonics
on $S^3$, and ${\bf k}$ are the corresponding eigenvalues. We shall be interested in modes whose wavelength is much shorter than the radius of the $S^3$, {\it i.e.}, $R_{AdS}$, so that we can approximate the spherical harmonics with plane waves $e^{i{\bf k \cdot x}}$. As explained in Appendix~\ref{renormv}, our calculational procedure is to integrate out the oscillatory modes, {\it i.e.}, those for which $|k t| > 1 $, and to study the long-wavelength modes, {\it i.e.}, those for which $|k t| < 1 $, in the linearized approximation where we neglect their backreaction on the homogeneous mode. 


We want to describe the homogeneous mode, and the ${\bf k} \neq 0$ modes, in a single semiclassical wavefunction, obtained from complex, spatially inhomogeneous solutions to the classical equation of motion for the field. In the case where the quantum spreading of the wavefunction is negligible for the homogeneous background, this treatment will reproduce the usual description of quantum field theory in time-dependent backgrounds, where one uses the Heisenberg picture to compute Bogolubov coefficients, and then studies particle creation etc. We shall describe the same phenomena in the more general context of a quantum mechanical background, using the Schr\"odinger wavefunction, calculated from complex classical solutions. Before proceeding with our specific case, let us then briefly review how a quantum field in a time-dependent background can be described in the Schr\"odinger picture (see, for example, Ref. \cite{ratra}). 

\subsection{Particle Production Using Complex Classical Solutions}

The description of linearised perturbations around a time-dependent, spatially homogeneous background can always be framed (after a possible field redefinition to put the kinetic term into canonical form) in terms of a quadratic action
\be
{\cal S} = \int dt {1\over 2} \left(\dot{q}^2 - \omega^2(t) q^2 \right),
\label{actquad} 
\ee
where $q$ is the coordinate of the particular linearized mode being considered (for example, the real or imaginary part of the $\phi_{\bf k}$ mode coefficients in (\ref{modes}) above) and $\omega^2(t)$ is a time-dependent frequency squared. 

We would like to describe the situation where the inhomogeneous modes start out in the incoming adiabatic vacuum. That is, we assume that for the modes of interest, $\omega^2$ is positive, and $\dot{\omega}/\omega^2$ is far smaller than unity, so that the mode behaves like a harmonic oscillator with a slowly-varying frequency. We also assume that every mode starts out in its adiabatic ground state. This is a special case of the formalism described above, for describing the time evolution of Gaussian wavepackets. In the present case, the initial width of the Gaussian is determined in terms of the frequency $\omega$. It is straightforward to determine the classical action, including the boundary term needed to describe the incoming adiabatic vacuum: in canonical Hamiltonian form,
\be
{\cal S}_{Cl} = \int dt \left(p \dot{q} -{1\over 2}( p^2 + \omega^2(t) q^2) \right)\quad  + {p_i^2 \over 2 i \omega_i},
\label{hamact} 
\ee
where $p_i$ and $\omega(t_i)$ are the initial momentum and frequency. In order to calculate the Schr\"odinger wavefunction for the mode, the correct initial and final conditions are:
\be
p=\dot{q}=i \omega(t_i)q,\quad  t=t_i; \qquad q=q_f,  \quad t=t_f,
\label{ICSSch} 
\ee
with $q_f$ real.
The equation of motion is just $\ddot{q}= -\omega^2 q$, and we assume that at $t_i$ the evolution is adiabatic. Hence the two independent solutions are proportional to $e^{+i \int \omega dt}$ and $e^{-i \int \omega dt}$ respectively. Clearly, the initial condition (\ref{ICSSch}) selects the former -- the incoming negative frequency mode, which we shall denote $R^{(-)}_{in}$, while the final condition specifies its amplitude. The desired solution is then just
\be 
q(t)= {R^{(-)}_{in}(t) \over R^{(-)}_{in}(t_f)} q_f.
\label{cclasssol}
\ee
We then easily compute the classical action (\ref{hamact}), using integration by parts and the equation of motion to obtain
\be
\Psi \sim e^{i {{\cal S}_{Cl}\over \hbar}}, \qquad {\cal S}_{Cl} = {1\over 2} {\dot{R}^{(-)}_{in} \over R^{(-)}_{in}} q_f^2.
\label{evalhamact} 
\ee
This formula agrees with that obtained by solving the time-dependent Schr\"odinger equation, given for example in Ref.~\cite{ratra}.
 
Having obtained the complete $q_f$ dependence of the Schr\"odinger wavefunction, we would like to calculate the number of particles in the final state. The calculation is straightforward. First, we construct the creation and annihilation operators in the Schr\"odinger picture, at the final time $t_f$: 
\be 
a= {ip_f+\omega_f q_f\over \sqrt{2 \omega_f \hbar}} = 
{1\over \sqrt{2 \omega_f \hbar}} \left(\hbar {d \over d q_f} + \omega_f q_f \right), \ \ 
a^\dagger= {-ip_f+\omega_f q_f  \over \sqrt{2 \omega_f \hbar}} = {1 \over \sqrt{2 \omega_f \hbar} }\left(-\hbar {d \over d q_f} + \omega_f q_f \right).
\label{caops}
\ee
Since $\omega_f\equiv \omega(t_f)$ is real, $a^\dagger$ is the Hermitian conjugate of $a$, and the number of particles in this mode in the final state is 
\be
\langle n \rangle = 
{\int dq_f \Psi^* a^\dagger a \Psi \over \int dq_f \Psi^* \Psi }=
 {|\dot{R}^{(-)}_{in}-i\omega_f R^{(-)}_{in}|^2 \over (-2 i \omega_f) \left( R^{(-)*}_{in}\dot{R}^{(-)}_{in} - R^{(-)}_{in} \dot{R}^{(-)*}_{in}\right)}.
\label{expecn}
\ee  
This is our final result, giving the expected number of particles in mode $k$ in terms of the behavior of the incoming negative frequency mode at late times. In the next subsections, we shall use this formula to derive the spectrum of produced particles.

In passing, let us note the relation between (\ref{expecn}) and the usual Bogolubov coefficients, calculated by evolving the inhomogeneous field modes in a real classical background. If the mode evolution is adiabatic at the final time $t_f$, we can express the incoming negative frequency mode in terms of the outgoing negative and positive frequency modes,
\ba 
{R}^{(-)}_{in} 
&\rightarrow& {e^{+i\int^t \omega dt} \over \sqrt{2 \omega}}, \qquad \qquad \qquad \qquad t \rightarrow t_i, \cr
&\rightarrow& 
{\alpha e^{+i\int^t \omega dt} +\beta e^{-i\int^t \omega dt} \over \sqrt{2 \omega}}, \qquad t \rightarrow t_f, 
\label{bogcos} 
\ea
with $\omega$ nearly constant and real. Substituting (\ref{bogcos}) into (\ref{expecn}) we obtain
\be
\langle n \rangle = {|\beta|^2\over |\alpha|^2-|\beta|^2}.
\label{expn} 
\ee
If $\omega$ is always real, then the conservation of the Wronskian $R_{in}^{(-)*}\dot{R}_{in}^{(-)}-\dot{R}_{in}^{(-)*} R_{in}^{(-)}$ yields $|\alpha|^2-|\beta|^2 =1$. However, when $\omega$ is complex, we have to use the more general result (\ref{expn}).

In this subsection we have shown how complex solutions of the classical field equations may be used to compute the Schr\"odinger wavefunction for the fluctuation modes. In the following subsections we shall solve for the positive frequency mode and explicitly determine $\langle n \rangle$ for various particle species in the dual theory.

Before doing so, let us make the following general point. When the Higgs field $\Phi_1$, whose magnitude we have parameterised by $\phi$, rolls to large values, certain gauge, Higgs and Fermi particles acquire time-dependent masses ${\cal M}$ of order $g\phi$, where $g$ is the Yang-Mills gauge coupling. A dimensionless measure of the slowness of increase of ${\cal M}$, {\it i.e.} the adiabaticity of the change in ${\cal M}$,  as the singularity approaches is $\dot{{\cal M}}/{\cal M}^2$. If this quantity is small, one expects particle production of particles of mass ${\cal M}$ to be exponentially suppressed. In our case, using the scaling solution $\dot{\phi} \sim (\lambda_\phi/2)^{1/2} \phi^2$ we have $\dot{{\cal M}}/{\cal M}^2 \sim \sqrt{\lambda_\phi}/g$, which tends to zero as the singularity approaches because $\lambda_\phi$ is asymptotically free (see \eq{logactphi}). The situation is further improved when one takes into account the dependence on $N$, arising from the gauge group $SU(N)$ of the dual theory. To study the theory at large $N$, one keeps the 't~Hooft coupling $g_t=g^2 N$ fixed, while $\lambda_\phi= 2a^2/N^2\ln (\phi/NM)$ (see \eq{logactphi} and \eq{lambda0}). Hence the parameter measuring the departure from adiabaticity in the massive fields is $\dot{{\cal M}}/{\cal M}^2 \sim 1/(g_t N\ln (\phi/NM))^{1/2}$.

Therefore, when the logarithm and $N$ are large, it should be a good approximation to integrate out the particles with masses of order $g \phi$ and consider only the remaining degrees of freedom with tachyonic masses (the fluctuations $\delta \phi$ in the rolling field itself), with light masses, like Higgs modes acquiring their mass only from the deformation, or zero mass, like the massless gauge bosons. 

In the following subsection, we shall derive the equations of motion for the relevant quantum field in each case and show that, in an expansion in inverse powers of the logarithm $l\equiv \ln(\phi/NM)$, to lowest order there is no particle creation in the vicinity of the singularity. Then, in the subsequent three subsections, we shall calculate the particle production in each species at the first nontrivial order in an expansion in inverse powers of $l$.

\subsection{Equations of Motion for Inhomogeneous Fluctuations}

In the following subsections, we shall compute the evolution of the inhomogeneous modes and the resulting particle production. We focus on those wavenumbers which depart from adiabatic oscillatory behavior (or, in more colloquial language, which ``freeze out"), in the regime of large $\phi$ where the classical background is well-described by the zero-energy scaling solution. 

Let us start by recalling the zero energy real scaling solution,
\be
\dot{\phi}=-\sqrt{{\lambda_\phi\over 2}} \phi^2, \qquad \lambda_\phi={\lambda_0\over l}, \qquad l= \ln (\phi/NM), \qquad \lambda_0={2a^2\over N^2}.
\label{bges}
\ee
Note that for the final times we are considering, roughly the time taken for the homogeneous field to roll classically to infinity and bounce back, the quantum wavefunction is dominated by the second, ``image'' complex classical solution in which $\phi$ emerges from ``behind'' the singularity. We can integrate (\ref{bges}) to obtain
\be 
\sqrt{{\lambda_0\over 2}} |t|= \int_\phi^\infty {d \phi \, l^{1\over 2} \over \phi^2} = {l^{1\over 2} \over \phi}\left(1  +{1\over 2 l} +\dots\right)
\label{bges1}
\ee
where the singularity has been placed at $t=0$ and the series in $l^{-1}$ is obtained by repeated integration by parts. From this we obtain
\be 
\phi=\sqrt{2\over \lambda_0} {l^{1\over 2}\over |t|} \left(1+{1\over 2 l} +\dots\right)
\label{phiseries}
\ee
It is also useful to calculate the time derivative of $l$, 
\be
\dot{l} = -{1\over t} (1+{1\over 2l}+\dots).
\label{tdl} 
\ee 
As we have explained at the end of Subsection~\ref{runcosec}, the complex solution which describes the homogeneous mode in the vicinity of the singularity is obtained by replacing $t$ with $t-i\epsilon$ in the above expressions.

\begin{figure}
\begin{center}
\epsfig{file=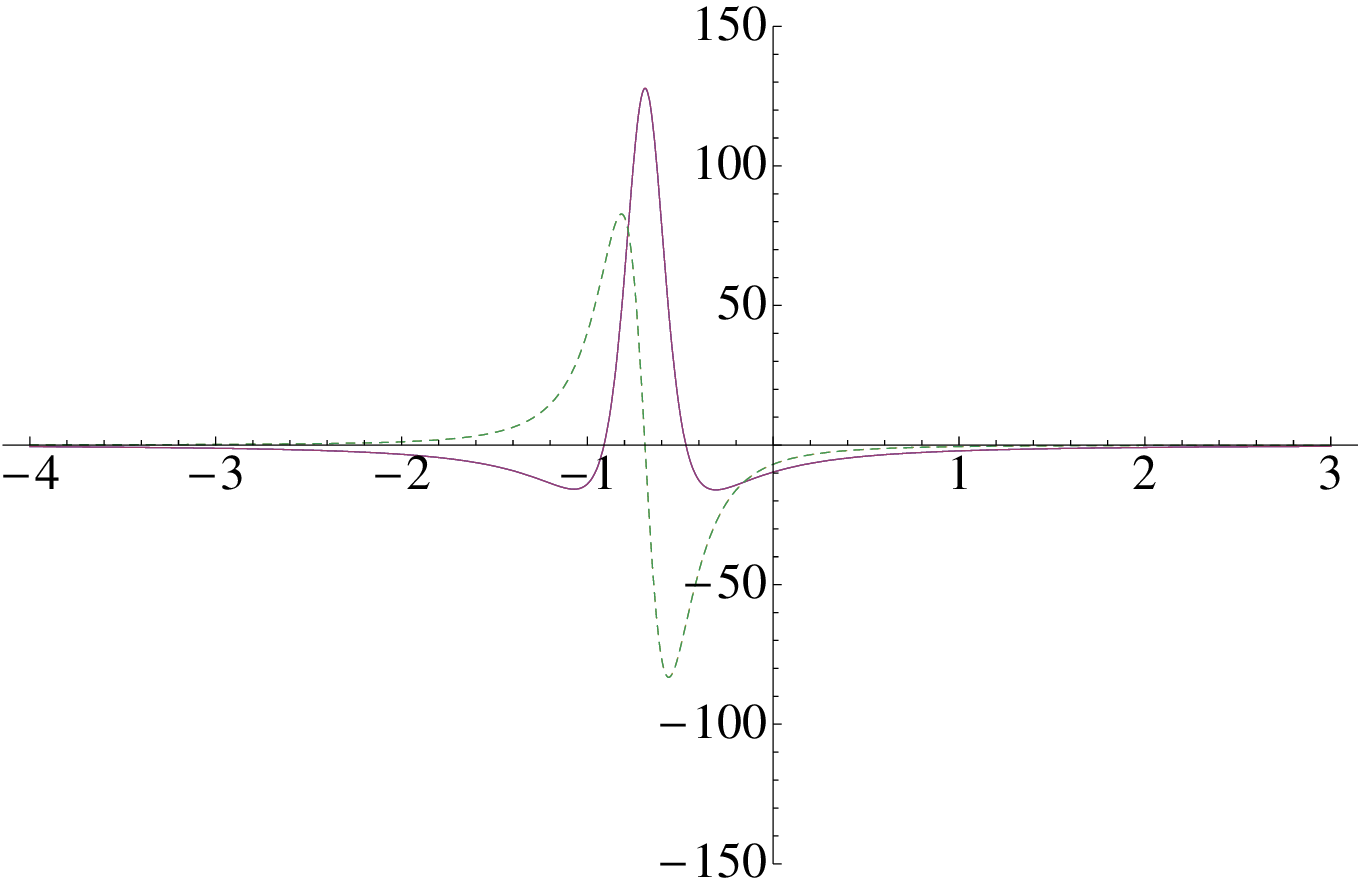, width=12cm}
\end{center}
\caption{
The frequency squared $\omega^2$ (see equation (\ref{dpeq})) from the full complex numerical solution compared with  the approximation
$\chi_l= t -i\epsilon$. The two sets of curves show the real part of $\omega^2$ and its approximation (solid curves, red and blue), and the imaginary part of $\omega^2$ and its approximation (dashed curves, green and yellow). The curves corresponding to numerical and approximate solutions are indistinguishable in the figure. Parameters are as in Figure~\ref{hombg}.
}
\label{wapp}
\end{figure}

The equation of motion for the inhomogeneous perturbation modes is obtained by computing $V_{,\phi\phi}$ and substituting (\ref{phiseries}). One finds 
\ba
\ddot{\delta \phi} &=& \left(-k^2 +{3 \lambda_0 \phi^2 \over {l}} \left(1- {7\over 12} {l}^{-1} + {1\over 6} {l}^{-2} \dots\right)\right) \delta \phi \cr
&\approx & \left(-k^2 +{6\over \chi_l^2} \left(1+{5\over 12} {l}^{-1} -{2\over 3} {l}^{-2}\dots \right)\right)\delta \phi \cr
&\equiv& - \omega^2 \delta \phi,
\label{dpeq} 
\ea
where in obtaining the second line we have used the definition \eq{chidef} of $\chi_l$, which is well-approximated by $t-i\epsilon$ in the complex classical background of interest (see Subsection~\ref{runcosec}). A check on the accuracy of this substitution is given in Figure~\ref{wapp}, where we plot the numerically calculated complex frequency squared, $\omega^2$ in (\ref{dpeq}), versus the approximation obtained by setting $\chi_l =t -i\epsilon$, with $\epsilon$ a constant. Clearly, this is an excellent approximation over a wide range of times, and hence for a wide range of wavenumbers $k$. 

We wish to solve equation (\ref{dpeq}) as a series in $l^{-1}$. At lowest order, (\ref{dpeq}) becomes 
\be\label{feq}
\ddot{\delta \phi} = {6\over (t-i\epsilon)^2} \delta \phi - k^2 \delta \phi,
\ee
which is regular for real $t$. The solutions are $(t-i\epsilon)^{1\over 2}$ times Bessel functions of order $\nu=5/2$. The two linearly independent solutions may be taken to be:
\be
f^{(1)}= \cos k\tilde{t} \left({1\over (k\tilde{t})^2}-{1\over 3}\right) + {\sin k\tilde{t} \over k\tilde{t}}, \qquad
f^{(2)}= \sin k\tilde{t} \left({1\over 3} -{1\over (k\tilde{t})^2}\right) +{\cos k\tilde{t} \over k\tilde{t}}.   
\label{fsols}
\ee
where $\tilde{t}=t-i\epsilon$. The incoming negative frequency mode, {\it i.e.}, the solution tending to $e^{ikt}$ at large negative times, is just $-3 (f^{(1)}-i f^{(2)})e^{-k\epsilon}$. Evolving this to large positive times, we find there is no positive frequency component and, from (\ref{expecn}), the net particle production is zero. To discuss the next order in $l^{-1}$ we shall have to solve (\ref{dpeq}) to this order and we shall do so in the following subsection.

Let us now work out the equation of motion for the light Higgs particles, {\it i.e.},  those acquiring masses solely from the deformation. 
The latter contributes a time-dependent positive mass squared ${1\over 5} \lambda_\phi {\rm Tr}(\Phi_1^2)= {1\over 5} \lambda_\phi \phi^2$ to each of the Higgs fields $\Phi_i$, $i=2,\dots 6$. In the scaling solution for $\phi$, the linearized equations of motion for the light Higgs fields are
\be
 \ddot \Phi_i = -{2\over 5 (t-i\epsilon)^2} (1+ l^{-1} +\dots) \Phi_i -k^2 \Phi_i, \qquad i=2,\dots 6.
\label{lhiggseom}
\ee
To lowest order in $l^{-1}$, the solutions are $(t-i\epsilon)^{1\over 2}$ times Bessel functions of imaginary order $\nu = i\sqrt{3/20}$. As for the tachyonic modes, an incoming negative frequency mode evolves to an outgoing negative frequency mode and there is no particle production. We solve (\ref{lhiggseom}) to next order in $l^{-1}$ in Subsection~\ref{secrunning}.

\begin{figure}
\begin{center}
\epsfig{file=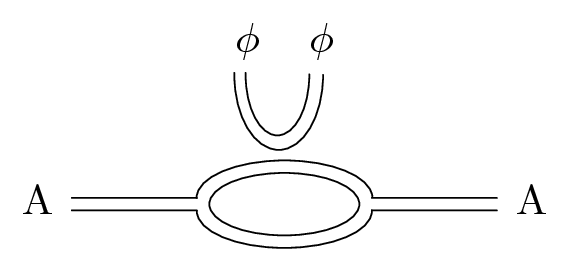, width=12cm}
\end{center}
\caption{
Feynman diagram for the one-loop correction to the kinetic term of the massless gauge bosons.
}
\label{figFeynman}
\end{figure}

Finally, let us consider the gauge bosons that do not acquire a mass from $\Phi_1$. Even with the deformation, there is no coupling between these light gauge fields and $\Phi_1$ in the classical Lagrangian. The rolling field generates no gauge current, and therefore does not directly source the gauge fields. However, a coupling does arise at one loop order. As can be seen from the results of Ref.~\cite{ball} for example, integrating out the massive Higgs, Fermi and gauge bosons results in an effective Lagrangian for the remaining massless gauge bosons of the form $-{1\over 4} F^2 \left(1 + \alpha \sum_{i=S,F,V} c_i{\rm Tr}(M_i^2/\mu^2)\right)$, where $\alpha=g^2/(16 \pi^2)$, $\mu$ is a renormalization scale and the trace is taken over scalars (with $c_S={1\over 6}$), fermions (with $c_F= {4\over 3}$ per Dirac fermion) and gauge bosons (with $c_V=-{11\over 3}$). The mass matrices for the heavy particles depend on the orientation of the Higgs field $\Phi_1$ in $SU(N)$ space. In general position, the simplest case to consider, the gauge group breaks to $U(1)^{N-1}$ and all but $N-1$ gauge, Higgs or fermi particles acquire masses of order $g \phi$. In the undeformed theory, there are six scalars, two Dirac fermions and one vector so  the correction vanishes by supersymmetry. The deformation alters the scalar mass matrix, $M_S^2 \sim g^2 \phi^2 \rightarrow g^2 \phi^2 + \lambda_\phi \phi^2$, up to numerical factors, leaving the other terms unchanged. The effective massless gauge boson Lagrangian is then $-{1\over 4} F^2 (1+C l^{-1})$, with $C$ a constant. The second term can also be derived from the finite Feynman diagram shown in Figure~\ref{figFeynman}. 

The equation of motion for the transverse massless gauge bosons acquires a damping correction term as follows, to leading nontrivial order,
\be
\ddot{A}_\perp - C l^{-2}{1\over (t-i\epsilon)} \dot{A}_\perp +k^2 A_\perp =0.
\label{gbe} 
\ee
Ignoring the $l^{-2}$ term, the solutions are trivial and a negative frequency incoming mode evolves to a negative frequency outgoing mode. We solve (\ref{gbe}) to order $l^{-2}$ in Subsection~\ref{subsec:gluons}.

In summary, in this subsection we have derived the equations of motion for the quadratic fluctuations which cannot be integrated out and which suffer significant particle production. In each case, to zeroth order in $l^{-1}$, the incoming negative frequency mode is proportional to $(t-i\epsilon)^{1\over 2} H_\nu^{(2)}\left(k(t-i\epsilon)\right)$ and it evolves to an outgoing negative frequency mode, so there is no particle production across the singularity. Similar behavior was encountered in Ref.~\cite{Tolley:2002cv}, which studied quantum fields on compactified Milne spacetime. In that case, $\epsilon$ was introduced to define a particular choice of continuation across the singularity. Here, we are in a better situation: $\epsilon$ is a physical parameter determined by the final state of the homogeneous ``background'' and for $\epsilon \neq 0$ the time evolution of the linearized perturbations is completely unambiguous. 


In the next three subsections, we compute the particle production to leading nontrivial order in the asymptotically free coupling $\lambda_\phi$. 

\subsection{Production of $\delta \phi$ Excitations with a Running Coupling}\label{secrunning}

In this subsection, we shall compute the evolution of the inhomogeneous modes when the logarithmic running is included in an expansion in $l^{-1}$.  It is convenient to ignore $\epsilon$ at first: having obtained the solution of (\ref{dpeq}) we can then simply replace $t$ with $t-i\epsilon$ and continue unambiguously from negative to positive times. 

The effect of the logarithm on the linearized perturbations at long wavelengths $|kt|\ll 1$ and at leading order in $l^{-1}$ is easily seen. A constant time delay $t_s({\bf x})$ in the classical background leads to a perturbation of $\phi$ of the form
\be
\label{tdelog} 
\delta \phi^{(1)} \sim {\delta \chi \over \chi^2} \sim  l^{1\over 2} h^2 {\delta 
\chi_l \over \chi_l^2} \sim l^{1\over 2} \, t_s({\bf x}) \, t^{-2}
\ee
to leading order in $l^{-1}$ as $t\rightarrow 0$. Whereas a local perturbation of the Hamiltonian density, $\delta {\cal H}({\bf x},0)\sim \rho({\bf x})$, leads from (\ref{enpert}) to a linearized perturbation 
\be
\label{enpertlog} 
\delta \phi^{(2)} \sim \rho({\bf x}) \dot{\phi} \int^t \dot{\phi}^{-2 } \sim l^{-{1\over 2}} \rho({\bf x}) t^3,
\ee
again to leading order in $l^{-1}$ as $t \rightarrow 0$. Here, we used $\dot{\phi} \propto -\dot{\chi}/\chi^2 = - l^{{1\over 2}}\, h^2 \,\dot{\chi}_l/\chi_l^2 \sim l^{{1\over 2}} t^{-2}$ at leading order in $l^{-1}$.  As we shall see, we need to include these logarithmic corrections in order to compute the production of particles and stress-energy perturbations as we follow the system across the bounce. At zeroth order in $l^{-1}$, the particle production vanishes - the in vacuum maps to the out vacuum at this order. But at first order in $l^{-1}$, there is particle production. 

The physical evolution of modes on different wavelengths may be pictured as follows. When modes of a given wavenumber $k$ are inside the effective horizon, {\it i.e.}, $|k t| \gg 1$, they are oscillatory and one must study their evolution using quantum field theory. However, once they have passed into the regime where $|k t| \ll 1$, spatial gradients become unimportant and the evolution is ultralocal. At this point, one can switch to a local, quantum mechanical description in which the field at every spatial point evolves independently, and in which, after carefully specifying the self-adjoint extension, the quantum wavefunction may be matched across the singularity.

From (\ref{tdelog}) and (\ref{enpertlog}), and taking into account the corrections in inverse powers of $l$, we expect to obtain one solution to (\ref{dpeq}) representing a time delay perturbation, of the form 
\be
\label{firstsol}
\delta \phi^{(1)} = l^{1\over 2} f^{(1)}(kt) + l^{-{1\over 2}} g^{(1)}(kt) +\dots, 
\ee
where $f^{(1)} \propto t^{-2}$ as $|kt| \rightarrow 0$ and subsequent terms in the series are suppressed by further powers of $l^{-1}$. We also expect a second solution, representing a local perturbation to the Hamiltonian density, of the form 
\be
\label{secondsol}
\delta \phi^{(2)} = l^{-{1\over 2}} f^{(2)}(kt) + l^{-{3\over 2}} g^{(2)}(kt) +\dots, 
\ee
with $f^{(2)} \propto t^{3}$ at small $|kt|$. Substituting these ans\"atze into (\ref{dpeq}), $f^{(1),(2)}$ obey \eq{feq}, as in the previous subsection, which is again solved by \eq{fsols}.

We can correct both these mode function solutions to next order in $l^{-1}$ by substituting (\ref{firstsol}) and (\ref{secondsol}) into (\ref{dpeq}) and equating the coefficients of the first subleading power in $l^{-1}$. We find
\ba
\label{geqs}
\ddot{g}^{(1)}+\left(k^2-{6\over t^2}\right) g^{(1)}&=& 2 {f^{(1)}\over t^2}+{\dot{f}^{(1)} \over t},\cr
\ddot{g}^{(2)}+\left(k^2-{6\over t^2}\right) g^{(2)}&=& 3 {f^{(2)}\over t^2}-{\dot{f}^{(2)} \over t},
\ea
whose particular solutions are 
\ba
\label{gsols}
g^{(1)}&=& {1\over 2} {\rm Si}(2kt) f^{(2)}(kt) +{1\over 2} {\rm Cin}(2kt) f^{(1)}(kt)
+{4\over 3} {\cos kt \over (kt)^2} + {1\over 3} {\sin kt \over kt},\cr 
g^{(2)}&=& {1\over 2} {\rm Si}(2kt) f^{(1)}(kt) -{1\over 2} {\rm Cin}(2kt) f^{(2)}(kt)
+{1\over 3} {\cos kt \over kt} - {4\over 3} {\sin kt \over (kt)^2},
\ea
where it is conventional to define the integrals
\be
{\rm Si}(y)\equiv \int_0^y dx {\sin x \over x}, \qquad {\rm Cin}(y)\equiv \int_0^y dx {(1- \cos x) \over x}.
\ee
In fact, the solutions (\ref{gsols}) are only determined up to an arbitrary multiple of $f^{(1)}$ and $f^{(2)}$, respectively (note that only one of them is consistent with the parity of either solution), but such an extra contribution shall not be important to us since it does not contribute to the particle number $\langle n \rangle $ at the leading order in $l^{-1}$ to which we calculate. 

Equations (\ref{firstsol}) and (\ref{secondsol}), with $t$ replaced by $t-i\epsilon$ throughout, solve (\ref{dpeq}) in our complex classical background.  Since the background avoids the singularity at $\chi=0$, there is no ambiguity in the evolution of the perturbation modes. 
We now want to take the linear combination of $\delta \phi^{(1)}$
and $\delta \phi^{(2)}$ that behaves as the incoming negative frequency mode, and evolve it to a time well after the quantum ``bounce," when it is undergoing oscillations with an adiabatically changing frequency once more, and it makes sense to calculate the particle number. 

We work to leading order in an expansion in inverse powers of the logarithm, which we assume is large. The classical background solution which dominates is the ``image" solution starting behind
the singularity: according to the brief discussion in Subsection~\ref{bcsect}, we must define the logarithm
to be real below the positive real $\chi$ axis (where $\chi=1/\phi$) , {\it i.e.}, as ${1\over 2} \left(\ln(\chi)+\ln({-\chi})-i\pi\right)$. Recall that at leading order in the large logarithm, {\it i.e.} ignoring $\ln(\phi/NM)$ where it occurs inside the logarithm, we have $\chi \propto t-i\epsilon = \tilde{t}$ in the classical  background. 

We start by extracting the time-dependence in the logarithm, as 
follows:
\be
\label{logapprox}
\ln (\phi/NM)\equiv l \rightarrow {\rm ln} ({k \over M})  - {1\over 2} \left(\ln(-k \tilde{t})+\ln(k \tilde{t})-i\pi\right) \equiv  l_0 -{1\over 2} \left(\ln(-k\tilde{t})+\ln(k\tilde{t})-i\pi\right),
\ee
where we ignored $\ln(\phi/NM)$ inside the logarithm. Our asymptotic ``in" and ``out" regions will be defined where $|kt|$ is large, so that the mode evolution is oscillatory, but $|Mt|$ is small so that $l$ is still large. We now identify the negative frequency mode in an expansion in $l_0^{-1}$.

At zeroth order in $l_0^{-1}$, the incoming negative frequency mode is just  $-3 l_0^{-{1\over 2}} \delta \phi^{(1)}+3i l_0^{1\over 2} \delta \phi^{(2)}$ and this evolves across $t=0$ to the negative frequency outgoing mode. At next order, we must include the corrections $g^{(1)}$ and $g^{(2)}$ derived in (\ref{gsols}). When taking the large $|kt|$ limit, we use 
\be
\label{CSin}
{\rm Si}(2kt)\rightarrow {\pi \over 2}-\int_{2kt}^\infty {dx \sin x \over x} ,\qquad {\rm Cin}(2kt)\rightarrow {\rm ln}(kt) +\gamma +\int_{2kt}^\infty {dx \cos x \over x},
\ee
where the integrals are easily expanded as a series in $1/(kt)$ by repeated integrations by parts. Notice that in the complex $t$-plane, ${\rm Si}(2kt)$ is an odd function of $t$ whereas ${\rm Cin}(2kt)$ is an even function. 

For large negative times $ -(k/M) \ll tk \ll -1$, we find 
\ba
\label{finalsol1}
\delta \phi^{(1)} &\rightarrow& l_0^{1\over 2} \left(1-{1\over 2}{{\rm ln} (-kt) \over l_0} +i{\pi \over 2 l_0}\right) \left(-{1\over 3} \cos (kt) \right)\cr
&& +  l_0^{-{1\over 2}}\left(-{1\over 6} \cos kt \,({\rm ln} (-2kt) +\gamma) -{\pi \over 12} \sin kt\right)\cr  
&=& l_0^{1\over 2}\left( -{1\over 3} \cos kt  -{\pi \over 12 l_0} \sin kt \right) -{1\over 6} l_0^{-{1\over 2}} \cos kt (i \pi +\gamma +\ln 2).
\ea
Notice that the time-dependent logarithm from the first term has canceled with the logarithm in the correction $g^{(1)}$ to the second term. A similar cancellation occurs in $\delta \phi^{(2)}$:
\ba
\label{finalsol2}
\delta \phi^{(2)} &\rightarrow&  l_0^{-{1\over 2}}\left( {1\over 3} \sin kt  +{\pi \over 12 l_0} \cos kt \right) -{1\over 6} l_0^{-{3\over 2}}\sin kt (i\pi +\gamma+\ln 2).
\ea
Similarly, at large $kt$ we obtain 
\be 
\label{finalsolpos}
\delta \phi^{(1)} = l_0^{1\over 2}\left( -{1\over 3} \cos kt  +{\pi \over 12 l_0} \sin kt \right) -{1\over 6} l_0^{-{1\over 2}} \cos kt (\gamma +\ln 2),
\ee
and
\ba
\label{finalsolpos2}
\delta \phi^{(2)} &\rightarrow&  l_0^{-{1\over 2}}\left( {1\over 3} \sin kt  -{\pi \over 12 l_0} \cos kt \right) -{1\over 6} l_0^{-{3\over 2}}\sin kt (\gamma+\ln 2).
\ea
We now replace $t$ by $t-i\epsilon$ and read off the incoming negative frequency mode
\ba 
&& R^{(-)}_{in}\cr
&=&  e^{-k\epsilon} \left( -3 l_0^{-{1\over 2}} \delta \phi^{(1)} +3 i l_0^{1\over 2} \delta \phi^{(2)} \right)-e^{-k \epsilon} l_0^{-1} \left({3 i\pi \over 4} +{\gamma+\ln 2 \over 2} \right) \left( -3 l_0^{-{1\over 2}} \delta \phi^{(1)} -3 i l_0^{1\over 2} \delta \phi^{(2)} \right) \cr
&\rightarrow& e^{ikt} \qquad \qquad  \qquad \quad        -(k/M) \ll kt \ll -1 \cr
&\rightarrow& e^{ikt}  -{i\pi \over l_0} e^{-2 k\epsilon} e^{-ikt}     \qquad 1\ll kt \ll k/M.
\label{posfreqmode}
\ea
The last term describes the particle production, from (\ref{expecn}) we have 
\be 
\langle n \rangle = {\pi^2 \over l_0^2} e^{-4k \epsilon},
\label{partino}
\ee
explicitly revealing the exponential UV cutoff at large $k$ due to the complex classical solution for the background missing the singularity at $\chi=0$ by the finite quantity $\epsilon$. 

\begin{figure}
\begin{center}
\epsfig{file=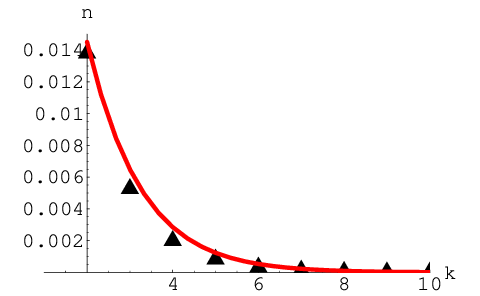, width=10cm}
\end{center}
\caption{Numerical calculation of particle production ($\langle n \rangle$ versus $k$) in the complex background of Figure 9, compared with the analytical calculation given in (\ref{partino}).}
\label{anvsnum}
\end{figure}

As a check of the validity of our approximations, we have computed $\langle n \rangle $ numerically, using the full complex solution for the background, shown in Figure~\ref{hombg}, and a numerical solution of the mode equations at each $k$ with the field fluctuation in the incoming negative frequency mode. Figure \ref{anvsnum} compares the analytical result (\ref{partino}), shown as the solid curve,  with the numerical result of the full computation. 

We can now write down the energy density in $\delta\phi$ particles produced:
\be
\rho_{c, \delta\phi}= \int {d^3{\bf k} \over (2 \pi)^3} k {\pi^2 \over \ln^2(k/M)} e^{-4k \epsilon},
\label{parts}
\ee
where, as we have emphasized, the integral is only to be taken over those values of $k$ so large that the transition from oscillatory to ultralocal evolution occurs while the homogeneous mode was accurately following the zero-energy scaling solution.

\subsection{Production of Light Higgs Particles}\label{subsec:higgs}

To compute the quantum production of light Higgs particles, namely the excitations of $\Phi_i$, $i=2,\dots 6$, which commute with $\Phi_1$ and whose mass is proportional to $\lambda_\phi$, we must solve (\ref{lhiggseom}) to first order in $l^{-1}$. We proceed similarly to the previous subsection. Setting $\tilde{t}=t-i\epsilon$, we set 
\be 
\Phi_i= l^{\beta} (f(\tilde{t})+ g(\tilde{t})l^{-1} +\dots),
\label{Phians}
\ee 
 and solve (\ref{lhiggseom}) order by order in $l^{-1}$. The two linearly independent solutions for $f$ at small $\tilde{t}$ are easily identified: $f_\pm \sim \tilde{t}^{{1\over2}\pm ia}$, with $a=\sqrt{3/20}$. At next order in $l^{-1}$, (\ref{lhiggseom}) yields $\beta_\pm= \mp 2i/\sqrt{15}$. These solutions are extended to large times as follows. Solving (\ref{lhiggseom}) to lowest order in $l^{-1}$, we find $f_\pm \sim \tilde{t}^{1\over 2} J_{\pm ia}(k\tilde{t})$. At next order, we obtain 
\be
 \ddot{g}_\pm +{2\over 5 \tilde{t}^2} g_\pm +k^2 g_\pm =  {2 \beta_\pm \over \tilde{t}} \dot{f}_\pm - {\beta_\pm+{2\over 5} \over \tilde{t}^2} f_\pm.
\label{lhiggseoma}
\ee
The right hand side is easily evaluated using standard Bessel function relations to be $-2 \beta_\pm k \tilde{t}^{-1/2} J_{1\pm ia}(k\tilde{t})$. We now solve (\ref{lhiggseoma}) using a Green's function, obtaining
\be
g_\pm(\tilde{t})= \pm {\pi \beta_\pm k i \over \sinh a\pi} \tilde{t}^{1\over 2} 
\left( J_{\pm ia } (k\tilde{t}) \int_0^{\tilde{t}} dt' J_{1\pm ia}(kt') J_{\mp ia}(kt') -
J_{\mp ia } (k\tilde{t}) \int_0^{\tilde{t}} dt' J_{1\pm ia}(kt') J_{\pm ia}(kt')\right),
\label{gsol}
\ee
where we used the Wronskian $t^{1/2}\left(\dot{J}_{ia}(kt) J_{-ia}(kt)-\dot{J}_{-ia}(kt) J_{ia}(kt)\right)=(2 i \sinh a \pi) /\pi$. The first term in (\ref{gsol}) is proportional to $f_\pm(\tilde{t})$. Using the large-argument asymptotic form for Bessel functions,
\be
J_{\nu}(kt) \sim \sqrt{2 \over \pi k t} \cos (kt - \nu {\pi\over 2} -{\pi \over 4}), \qquad kt \gg 1,
\label{asymbess} 
\ee
we see the first (second) integrand in (\ref{gsol}) tends to $\mp i (\sinh a \pi)/(\pi kt')$ plus oscillatory terms which integrate to a constant. The first integral in (\ref{gsol}) thus diverges logarithmically, giving a contribution to $g_\pm$ of the form $\beta_\pm \ln (k\tilde{t}) \tilde{t}^{1/2}  J_{\pm ia}(k\tilde{t})$. This is just what is required to cancel the time dependence of the overall coefficient $l^{\beta}$ in (\ref{Phians}) to first order in $\ln(k\tilde{t})/l_0$, where $l_0 \equiv \ln(k/M)$ and we are interested in the regime $1 \ll |k\tilde{t}| \ll k/M$ as in the previous subsection. Recalling that $l \approx l_0 -\ln (-k \tilde{t})+i\pi$ for $-k/M \ll k t \ll -1$ and $l \approx l_0 -\ln (k \tilde{t} )$ for $1\ll kt \ll k/M$, we can check that the same cancellation occurs for large negative $t$ using $J_{-ia}(kt) = e^{\pi a} J_{-ia}(-kt)$ and $J_{1+ia}(kt) = - e^{-\pi a} J_{1+ia}(-kt)$, which hold when $t$ is analytically continued to negative values by passing below the origin. The coefficient of $J_{\pm ia}(k\tilde{t})$  in the first term of (\ref{gsol}) is hence an even function of $\tilde{t}$, tending to $\beta_\pm \ln (|kt|)$ for large $|kt|$. 

The second term in (\ref{gsol}) involves an integral which converges at large $|t'|$. Here, we can make use of the formula $\int_0^\infty dx J_{1+\xi}(kx) J_\xi(kx)= 1/(2k)$,~\cite{gradshteyn} to obtain for our two linearly independent modes, to first order in $l_0^{-1}$ and at large $|t|$ (still $\ll M^{-1}$),
\ba 
 \hskip-.2in \Phi_{i,\pm} 
\rightarrow && \hskip-.2in i e^{\mp \pi a} l_0^{\beta_\pm} (-\tilde{t})^{1/2}\left( J_{\pm ia}(-k\tilde{t})(1+{i \pi \beta_\pm \over l_0})   \mp {i \pi \beta_\pm \over l_0 \sinh a \pi} J_{\mp ia} (-k\tilde{t})\right), \, kt \ll -1 \cr 
\rightarrow && \hskip-.2in l_0^{\beta_\pm} \tilde{t}^{1/2}\left( J_{\pm ia}(k\tilde{t}) \mp {i \pi \beta_\pm \over l_0 \sinh a \pi} J_{\mp ia} (k\tilde{t})\right), \qquad \qquad kt \gg 1.
\label{phiplus}
\ea
Using (\ref{asymbess}) we now construct the incoming negative frequency mode as a linear combination of $\Phi_{i,\pm}$, and then use (\ref{phiplus}) to evolve it to large positive $kt$. Explicitly, at large negative $kt$, we have 
\be 
R_+ \sim e^{ik t} \sim e^{-\epsilon k} e^{-i \pi/4} {1\over 2} \sqrt{\pi (-k\tilde{t}) \over 2} \left( J_{ia}(-k\tilde{t}) ({1\over c}-{1\over s}) +J_{-ia}(-k\tilde{t}) ({1\over c}+{1\over s})\right), kt \ll -1,
\label{posf}
\ee 
where $c\equiv \cosh(\pi a/2)$ and $s=\sinh (\pi a/2)$. Evolving to large positive $kt$ using (\ref{phiplus}) and working to first order in $l_0^{-1}$, we find
\be
R_+ \sim e^{i \pi/4} \left( e^{ikt} (1+2 i {\beta_+ \pi \over l_0}) +i e^{-ikt} e^{-2 \epsilon k} {\pi \beta_+ \over l_0} \coth \pi a \right). 
\label{finalmatch} 
\ee
The last term yields our final result for the mode-mixing coefficient. Using $\beta_+= -2 i/\sqrt{15}$, $a=\sqrt{3/20}$, we infer the energy density in light Higgs particles
\be
\rho_{c, \Phi}= \int {d^3{\bf k} \over (2 \pi)^3} k {4 \pi^2 \coth^2 (\pi \sqrt{3/20}) \over 15  \ln^2(k/M)} e^{-4k \epsilon},
\label{partsh}
\ee
which is parametrically of the same order as the result (\ref{parts}) for $\delta \phi$ particles.
However, we have still to multiply by the total number of light Higgs particles. The simplest case to consider is where the $SU(N)$ symmetry is broken by $\Phi_1$ to $U(1)^{N-1}$. In this case, there are of order $N$ surviving light Higgs bosons whose mass will be due solely to the deformation term. Hence the result (\ref{partsh}) is multiplied by $N$. (In more general cases, there could be of order $N^2$ such light Higgs modes.) 

\subsection{Production of Massless Gauge Bosons}\label{subsec:gluons}

To calculate the production of remaining light gauge bosons, we need to solve equation (\ref{gbe}) to the first nontrivial order in $l^{-1}$. The effect of the $l^{-2}$ correction is more modest in this case and a suitable ansatz for the transverse mode functions is
\be 
A_\perp = A_0 \sin kt \,(1- C l^{-1}) +A_1 \cos kt + g l^{-2} +\dots.
\ee 
Substituting into (\ref{gbe}) we obtain 
\be 
\ddot{g}+k^2 g = A_0 C( k t^{-1} \cos kt -t^{-2} \sin kt) +A_1 C kt^{-1} \sin kt,
\ee
whose solution is 
\bea 
g= &&{1\over 2} C A_0\left[ \sin kt \, ({\rm Cin}(2kt)-1) +\cos kt \,{\rm Sin} (2kt) \right]\cr
&&+{1\over 2} C A_1\left[ \sin kt \,\, {\rm Sin}(2kt) -\cos kt \,\,{\rm Cin} (2kt) \right]. 
\label{gasol}
\eea
Using the large time limits of the Cin and Sin functions, as given in Subsection~\ref{secrunning}, we may obtain the behavior of an incoming negative frequency mode $A_\perp^{(+),in} \sim e^{ikt} $, $ k/M \ll kt \ll -1$, including the effect arising from the continuation of $l$, to obtain
\be 
A_\perp^{(+),in} \rightarrow e^{ikt} +{i \pi C \over 2 l_0^2} e^{-ikt}.
\ee 
From this we read off the mode mixing coefficient, and hence the energy density in created light gauge bosons:
\be
\rho_{c, A}= \int {d^3{\bf k} \over (2 \pi)^3} k {\pi^2 C^2 \over 4 
\ln^4(k/M)} e^{-4k \epsilon},
\label{partsa}
\ee
which is suppressed relative to the other cases by the square of the logarithm. 

As for the light Higgs particles, we can discuss the dependence on $N$. If the $SU(N)$ symmetry breaks to $U(1)^{N-1}$, then there are of order $N$ massless bosons remaining, so the result (\ref{partsa}) is multiplied by $N$. However, the diagram which gives rise to the coupling $C$ involves only one closed line (hence one power of $N$) and is proportional to $\lambda_\phi \sim N^{-2}l^{-1}$. Hence $C$ scales as $N^{-1}$. (For symmetry breaking to a larger group, there can be of order $N^2$ massless gauge bosons remaining.) The quantum production of massless gauge bosons is always subdominant to light Higgs production, by two powers of $1/\ln(k/M)$ and by two powers of $1/N$.

\setcounter{equation}{0}
\section{Backreaction} 

We now want to estimate the backreaction effect of particle production on the homogeneous mode of the field $\phi$. We do so using energy conservation: the total energy is always zero so any energy which goes into particles will lower the energy available in the homogeneous mode and therefore cause it to sink lower down the negative potential. 

One possible criterion for the backreaction to be small is that the energy density $\rho_c$ in created particles should be small compared to (minus) the potential energy density reached at times $|t|\sim\epsilon$. In the previous subsections, we have shown that most of the energy density goes into light Higgs particles. Keeping in mind that the number of light Higgs species is of order $N$ and that for most of the final wavefunction of a minimal spread wavepacket $\epsilon\sim R_{AdS}/(N|\ln(MR_{AdS})|^{1/2})$, we find from \eq{partsh} that 
\be\label{rhochiggs}
\rho_c\sim{N\over\epsilon^4\,|\ln(MR_{AdS})|^2}\sim {N^5\over R_{AdS}^4}.
\ee 
Comparing this with (minus) the potential energy at $|t|\sim\epsilon$,
\be
{1\over4}\lambda_\phi\phi^4\sim {N^2|\ln(M\epsilon)|\over \epsilon^4}\sim {N^2|\ln(MR_{AdS})|\over \epsilon^4},
\ee
we see that the energy density in created particles is smaller by powers of $1/N$ and $1/|\ln(MR_{AdS})|$, suggesting that backreaction is small. (We are assuming $MR_{AdS}\ll 1$, so that the quartic coupling near the starting value of $\phi$, $f\sim 1/|\ln(MR_{AdS})|$, is small and perturbation theory reliable.)

However a more conservative criterion for the backreaction to be small is that the homogeneous mode should keep enough energy to roll back up the hill to nearly its starting value. The potential for $\phi$ is
\be
V={\phi^2\over 6 R_{AdS}^2}-{\lambda_0\phi^4\over4\ln(\phi/NM)}.
\ee
Setting $V=0$ to obtain the starting value of $\phi$, we find
\be
\phi_{start}=\sqrt{2\over3\lambda_0}{1\over R_{AdS}\ln(\phi_{start}/NM)^{1/2}}\approx\sqrt{2\over3\lambda_0}{1\over R_{AdS}|\ln(MR_{AdS})|^{1/2}}.
\ee
The minimal value of $\phi$ reached after the bounce is obtained by setting $V(\phi_{min})=-\rho_c$. Demanding that $\phi_{min}-\phi_{start}\ll\phi_{start}$ in order to have small backreaction, one finds the condition $\rho_c\ll N^2|\ln(MR_{AdS})|/R_{AdS}^4$. Using \eq{rhochiggs}, this becomes
\be\label{logN3}
|\ln(MR_{AdS})|\gg N^3
\ee
which does not hold.

Hence for typical final values of $\phi$ carrying most of the probability, backreaction is large.
The fraction of the probability associated with final $\phi$ values for which $\epsilon$ is sufficiently large so that backreaction is negligible, and hence our perturbative calculation applies, is parametrically small. Therefore we conclude that in this class of models, a bounce is highly improbable.

Now we would like to tie up some loose ends from previous (sub)sections. First, are non-linear corrections to the fluctuation equation \eq{dpeq} really small? It is easy to see from the full equation of motion that non-linear corrections will be small as long as $|\delta\phi|\ll|\bar\phi|$. In our setup, the perturbations $\delta\phi$ have a quantum mechanical origin: as the singularity is approached, more and more Fourier modes ``freeze'' and start behaving classically. To estimate the typical magnitude of the perturbation $\delta\phi$, let us compute the contribution of frozen modes to the expectation value of $\delta\phi^2$. {}From \eq{fsols}, we see that a properly normalized frozen negative frequency mode with wavenumber ${\bf k}$ behaves like
\be
\delta\phi_{\bf k}\sim {e^{i{\bf k}\cdot{\bf x}}\over k^{5/2}\tilde t^2},
\ee
so that 
\be
\langle \delta\phi({\bf x})\delta\phi({\bf x})\rangle|_{\rm frozen}\sim{1\over \tilde t^4}\int {d^3k\over k^5}\sim {R_{AdS}^2\over \tilde t^4}, 
\ee
with $\tilde t=t-i\epsilon$ (see Subsection~6.2) and where in the last step we have used the infrared cutoff $k \ge 1/R_{AdS}$ due to the finite volume of $S^3$. So the typical size of fluctuations at time $t$ is
\be
\delta\phi \sim {R_{AdS}\over \tilde t^2}.
\ee
This we need to compare to the homogeneous background
\be
\bar\phi\sim{1\over\sqrt{\lambda_\phi}\tilde t}.
\ee
The most dangerous region is around $t=0$, where 
\be
|\tilde t|= \epsilon \sim \sqrt{\lambda_\phi}R_{AdS}.
\ee
Therefore, nonlinearities will be important for most of the wavefunctions very near the singularity where
$|\delta\phi|  \sim  |\bar\phi| $. Similarly, one finds that near $t=0$, the fluctuations may have a significant effect on the homogeneous mode except in the tail of the wavefunction.


Finally, in Subsection~4.2 we assumed that the surface $\Sigma$ where the classical field $\phi$ reaches infinity, given by $t=t_s({\bf x})$, was smooth and space-like. Now we would like to verify that $|\nabla t_s|<1$ so that $\Sigma$ is indeed space-like, for typical classical solutions $\phi$ generated by the quantum mechanical initial state we have considered. Below \eq{chilin}, we concluded that $t_s$ is the time delay mode. In the linearized approximation, close to the singularity and to leading order in gradients, it can be related to the field fluctuation $\delta\phi$:
\be
t_s({\bf x})= {\delta\phi \over \dot{\bar\phi}}|_{t=0}\sim \sqrt{\lambda_\phi}\,\tilde{t}^2\delta\phi|_{t=0}.
\ee
To estimate  $\nabla t_s$, we compute the contributions of the frozen modes%
\footnote{
These are the only modes that have become classical and thus contribute to the classical perturbation that determines the shape of $\Sigma$.
} 
to the two point function
\be\label{gradts}
\langle(\nabla t_s)^2\rangle |_{\rm frozen}\sim \lambda_\phi\tilde t^4\int d^3k{k^2\over k^5\tilde t^4}\sim \lambda_\phi\int {dk\over k}.
\ee
This is indeed small compared to 1 since $\lambda_\phi$ is very small and the integral has both an IR cutoff due to the finite volume of $S^3$, and a UV cutoff because modes with $k>1/\epsilon$ never freeze: we conclude \eq{gradts} is suppressed by inverse powers of the large parameters $N$ and $\ln(MR_{AdS})$. Thus typical surfaces $\Sigma$ generated from our quantum initial conditions are indeed smooth and spacelike, as assumed in Subsection~4.2.

\setcounter{equation}{0}
\section{Conclusions}

We have studied the AdS/CFT dual description of solutions of ${\cal N}=8$ gauged supergravity in five dimensions where smooth asymptotically AdS initial data evolve to a big crunch singularity in the future. At the singularity the classical supergravity description breaks down, but we have explored the possibility that the dual field theory evolution provides a consistent quantum description of the singularity. 

The dual theory is an unstable deformation of $\N=4$ supersymmetric $SU(N)$ gauge theory on $\Rbar\times S^3$, and the big crunch singularity in the bulk occurs when a boundary scalar field runs to infinity, in finite time. One can define consistent quantum evolution in the boundary theory by considering a self-adjoint extension of the system. We have developed a method that enables one to calculate whether the boundary scalar bounces back from infinity, which would predict a quantum transition from a big crunch to a big bang in the bulk.

Specifically we have calculated the semiclassical wavefunction for the unstable boundary scalar, including the effect of inhomogeneous perturbations in all fields to leading order.
To leading order in the semiclassical expansion, the wavefunction is determined by (spatially inhomogeneous) complex classical solutions. The self-adjoint boundary conditions can be implemented in the semiclassical expansion by the method of images. The resulting wavefunction for the homogeneous wave packet rolls down the potential and bounces back. 

The homogeneous mode of the boundary scalar is quantum mechanical because the boundary theory is defined on a finite space. This leads to a natural UV cutoff in particle production as the wavefunction for the homogeneous mode bounces back from infinity. 
But in order for the expectation value of  the boundary scalar to return close to its original form the backreaction of particle production across the bounce must be sufficiently small. 
We find that for most final values of the homogeneous mode $\bar\phi$, the logarithmic running of the boundary coupling governing the instability leads to significant particle production. In particular for most of the wave packet our perturbative calculation of particle creation indicates that the backreaction of particles appears to prevent the homogeneous mode from rolling back up the potential. Hence we find that a transition from a big crunch to a big bang is rather improbable in this class of models. 

At the same time, however, our calculations indicate that the situation may be rather different in models where conformal invariance remains unbroken at the quantum level. In future work we hope to present a concrete model where this is realized.

\section*{Acknowledgments}
We would like to thank N.~Arkani-Hamed, O.~Evnin, D.~Gross, G.~Horowitz, S.~Kachru, S.~Kovacs, J.~Maldacena, R.~Myers, E.~Silverstein, K.~Skenderis, D.~Tong, H.~Verlinde and especially N.~Dorey for useful discussions. B.C.\ and T.H.\ thank the Centre for Theoretical Cosmology in Cambridge for its hospitality at various stages of this project. T.H.\ and N.T.\ are grateful for the hospitality of the Mitchell family at their Cook's Branch Conservancy and to Chris Pope for hospitality at the Mitchell Institute at Texas A\&M University. T.H.\ thanks the KITP in Santa Barbara for support. The work of B.C.\ was supported in part by the Belgian Federal Science Policy Office through the Interuniversity Attraction Poles IAP V/27 and IAP VI/11, by the European Commission FP6 RTN programme MRTN-CT-2004-005104 and by FWO-Vlaanderen through projects G.0428.06 and G011410N. N.T.\ acknowledges the support of STFC(UK) and of the Centre for Theoretical Cosmology in Cambridge.

\appendix

\section{More on the Bulk Theory}

In this appendix, we first show that the bulk theory discussed in Section~2, with action
\be \label{lagra}
S = \int \sqrt{-g}\left[ \frac{1}{2} R -\frac{1}{2}(\nabla \phi )^2 +
\frac{1}{4 R_{AdS}^2}\(15e^{2\gamma \phi}+10e^{-4\gamma \phi}-e^{-10\gamma\phi} \)
\right ],
\ee
admits an $O(5)$-invariant Euclidean instanton solution of the form
\be \label{inst5d}
ds^2 = R_{AdS}^2\left(\frac{d\rho^2}{ b^2(\rho)} +\rho^2 d\Omega_4\right)
\ee
with $\phi=\phi(\rho)$ and for scalar boundary condition $\a_f = f\b$.

The field equations determine $b$ in terms of $\phi$. Asymptotically one finds
\be \label{asb5d}
b^2=  \rho^2  +1 +\frac{ 2\alpha^2 (\ln \rho)^2}{ 3 \rho^{2}}
+\frac{\alpha(4 \beta -\alpha) \ln \rho }{3 \rho^{2}}
+\frac{8 \beta^2 -4 \alpha \beta +\alpha^2 }{ 12 \rho^{2}},
\ee
and the scalar field $\phi$ itself obeys
\be\label{inst2}
b^2 \phi'' + \left( \frac{4 b^2 }{\rho} +bb' \right) \phi' - R_{AdS}^2 V_{,\phi} =0,
\ee
where prime denotes $\partial_{\rho}$.

\begin{figure}[t]
\includegraphics[width=3in]{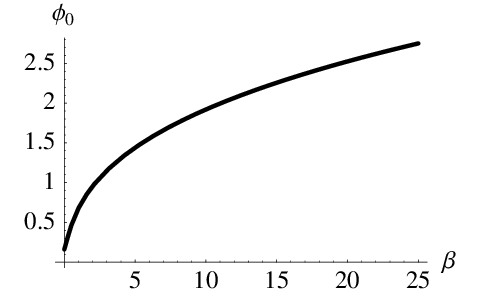} \hfill
\includegraphics[width=3in]{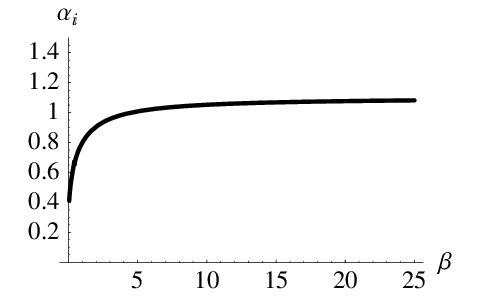}
\caption{The value of $\phi$ at the origin as a function of $\b$ (left), and the function $\a_{i}(\b)$ (right) obtained by integrating the field equations outward starting with different values for $\phi_0$.}
\label{inst}
\end{figure}

Regularity at the origin requires $\phi'(0)=\phi_0' =0$. Thus the instanton solutions can be labeled by $\phi_0$, the value of $\phi$ at the origin. For each $\phi_0$, one can integrate (\ref{inst2}) and get an instanton. Asymptotically one finds $\phi(\rho)= \alpha \ln \rho/\rho^2 +\beta/\rho^2$, where $\a$ and $\b$ are now constants. Hence for each $\phi_0$ one obtains a point in the $(\a,\b)$ plane. Repeating for all $\phi_0$ yields a curve $\a_i(\b)$ where the subscript indicates this is associated with instantons. This curve is plotted in Figure~\ref{inst} (right panel). The left panel of Figure~\ref{inst} shows $\phi_0$ as a function of $\b$. 

The slice through the instanton obtained by restricting to the equator of the $S^4$ defines time symmetric initial data for a Lorentzian solution. The Euclidean radial distance $\rho$ simply becomes the radial distance $r$ on the initial data slice. So given a choice of boundary condition $\a(\b)$, one can obtain suitable initial data by first selecting the instanton corresponding to a point where the curve $\a_i(\b) $ intersects $\a(\b)$, and then taking a slice through this instanton.

One immediately sees that all $\a_f$ boundary conditions, for $f>0$, admit precisely one instanton solution. When $f \rightarrow 0$ one has  $\phi_0 \rightarrow \infty$. The dual field theory is supersymmetric in this limit, so one expects there to be no regular instanton solutions that describe the decay of the AdS state in the bulk theory. For small positive $f$ one sees that $\b \sim 1/f$. This scaling is nicely reproduced by the solutions of the simplified model field theory discussed in the text. From the asymptotic form (\ref{asb5d}) of $b$ it follows that the conserved mass (cf.\ (\ref{mass})) of the resulting initial data is $M= -\pi^2 R_{AdS}^2 \alpha^2/4$, so $M \approx -2 R_{AdS}^2$ for small $f$. The instantons therefore specify initial data with {\it negative mass} in this theory. It is only for AdS-invariant boundary conditions discussed below that instantons of this type define initial data of exactly zero mass, in line with their
interpretation as the solution $AdS_5$ decays into.\footnote{One can also show that these instantons
have finite action, which is large for small $f$ and small for large $f$.}

\bigskip
{\it Lorentzian Evolution}
\bigskip

Analytic continuation of the instanton geometry yields a Lorentzian solution that describes the evolution 
of instanton initial data under AdS-invariant boundary conditions \cite{Coleman80,Hertog:2004rz}.
These differ slightly from $\a_f$ boundary conditions and are most conveniently expressed as \cite{Hertog:2004dr,Henneaux:2004zi}
\be\label{asads}
\a (1 - \frac{f }{ 2}  \ln \a)=f\b,
\ee
where $f$ is again an arbitrary constant. One sees in particular that rescaling $r$ leaves $f$ unchanged. One also sees that the $\a_f$ boundary conditions considered in the text are approximately AdS-invariantor when $f$ is small, which is the parameter regime we will concentrate on. This restricts us to positive $f$ since $f<0$ boundary conditions admit instanton solutions only for $\vert f \vert > {\cal O}(1)$.

After analytic continuation the origin of the Euclidean instanton becomes the lightcone of the Lorentzian solution. Inside the lightcone, the $SO(4,1)$ symmetry ensures that the solution evolves like an open 
FRW universe,
\be \label{ametrica}
ds^2 = -dt^2+ a^2(t) d\sigma_4,
\ee
where $d\sigma_4$ is the metric on the four-dimensional unit hyperboloid. Under evolution $\phi$ rolls down the negative potential. This causes the scale factor $a(t)$ to vanish in finite time, producing a big crunch singularity. Outside the lightcone, the solution is given by (\ref{inst5d}) with $d\Omega_4$ replaced by four-dimensional de Sitter space. The scalar field $\phi$ remains bounded in this region. On the light cone we have $\phi=\phi_0$ and $\partial_{t}\phi=0$ (since $\phi_{,\rho}=0$ at the origin in the instanton).

One can verify that this analytic continuation indeed satisfies the boundary conditions (\ref{asads}) by doing a coordinate transformation in the asymptotic region outside the light cone. The relation between the usual static coordinates (\ref{adsmetric}) for $AdS_5$ and the $SO(4,1)$ invariant coordinates,
\be
ds^2 = R_{AdS}^2\left(\frac{d\rho^2}{ 1+\rho^2} + \rho^2 (-d \tau^2 + \cosh^2\tau d\Omega_3)\right),
\ee
is
\be
\rho^2 = r^2 \cos^2 t -\sin^2 t.
\ee
Hence the asymptotic behavior of $\phi$ in global coordinates is given by
\be \label{assc}
\phi(r) = \frac{\tilde \alpha \ln r}{ r^2} +\frac{\tilde\b }{ 2r^2},
\ee
where $\tilde \alpha= \alpha_i/\cos^2 t$ and $\tilde \b$ is given by (\ref{asads}) with $\a$ replaced by $\tilde \a$. Hence $\tilde \a$ is now time dependent and  blows up as $t \rightarrow \pi/2$, when the singularity hits the boundary.

Although for $\a_f$ boundary conditions one can in general not obtain the solution by analytic continuation of the instanton, we have argued in Section~2 that when $f$ is small and positive, the effect of the change in the boundary conditions on the evolution is negligible except in the corners of the conformal diagram where the singularity hits the boundary at infinity. In particular, since the evolution of the initial data has trapped surfaces, a singularity will still form in the central region. However, a priori it is now possibe that the singularity is enclosed inside a large black hole and does not extend out to infinity. We now investigate this possibility.

\bigskip
{\it Black Holes with Scalar Hair}
\bigskip

We now numerically integrate the field equations derived for static spherical solutions to verify if the theory (\ref{lagra}) for $\a_f$ boundary conditions admits a class of static, spherical black holes with scalar hair outside the horizon. In light of the discussion above, we would like to know whether there are negative mass black holes with scalar hair.\footnote{Negative mass hairy black holes were found in closely related theories in four dimensions \cite{Hertog:2005hu}.} Writing the metric as
\be 
ds_5^2=R_{AdS}^2\left(-h(r)e^{-2\delta(r)}dt^2+h^{-1}(r)dr^2+r^2d\Omega_3^2\right),
\ee
the Einstein equations read
\be \label{field5d1}
h\phi_{,rr}+\left(\frac{3h}{r}+\frac{r}{3}\phi_{,r}^2h+h_{,r} 
\right)\phi_{,r} =R_{AdS}^2 V_{,\phi}, 
\ee
\be \label{field5d2}
2(1-h)-rh_{,r}-\frac{r^2}{3}\phi_{,r}^2h=\frac{2}{3}r^2 R_{AdS}^2 V(\phi),  
\ee
\be \label{field5d3}
\delta_{,r}=-\frac{r}{3}\phi_{,r}^2.
\ee
We integrate the field equations outward from the horizon.
Regularity at the event horizon $r=R_e$ requires
\be \label{horcon}
\phi_{,r}(R_{e}) =\frac {R_{e} R_{AdS}^2 V_{,\phi_{e}} }{ 2-2R_{e}^2 R_{AdS}^2 V(\phi_{e})/3}.
\ee

\begin{figure}[t]
\begin{center}
\includegraphics[width=3in]{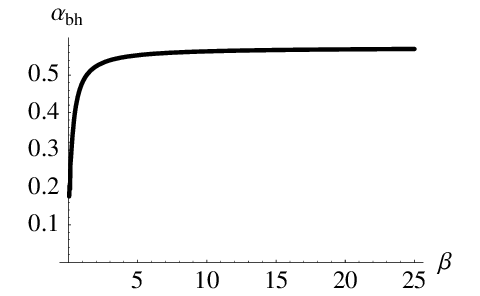} 
\end{center}
\caption{The function $\alpha_{bh}(\beta)$ obtained by integrating the field equations outward starting with different values for $\phi_e$ at the horizon $R_e=0.1$ of spherical hairy black holes.}
\label{bhalpha}
\end{figure}

The scalar field asymptotically behaves as (\ref{asscalar}), so we obtain a point in the $(\a,\b)$ plane for each combination $(R_e,\phi_e)$.  Repeating for all $\phi_e$ gives a curve $\a_{R_e}( \b)$. We show this curve in Fig \ref{bhalpha} for $R_e=.1$. Given a choice of boundary conditions $\a_f (\b)$, the allowed black hole solutions are simply given by the points where the black hole curves intersect the boundary condition curve: $\a_{bh}(R_e,\b) = \a_f (\b)$. One sees that for all values of $f>0$ there is precisely one hairy black hole solution of size $R_e=0.1$, and the same is true for different radii $R_e$. For $\alpha_f$ boundary conditions the mass of the hairy black holes is given by
\be \label{mass5dhair}
M_{h} =Q[\partial_{t}]=2\pi^2 R_{AdS}^2 \left(\frac{3 }{ 2}
M_0+\b^2\left(1-\frac{1}{2}f \right)\right),
\ee
where $M_0$ is the coefficient of the $O(1/r^6)$ term in the asymptotic expansion of the $g_{rr}$ component of the metric.

\begin{figure}
\begin{center}
\epsfig{file=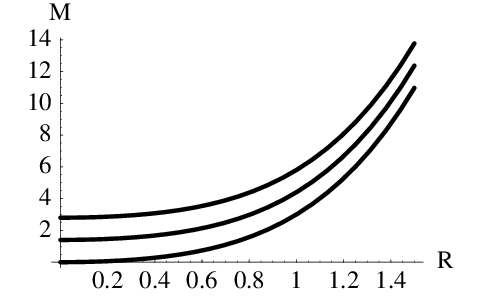, width=12cm}
\end{center}
\caption{The mass $M$ of static spherical hairy black holes (in units of $R_{AdS}^2$) as a function of horizon size $R$ (in units of $R_{AdS})$, for two different choices of $\a_f$ boundary conditions and for Schwarzschild-AdS. From top to bottom, the curves correspond to $f=0.05$, $f=0.1$ and Schwarzschild-AdS. One sees that all hairy black holes 
have positive mass.}
\label{bhmass}
\end{figure}

This is shown in Figure~\ref{bhmass}, as a function of horizon 
size $R_e$, for two different choices of $\alpha_f$ boundary conditions ($f=0.05$ (top), and $f=0.1$ (middle)) and for Schwarzschild-AdS. One sees that the scalar hair increases the mass; the hairy black holes are more massive than a Schwarzschild-AdS black holes of the same size. In particular, one sees that $M_h >0$ for all $R_e$. For large $R_e$ one has $E_h \sim R_{AdS}^2 R_e^4$, with $M_{h}/M_{s} >1$ for all $R_e$ and $M_{h}/M_{s} \rightarrow 1$ for large $R_e$. 
Furthermore, the mass (for given $R_e$) increases for decreasing $f$. 

The dual interpretation of the increase in mass relative to vacuum black holes is that whereas Schwarzschild-AdS black holes correspond to thermal states in the usual vacuum of the dual field theory, hairy black holes are described by typical excitations around the {\it local maximum} of the effective potential of the dual theory in the presence of the negative double-trace deformation \cite{Hertog:2005hu}
(see Fig \ref{dual} (left)). An important consequence is that the singularity that develops from initial data defined by the instanton cannot be hidden behind an event horizon, since there is no candidate black hole with the required mass. Instead the singularity probably extends all the way to the boundary, cutting of all space.

By contrast there do exist negative mass black holes for bulk boundary conditions that correspond to a deformation of the dual field theory that yields an effective potential with a global negative minimum, such as shown in Figure~\ref{dual} (right). These negative mass hairy black holes have a natural interpretation in the dual theory as typical excitations above the global vacuum state  \cite{Hertog:2005hu}.

For example, in the case at hand, for boundary conditions $\alpha_{f,\epsilon} = f\beta-\epsilon \beta^3$ with $\epsilon$ sufficiently small, the boundary condition curve $\alpha_{f,\epsilon}$ will generally have two intersection points with the black hole curves $\alpha_{bh}(R_e)$ of Figure~\ref{bhalpha}. The first set of intersection points specifies a branch of hairy black holes with rather small $\beta$ and hence with not much hair. Their mass is positive and they correspond to excitations around the local maximum of the effective potential in the dual theory, as before. On the other hand the solutions with the larger $\beta$, which are associated with the second intersection point, have much smaller mass and provided $\epsilon$ is sufficiently small a subset of these have negative mass. For example, for $f=1$ and $\epsilon=.05$ the theory admits an $M \approx -3 R_{AdS}^2$ black hole  of horizon size $R_e=0.1$ and with $\phi_e = 1.517$.  

The existence of an additional branch of hairy black holes - some with negative mass - for $\alpha_{f,\epsilon}$ boundary conditions also means that in this case there does exist a hairy black hole with the same mass as the instanton, $M \approx -\epsilon R_{AdS}^2$. This is therefore the natural end state of evolution of the initial data defined by the instanton, under the modified $\alpha_{f,\epsilon}$ boundary conditions.
In general, regularization of th
e dual theory that makes the effective potential bounded from below encloses the singularity in the bulk inside a horizon. Furthermore if one considers a series of dual theories where for decreasing values of $\epsilon$ one finds the horizon size of the black hole of the same mass as the instanton increases. In the limit $\epsilon \rightarrow 0$ the black hole becomes infinitely large and we recover the cosmological solutions we studied here.

\section{Renormalization of the Boundary Theory}
\label{renormv}

In this appendix, we discuss ${\cal N}=4$ super-Yang-Mills theory with gauge group $SU(N)$,
\be\label{SYM}
S_0=\int d^4x {\rm Tr}\left\{-{1\over4}F_{\mu\nu}F^{\mu\nu}-{1\over2}D_\mu\Phi^i D^\mu\Phi^i+{1\over4}g^2[\Phi^i,\Phi^j]\,[\Phi_i,\Phi_j]+\ {\rm fermions}\right\},
\ee
deformed by the double trace potential 
\be\label{double}
W=-{f\over2}\int d^4x\,{\cal O}^2,
\ee
see \eq{Ocubed}. In our conventions,
\be
F_{\mu\nu}=\partial_\mu A_\nu-\partial_\nu A_\mu+ig[A_\mu,A_\nu],\ \ D_\mu\Phi^i=\partial_\mu\Phi^i+ig[A_\mu,\Phi^i]. 
\ee
The operator ${\cal O}$ was defined in \eq{operator}: 
\be \label{operatorbis}
{\cal O}={a\over N}\,\Tr\left[\Phi_1^2 - {1\over5} \sum_{i=2}^6 \Phi_{i}^2\right].
\ee

As explained in \cite{Witten:2001ua} and reviewed in Section~3, the computation of amplitudes at order $f^2$ involves matrix elements of
\be
{f^2\over8}\int d^4x d^4y {\cal O}^2(x){\cal O}^2(y).
\ee
From conformal invariance, $\langle{\cal O}(x){\cal O}(y)\rangle=v/|x-y|^4$ on flat $\Rbar^4$. This leads to a short distance divergence that renormalizes $f$ and survives in the large $N$ limit (where the 't~Hooft coupling $g_t\equiv g^2N$ as well as $f$ are kept fixed):
\be
{f^2\over2}\int d^4x d^4y {\cal O}(x){\cal O}(y)\langle{\cal O}(x){\cal O}(y)\rangle\sim \pi^2 f^2  \ln\Lambda \int d^4x {\cal O}^2(x),
\ee
with $\Lambda$ an ultraviolet cutoff. This leads to a one-loop beta function for $f$, which does not receive higher loop corrections in the large $N$ limit \cite{Witten:2001ua}. 

\begin{figure}
\begin{center}
\epsfig{file=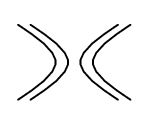, width=5cm}
\end{center}
\caption{The $({\rm Tr}\Phi^2)^2$ vertex in double line notation.}\label{figVertex}
\end{figure}

\begin{figure}
\begin{center}
\epsfig{file=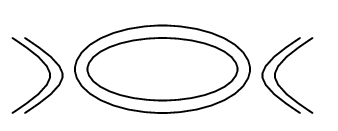, width=5cm}
\end{center}
\caption{The one-loop diagram that renormalizes the coupling $f$ at large $N$.}\label{figOneLeading}
\end{figure}

\begin{figure}
\begin{center}
\epsfig{file=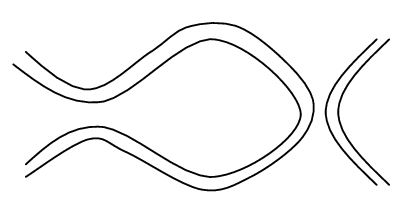, width=5cm}
\end{center}
\caption{A diagram that does not survive the large $N$ limit.}\label{figOneNonleading}
\end{figure}

\begin{figure}
\begin{center}
\epsfig{file=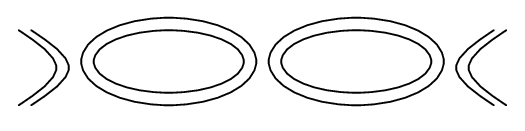, width=10cm}
\end{center}
\caption{Higher order diagrams that survives the large $N$ limit are factorizable.}\label{figHigher}
\end{figure}

Diagrammatically, we represent the $({\rm Tr}\Phi^2)^2$ vertex in double line notation as in \fig{figVertex}. 
The diagram in \fig{figOneLeading} 
then leads to the renormalization discussed above. To see that it survives the large $N$ limit, note that both vertices come with a factor $f/N^2$, while the two index loops each contribute a factor of $N$, so that the one-loop diagram in \fig{figOneLeading} is of the same order in $N$ as the vertex \fig{figVertex} (if $f$ if kept fixed in the large $N$ limit). This also makes it clear that diagrams such as \fig{figOneNonleading} 
do not survive the large $N$ limit: there are no index loops that can provide factors of $N$. By the same token, the higher-loop diagrams that survive the large $N$ limit are those like the one in \fig{figHigher}, 
which are factorizable.

\begin{figure}
\begin{center}
\epsfig{file=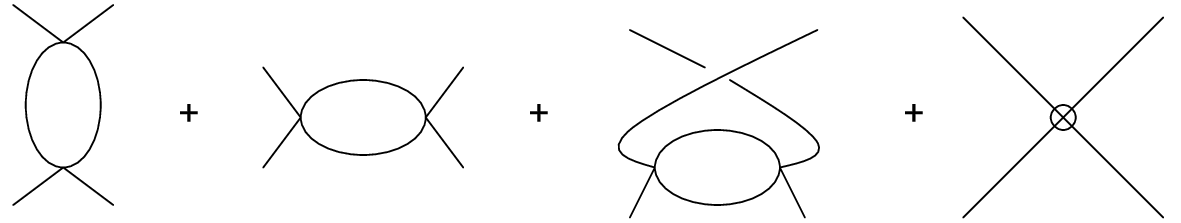, width=10cm}
\end{center}
\caption{One-loop diagrams and a counterterm for a single scalar field.}\label{figOneSingle}
\end{figure}

To verify the above one-loop result and to show the absence of higher-loop corrections to the beta function for $f$, we wish to compute the diagrams that survive the large $N$ limit. Before doing so, let us briefly review the one-loop renormalization of the theory of a single (massive or massless) scalar field with a $-\lambda\phi^4/4$ potential. We use renormalized perturbation theory (see, for instance, \cite{Peskin:1995ev}):
\be
{\cal L}=-{1\over2}(\partial_\mu\phi)^2-\half m^2\phi^2+{\lambda\over4}\phi^4-\half\delta_Z(\partial_\mu\phi)^2-\half\delta_m\phi^2+{\delta_\lambda\over 4}\phi^4,
\ee
where all the fields and parameters are renormalized ones.  To one-loop order, the two to two scattering amplitude gets contributions from the diagrams in \fig{figOneSingle}. 
The result is
\be
i{\cal M}(p_1p_2\rightarrow p_3p_4)=6i\lambda+(6i\lambda)^2\left[iV(s)+iV(t)+iV(u)\right]+6i\delta_\lambda^{(1)},
\ee
where $s,t$ and $u$ are Mandelstam variables,
\be
iV(p^2)=\half\int{d^4k\over(2\pi)^4}{-i\over k^2+m^2}{-i\over (k+p)^2+m^2},
\ee
and $\delta_\lambda^{(1)}$ is the order $\lambda$ contribution to $\delta_\lambda$. We impose the renormalization condition
\be
i{\cal M}(p_1p_2\rightarrow p_3p_4)=6i\lambda\ \ \ {\rm at}\ \ \ (p_1+p_2)^2=(p_1+p_3)^2=(p_1+p_4)^2=\mu^2.
\ee
This implies
\be
\delta_\lambda^{(1)}=18\lambda^2V(\mu^2)=-{9\lambda^2\over16\pi^2}\left({2\over\epsilon}-\ln \mu^2+\ {\rm finite}\ \right),
\ee
where in the last equality we used dimensional regularization. To identify the beta function, consider the Callan-Symanzik equation
\be
\left[\mu{\partial\over\partial\mu}+\beta(\lambda){\partial\over\partial\lambda}+n\gamma(\lambda)\right]\,G^{(n)}(\{x_i\};\mu,\lambda)=0,
\ee
where $\mu$ is the renormalization scale and $G^{(n)}(x_1,\ldots,x_n)$ is the connected $n$-point function computed in renormalized perturbation theory. At one-loop order, $\gamma(\lambda)=0$. For the 4-point function,
\be
G^{(4)}(p_1,p_2,p_3,p_4)=i{\cal M}(p_1p_2\rightarrow p_3p_4)\prod_{j=1}^4{-i\over p_j^2},
\ee
we find
\be
\mu{\partial\over\partial\mu}G^{(4)}={27i\lambda^2\over4\pi^2}\prod_{j=1}^4{-i\over p_j^2}.
\ee
Therefore the Callan-Symanzik equation is satisfied to order $\lambda^2$ if
\be
\beta(\lambda)=-{9\lambda^2\over8\pi^2}+O(\lambda^3),
\ee
which is the familiar expression for the one-loop beta function.

\begin{figure}
\begin{center}
\epsfig{file=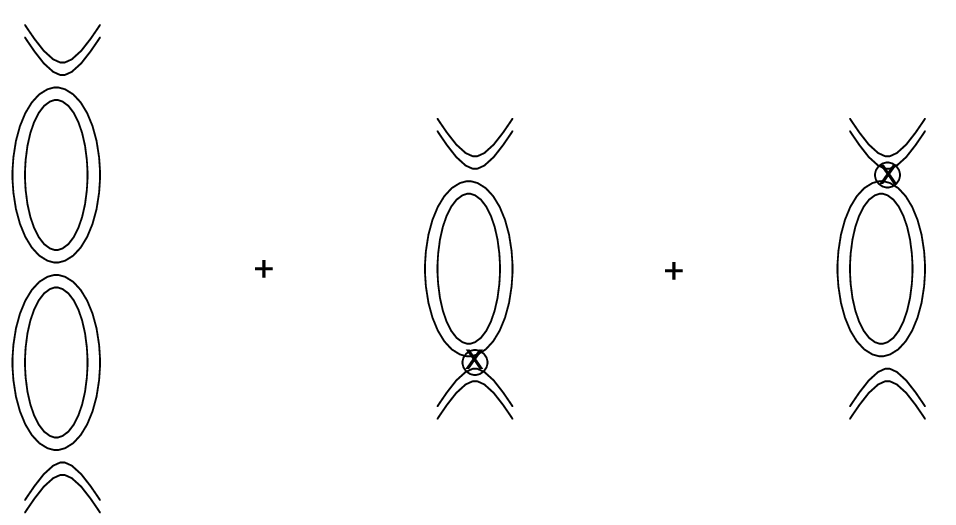, width=10cm}
\end{center}
\caption{Two-loop diagrams in the large $N$ limit.}
\label{figTwoLoop}
\end{figure}

In the analogous computation for the renormalization of the double trace deformation, only the ``s-channel'' part of the above computation survives in the large $N$ limit, as we have argued before. We write
\be
\delta_\lambda^{(1)}=\delta_\lambda^{(1)}(s)+\delta_\lambda^{(1)}(t)+\delta_\lambda^{(1)}(u)
\ee
with
\be
\delta_\lambda^{(1)}(s)=6\lambda^2V(\mu^2)=-{3\lambda^2\over16\pi^2}\left({2\over\epsilon}-\ln \mu^2+\ {\rm finite}\ \right).
\ee
Also at two-loop order, only ``s-channel'' diagrams will be important to leading order in $1/N$, as we have seen before; the relevant diagrams are in \fig{figTwoLoop}, 
where the insertion of a cross denotes the one-loop counterterm $\delta_\lambda^{(1)}(s)$. We find (suppressing external propagators)
\be
G^{(4)}=6i\lambda+(6i\lambda)^2\left[iV(p^2)-iV(\mu^2)\right]-(6i\lambda)^3\left[V(p^2)-V(\mu^2)\right]^2+\ldots,
\ee
where $p=p_1+p_2$. The various $V(\mu^2)$ come from one-loop counterterms; a two-loop counterterm has merely cancelled a momentum-independent divergence $(6i\lambda)^3[V(\mu^2)]^2$. Computing
\be
\mu{\partial\over\partial \mu}G^{(4)}={9i\lambda^2\over4\pi^2}-{27i\lambda^3\over \pi^2}\left[V(p^2)-V(\mu^2)\right]+\ldots
\ee
and
\be
{\partial\over\partial\lambda}G^{(4)}=6i-72i\lambda\left[V(p^2)-V(\mu^2)\right]+\ldots,
\ee
it is easy to see that the Callan-Symanzik equation is satisfied for
\be
\beta(\lambda)=-{3\lambda^2\over8\pi^2}.
\ee
It is easy to check that this remains the case at higher order in $\lambda$. Thus we have verified that the beta function is one-loop exact.

We will be interested in an approximation of the quantum effective action that is valid for a large range of field values, in particular for large field values. An appropriate framework is that of \cite{Coleman:1973jx}, where the standard Feynman diagram expansion is resummed and the theory is organized in a derivative expansion,
\be\label{gamma_exp}
\Gamma=\int d^4x [-V(\phi)-\half \partial_\mu\phi\,\partial^\mu\phi\,Z(\phi)+\ldots],
\ee 
and an expansion in the number of loops. First, we focus on the effective potential, {\it i.e.}, (minus) the term without derivatives, in the one-loop approximation.

Starting from the classical potential
\be
V_{\rm class}=-{\lambda\over 4}\phi^4,
\ee
one computes \cite{Coleman:1973jx}
\be
V_{\rm 1-loop}=\half\int {d^4k\over(2\pi)^4}\ln\left(1-{3\lambda\phi^2\over k^2}\right),
\ee
where the integral is over Euclidean four-momenta. This can be rewritten in terms of $x=k^2$:
\be
V_{\rm 1-loop}={1\over32\pi^2}\int_0^\infty dx\, x\ln\left(1-{3\lambda\phi^2\over x}\right).
\ee
Since the integral diverges, we introduce a UV cutoff $x\le\Lambda^2$. For $\lambda>0$, we also need an IR cutoff $x\ge\mu_{IR}^2$ if we want a real effective potential (see \cite{Weinberg:1987vp}). Thus we compute, dropping terms that vanish as $\Lambda\to\infty$:
\bea
\int_{\mu_{IR}^2}^{\Lambda^2} dx\, x\ln\left(1-{3\lambda\phi^2\over x}\right)
&=&-{3\lambda\Lambda^2\phi^2\over2}-{9\lambda^2\phi^4\over4}-{\mu_{IR}^4\over2}\ln\left(1-{3\lambda\phi^2\over\mu_{IR}^2}\right)\nonumber\\
&&-{3\lambda\phi^2\over2}(\Lambda^2-\mu_{IR}^2)-{9\lambda^2\phi^4\over2}\ln\left({\Lambda^2\over\mu_{IR}^2-3\lambda\phi^2}\right).
\eea
For $\lambda<0$, {\it i.e.}, a positive quartic potential, we can take $\mu_{IR}=0$ and recover the standard result \cite{Coleman:1973jx}
\be
V_{\rm 1-loop}={1\over32\pi^2}\left\{-3\lambda\Lambda^2\phi^2-{9\lambda^2\phi^4\over4}+{9\lambda^2\phi^4\over2}\ln\left({-3\lambda\phi^2\over\Lambda^2}\right)\right\}.
\ee
For $\lambda>0$, the case of interest in the present paper, choosing $\mu_{IR}=0$ would give a complex effective potential, signaling the fact that localized wavefunctions are unstable to spreading while keeping the expectation value of $\phi$ fixed \cite{Weinberg:1987vp}. Therefore we choose instead to integrate out only the ``unfrozen'' (or ``non-tachyonic'') modes; thus we choose 
\be
\mu_{IR}^2=3\lambda\phi^2+\epsilon^2
\ee
and let $\epsilon\to0$ at the end of the computation. This results in
\be
V_{\rm 1-loop}={1\over32\pi^2}\left\{
-3\lambda\Lambda^2\phi^2+{9\lambda^2\phi^4\over4}+{9\lambda^2\phi^4\over2}\ln\left({3\lambda\phi^2\over\Lambda^2}\right)
\right\}.
\ee
Now we add counterterms, imposing that the renormalized mass should vanish and defining the renormalized coupling $\lambda_\mu$ by the renormalization condition
\be
V(\mu)=-{\lambda_\mu\over 4}\,\mu^4.
\ee
This corresponds to choosing for the sliding scale $\mu$ the value of a spacetime-independent external scalar field $\phi$ rather than, say, the momentum of an external line. This has the virtue that no large logarithms will interfere with perturbation theory, which will thus be reliable as long as the renormalized coupling $\lambda_\mu$ is small. The result is
\be\label{pot_mu}
V(\phi)=-{\lambda_\mu\over 4}\phi^4+{9\lambda_\mu^2\phi^4\ln(\phi^2/\mu^2)\over 64\pi^2}.
\ee
The renormalization group equation can be obtained by demanding that $V(\phi)$ be independent of $\mu$:
\be\label{eqren}
\mu{d\lambda_\mu\over d\mu}=-{9\lambda_\mu^2\over 8\pi^2}.
\ee
Here, we have ignored a contribution from $d/d\mu$ hitting the $\lambda_\mu^2$ in the second term of \eq{pot_mu}, which is justified as long as $|\lambda_\mu\ln(\phi/\mu)|\ll 1$. 
Equation \eq{eqren} is solved by
\be\label{runningcoupling}
\lambda_\mu={16\pi^2\over9\ln(\mu^2/M^2)},
\ee
with $M$ an arbitrary scale (this implements dimensional transmutation). Choosing $\mu=\phi$, {\it i.e.}, the renormalization scale is set by the value of the field $\phi$, the Coleman-Weinberg potential can then be written as
\be\label{CWphi1}
V(\phi)=-{4\pi^2\phi^4\over 9\ln(\phi^2/M^2)}.
\ee
Now suppose that for some value $\phi_0$, the coupling is small,
\be
0<\lambda_{\phi_0}\ll 1,
\ee
then $M<|\phi_0|$ and \eq{CWphi1} is trustworthy ({\it i.e.}, higher order corrections can be ignored) for any $\phi$ such that $|\phi|>|\phi_0|$. As a result, we can conclude that in this case 
\be
V(\phi)\rightarrow -\infty\ \ \ {\rm for}\ \ |\phi|\rightarrow\infty.
\ee
(This also holds for the massive theory, as long as $m\ll|\phi_0|$.) This analysis was for a single scalar field; as can be seen from the above discussion, it extends to the large $N$ adjoint theory with only very minor changes.

As we have seen, the one-loop diagram \fig{figOneLeading} involving two double trace vertices gives an order $f^2$ contribution to the beta function for the coupling $f$, leading to the effective potential \eq{effpot}. One can ask whether one-loop diagrams with one double trace vertex and one commutator squared vertex (see \eq{SYM}) could give contributions of order $fg^2$ to the beta function for $f$ and thus to the effective potential, which would be important at small $f$. The answer is that the $fg^2$ contributions to the effective potential cancel. To see this, let us compute the one-loop effective potential for $\phi$, where $\Phi^1(x)=\phi(x)U$ with ${\rm Tr}U^2=1$, in our deformed ${\cal N}=4$ SYM theory, see \eq{Ocubed} with \eq{SYM} and \eq{double}. We choose a constant background $\Phi^1=\phi U$ and compute the masses $M(\phi)$ of the various modes in this background; this involves fixing a gauge and introducing ghosts, see for instance \cite{ball}. Up to terms that can be absorbed in counterterms, every bosonic mode contributes
\be
{1\over 32\pi^2}\,{1\over2} M^4\ln\left({M^2\over\Lambda^2}\right)
\ee  
to the effective potential, where $\Lambda$ is a UV cutoff, while every fermionic mode contributes with a minus sign. In the undeformed ${\cal N}=4$ SYM theory, these contributions cancel exactly between bosons and fermions. Our double trace deformation changes the scalar masses but not the masses of gauge bosons and fermions, leading to a non-trivial effective potential. It is easy to see that the scalars $\Phi^i_{ab}$ have
\be
(M^i_{ab})^2=g^2\phi^2(U_{aa}-U_{bb})^2+{2fa^2\phi^2\over5N^2},
\ee
while after gauge fixing as in Ref.~\cite{ball}, p. 122, the scalars $\Phi^1_{ab}$ have
\be
(M^1_{ab})^2=g^2\phi^2(U_{aa}-U_{bb})^2-{2fa^2\phi^2\over N^2}(1+2\delta_{ab}U^2_{aa}).
\ee
The contributions of order $g^4$ are those of the undeformed SYM theory and cancel by supersymmetry against those of gauge fields and fermions. Contributions of order $fg^2$ would have to come from off-diagonal scalars ($a\neq b$), but it is easy to see that they cancel between $\Phi^1$ and $\Phi^i$ ($i=2,\ldots 6$), at least up to terms that can be absorbed in the counterterms. So we have verified that the one-loop contribution to the effective potential is at least of order $f^2$.

Since we will be interested in non-static solutions, knowledge of the potential alone is insufficient for our purposes: we also need to know the form of the derivative terms in the quantum effective action. Terms arising from non-divergent Feynman diagrams are suppressed by powers of the effective coupling $\lambda_\phi$, which is small in the regime where we trust our one-loop effective potential. It will turn out that such terms will be small for the configurations of interest. Logarithmically divergent Feynman diagrams induce, upon renormalization, terms involving $\ln(\phi/\mu)$, with $\mu$ the renormalization scale. Such terms disappear when we choose $\mu=\phi$, as we have done for the effective potential. From power counting, the only possible divergences involving momenta arise in graphs with two external legs, which lead to wave function renormalization (as it turns out, such divergences do not occur at one-loop order in this theory, but do occur at higher loop order). We choose the renormalization condition $Z(\mu)=1$ in \eq{gamma_exp}, which for our choice $\mu=\phi$ implies that the field $\phi$ has a canonical kinetic term. Strictly speaking, the last conclusion holds for the theory of a single scalar field; in the adjoint model, the standard kinetic term $-{\rm Tr}\partial_\mu\Phi\,\partial^\mu\Phi$ may be accompanied by terms with a different index structure, e.g.\ $-({\rm Tr}\Phi\partial_\mu\Phi)( {\rm Tr}\Phi\partial^\mu\Phi)/{\rm Tr}\Phi^2$, which will, however, be suppressed by powers of the effective coupling (these terms do not arise form ultraviolet divergent Feynman diagrams, but from resumming diagrams with more than two external legs).

\setcounter{equation}{0}
\section{Stress-Energy Correlators}\label{subsec:stress}

Although we have concluded that a bounce seems unlikely in the specific model under study, we now study quantities that would have been important if backreaction had been small, or more to the point, that would be important in related models with a bounce.

In addition to computing the quantum creation of particles, we would like to calculate the two-point correlation function of the stress-energy tensor for the scalar field fluctuations as a function of time. There is no gravity (and hence no backreaction) in our system, but stress-energy tensor correlators of the boundary CFT determine the metric perturbations in the bulk, {\it i.e.}, the cosmological spacetime, and are as such related to cosmological perturbations. In fact, the relation between the stress-energy tensor and the bulk metric is completely analogous to the relation between the operator ${\cal O}$ and the bulk scalar field $\phi$ in Section~\ref{sec:boundary}.

We shall consider the conformally improved stress tensor\cite{Coleman:1970je}, consisting of the canonical stress tensor plus an identically conserved quantity which has been added to make the stress tensor traceless in the limit when the theory is conformally invariant. In our case, this limit is where the logarithm is large so the running of $\lambda$ may be neglected. In four spacetime dimensions, the conformally improved stress tensor is
\be 
\label{energydensity}
T_{\mu \nu} = \left( \partial_\mu \phi \partial_\nu \phi - g_{\mu \nu} ({1\over 2} (\partial \phi )^2 - {1\over 4} \lambda) \right) -{1\over 6} \left( \nabla_\mu \partial_\nu -g_{\mu \nu} \partial^2 \right) \phi^2.
\ee
We shall calculate the correlators of $T_{\mu \nu}$ to linear order in the field fluctuation $\delta \phi$ about the background solution. We shall treat the latter as a classical background, although strictly speaking, as we detailed in Subsection~\ref{shortt}, the expectation value of the homogeneous background mode $\overline{\phi}$ does not exist. Nevertheless, at times well after the bounce, in the parameter regime where backreaction is small, we expect the description of $\overline{\phi}$ in terms of a localized wavepacket rolling back up the hill is accurate, so we can replace $\overline{\phi}$ with the appropriate classical solution and take expectation values of the inhomogeneous fluctuation modes $\delta \phi({\bf x})$ using the Schr\"odinger wavefunction, calculated above. As we shall discuss below, these linearized correlators are in principle sufficient to determine the correlators of the linearized metric perturbations in the bulk. 

Denoting the linearized perturbation of the stress tensor as  $\delta T_{\mu \nu}$, the linearized components at given wavenumber ${\bf k}$ are:
\ba
\delta T_{00} &=& \dot{\phi}^2 {d \over dt} \left({\delta \phi \over \dot{\phi}}\right) +{1\over 3} \phi k^2 \delta \phi \cr
\delta T_{0i} &=& i k_i \left( {2\over 3} \dot{\phi} \delta \phi -{1\over 3} \phi \dot{\delta \phi} \right)\cr
\delta T_{ij} &=& \delta_{ij}\left({1\over 3} \dot{\phi} \dot{\delta \phi} + {2\over 3} \ddot{\phi} \delta \phi -{1\over 3} \phi \ddot{\delta \phi} -{1\over 3} \phi k^2 \delta \phi \right) +{1\over 3} \phi k_i k_j \delta \phi.
\label{stren}
\ea
We want to compute the fluctuations in the stress tensor in the incoming adiabatic vacuum state for the fluctuations, denoted $|0,{\rm in}\rangle$, relative to the vacuum defined by the outgoing adiabatic vacuum state $|0,{\rm out}\rangle$. Equivalently, we compute the correlators of $T_{\mu \nu}$ normal ordered with respect to $|0,{\rm out}\rangle$. To compute the correlators, we first express the scalar field, and the linearized stress tensor, in terms of the adiabatic positive and negative frequency modes, and creation and annihilation operators, appropriate to $|0,{\rm out}\rangle$. We then normal order the expression and calculate the expectation value in $|0,{\rm in}\rangle$. This requires expressing the ``out" creation and destruction operators in terms of ``in" creation and destruction operators: 
\be
a_{\rm out}= \alpha a_{\rm in} + \beta^* a_{\rm in}^\dagger, \qquad a_{\rm out}^\dagger= \alpha^* a_{\rm in}^\dagger + \beta a_{\rm in}.
\label{cdops}
\ee
Consider some quantity linear in the quantum field fluctuation and its time derivatives: it may be expressed as 
\be
f(t,{\bf x}) = \sum_{\bf k} {1\over \sqrt{2 k V}} \left( a_{{\bf k},{\rm out}} \chi^{(+)}_{\rm out} e^{i {\bf k} \cdot {\bf x} } + a_{{\bf k},{\rm out}}^\dagger \chi^{(-)}_{\rm out} e^{-i {\bf k} \cdot {\bf x} }\right),
\label{expans}
\ee
where $\chi^{(+,-)}_{\rm out}$ are the positive and negative frequency parts with respect to the out vacuum. We normalize the creation and destruction operators so $\left[a_{\bf k}, a^\dagger_{{\bf k}'}\right] = \delta_{{\bf k,k}'}$ etc. We define the vacuum-subtracted expectation of some operator ${\cal O}$ as
\be
\langle {\cal O} \rangle \equiv \langle 0,{\rm in}| {\cal O} |0,{\rm in}\rangle  
- \langle 0,{\rm out}| {\cal O} | 0,{\rm out}\rangle.
\label{exdef}
\ee
Substituting (\ref{cdops}) into (\ref{expans}) and (\ref{exdef}), using the commutation relations for $a_{{\bf k}}$ and $a^\dagger_{{\bf k}}$ and their action on the ``in" and ``out" vacua, and replacing $\sum_{\bf k}$ with $V \int d^3 {\bf k}/(2 \pi)^3$, we obtain the two-point correlator
\be
\langle f(t,{\bf x}) f(t,{\bf 0}) \rangle = \int {d^3 {\bf k}\over (2 \pi)^3} {1\over 2 k} e^{i {\bf k} \cdot {\bf x} } \left(2 |\beta|^2 |\chi^{(+)}_{\rm out}|^2
+ \alpha^*\beta (\chi^{(-)}_{\rm out})^2 +\alpha \beta^* (\chi^{(+)}_{\rm out})^2 \right),
\label{fcorr} 
\ee
where we used $|\alpha|^2-|\beta^2|=1$. In the regime of interest, where $\beta$ is small, we can replace $\alpha$ with $-1$. 

Since we only want the leading order result, we can use the positive frequency mode functions to zeroth order, namely
\be 
\delta \phi^{(+)} = e^{-ikt} \left(1-{3 i\over kt} - {3\over (kt)^2} \right),
\label{posfreq}
\ee
which leads to the following positive frequency contributions to $\delta T_{\mu \nu}$: 
\ba
\sqrt{\lambda_\phi\over 2}\,\delta T_{00}^{(+)} &=&  {k^2 \over 3 t} e^{-ikt}\cr
\sqrt{\lambda_\phi\over 2}\,\delta T_{0i}^{(+)} &=& i k_i {1+ikt \over 3 t^2} 
e^{-ikt}\cr
\sqrt{\lambda_\phi\over 2}\,\delta T_{ij}^{(+)} &=& \delta_{ij}{1+ikt \over 3 t^3} +k_i k_j {1\over 3t} \left( 1- {3 i \over kt} - {3 \over (kt)^2} \right)e^{-ikt} .
\label{strenpos}
\ea
Notice that the spatial trace $\delta T_{ii}^{(+)} = \delta T_{00}^{(+)}$ as it has to be from the tracelessness of the stress tensor. Hence it is only the traceless part $\tilde{\delta T}_{ij}^{(+)} \equiv \delta T_{ij}^{(+)}-{1\over 3} \delta_{ij} \delta T_{kk}^{(+)}$ which needs to be considered:
\be 
\sqrt{\lambda_\phi\over 2}\tilde{\delta T}_{ij}^{(+)} = \left( k_i k_j - {1\over 3} k^2 \delta_{ij}\right){1\over 3t} \left( 1- {3 i \over kt} - {3 \over (kt)^2} \right)e^{-ikt}. 
\label{strentrless}
\ee

Equations (\ref{strenpos}) and (\ref{fcorr}) determine the two point correlation function (at equal times) of the stress-tensor in the boundary theory. Here we shall present only a partial discussion of the equal time correlators, postponing a fuller discussion to future work in which we attempt to use the boundary correlations to determine the fluctuations in the bulk.  

The first term in (\ref{fcorr}) provides the simplest contribution. In order to compute it, our strategy is to replace factors of $k^2$ under the integral by minus the spatial Laplacian outside the integral. We then perform the angular integrals in (\ref{fcorr}) to obtain a factor of $4 \pi ({\rm sin}kr)/kr$. The two powers of $k$ in the denominator then cancel the $k^2$ factor in the measure, leaving us with $r^{-1} \int_0^\infty dk {\rm sin} kr$.  When performing the integral over $|\beta|^2$, we replace $k$ with $r^{-1}$, the value of $k$ dominating the integral. For example, $\int_0^\infty dk ({\rm sin} kr)/\left({\rm ln}(k/M)\right)^2 \approx r^{-1}/\left({\rm ln}(1/Mr)\right)^2$.  These manipulations are sufficient to compute the first term in (\ref{fcorr}). The second and third terms involve additional oscillatory factors of $e^{\pm 2ikt}$ under the integral, which generates additional structure on the causal scale $r=2 t$, but which suppresses them for $r\ll t$. 

In this way, we obtain the equal time correlators of various scalar quantities constructed from the stress-tensor, for example we exhibit the detailed form for $r\ll t$,
\ba
&&\langle \delta T_{00} (r,t) \delta T_{00}(0,t) \rangle =   
{ {\rm ln}(1/Mt)\over \lambda_0\pi^2({\rm ln}Mr)^2} {2\over 3 r^6 t^2} \cr
&&\langle \partial_i\delta T_{0i} (r,t) \partial_i \delta T_{0i}(0,t) \rangle = 
{ {\rm ln}(1/Mt)\over \lambda_0 \pi^2({\rm ln}Mr)^2} \left({20 \over r^8 t^2}- {2\over 3 r^6 t^4}\right) \cr
&& \langle \partial_i\partial_j\tilde{\delta T}_{ij}(r,t) \partial_k\partial_l \tilde{\delta T}_{kl}(0,t) \rangle
={ {\rm ln}(1/Mt)\over \lambda_0 \pi^2({\rm ln}Mr)^2} \left({4480 \over 9 r^{10} t^2}+ {80\over 3 r^8 t^4}+{8\over 3 r^6 t^4} \right) 
\label{too}
\ea

The key point is that {\it all of these correlators are nearly scale-invariant} in form, that is they are all of the form $t^{-n} g(r/t)$, where the overall power of $t$ is determined on dimensional grounds. The only deviation from perfect scale-invariance is due to the logarithmic running of the coupling with scale. This behavior is the result of the classical scale-invariance, and the modest quantum breaking of scale-invariance, in the boundary conformal field theory.

Although the background energy density is zero, we can define a dimensionless perturbation by comparing the fluctuation $\delta T_{00}=\delta \rho$ with the background quantity $\rho+P = \dot{\phi}^2$. We find for example
\be 
\langle {\delta \rho (r,t) \over P+\rho} {\delta \rho (0,t) \over P+\rho}\rangle \sim {1 \over N^2 {\rm ln}^2 (1/Mr) {\rm ln}(1/Mt)} {t^6 \over r^6},
\ee
a slightly red spectrum, with the reddening arising from the running coupling as before. Furthermore, we note that the overall amplitude of the fractional energy density perturbation is proportional to $1/l^3$, where $1/l$ is the appropriate value of the running coupling, which has been our small parameter all along. To summarize, the fluctuations in $\phi$ give rise to stress-energy perturbations that are naturally small, being suppressed by powers of $1/N$ and $1/l$, approximately scale-invariant, with a slightly red tilt due to asymptotic freedom, scalar, adiabatic and nearly Gaussian in character, to leading order in $1/l$ and $1/N$.

\begin{figure}
\begin{center}
\epsfig{file=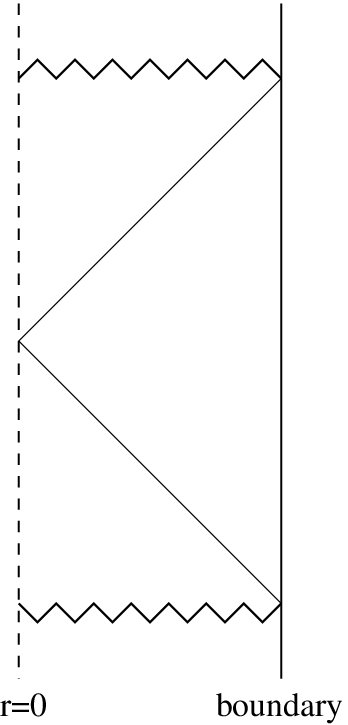, width=6cm}
\end{center}
\caption{The stress-energy correlators on the boundary may predict the spectrum of classical cosmological fluctuations on a spacelike surface in the bulk at late times.}
\label{fluct}
\end{figure}

The quantities of physical interest, however, are correlators of metric fluctuations in the bulk cosmology, which couple to the stress-energy tensor of the boundary theory. We would like to compute the bulk perturbations on a spacelike surface at late times, roughly halfway between the past and future singularities in our model, as shown schematically in Figure~\ref{fluct}. In general, while knowing the state of the boundary theory is in principle equivalent to knowing the state in the bulk theory, it is in practice not straightforward to extract local information about the bulk theory from the boundary theory (see \cite{Hamilton:2006az} for a recent discussion). Our specific problem might be simpler due to the fact that the metric fluctuations behave classically in the relevant regime. Furthermore, it is sufficient to treat them in a linearized approximation.  

One idea for how one might be able to compute the spectrum of bulk metric perturbations is as follows. The state of the boundary theory is encoded in the boundary correlators, some of which we have computed in this paper. Since the relevant modes are frozen in, it may be possible to view the boundary correlators as those of a classical ensemble of fluctuating fields. Each member of the ensemble could then be viewed as an expectation value of a boundary operator, which would determine a corresponding bulk solution (see for instance \cite{Balasubramanian:1998sn}). For instance, knowing the expectation value of the stress-energy tensor on the boundary determines a solution for the bulk metric. So we would end up with a classical ensemble of bulk metric perturbations, which we could use to compute bulk correlators.

Furthermore, since we are interested in wavenumbers $k\gg R_{AdS}^{-1}$, which is the only scale that enters in the correspondence, and since scale-invariance is holographic in the sense that the fluctuations on a planar subspace of a space with scale-invariant fluctuations are scale-invariant, it seems plausible that the bulk metric perturbations will turn out to be scale-invariant too.


\end{document}